\documentclass[a4paper,fleqn]{cas-sc}
\usepackage{graphicx,bm}
\usepackage{amssymb}
\usepackage[numbers,sort&compress]{natbib}
\usepackage{subfig,float}
\usepackage{ulem,verbatim}
\usepackage[figurename=Fig.]{caption}
\usepackage{titletoc}
\definecolor{prcolor}{RGB}{0,128,172}
\hypersetup{colorlinks=true, linkcolor=prcolor, citecolor=prcolor, urlcolor=prcolor}
\def\tsc#1{\csdef{#1}{\textsc{\lowercase{#1}}\xspace}}
\tsc{WGM}
\tsc{QE}
\tsc{EP}
\tsc{PMS}
\tsc{BEC}
\tsc{DE}

\def\bx{{\bf x}}
\def\bk{{\bf k}}

\newcommand{\ket}[1]{\left| {#1} \right\rangle}
\newcommand{\bra}[1]{\left\langle {#1} \right|}
\newcommand{\expect}[1]{\left\langle {#1} \right\rangle}
\newcommand{\ketb}[1]{\left| {#1} \right\rangle_{b}}
\newcommand{\keta}[1]{\left| {#1} \right\rangle_{a}}

\ExplSyntaxOn
\keys_set:nn { stm / mktitle } {nologo}
\ExplSyntaxOff

\begin{document}
\let\WriteBookmarks\relax
\def\floatpagepagefraction{1}
\def\textpagefraction{.001}
\shorttitle{Quantum  magnonics}
\shortauthors{\textit{Yuan et~al.}}

\title [mode = title]{Quantum  magnonics: when magnon spintronics meets quantum information science}
%
%

\author[1]{H. Y. Yuan}[orcid=0000-0003-0617-9489]
\address[1]{Institute for Theoretical Physics, Utrecht University, 3584 CC Utrecht, The Netherlands}
\ead{huaiyangyuan@gmail.com}

\author[2]{Yunshan Cao}[orcid=0000-0002-3409-2578]
\address[2]{School of Electronic Science and Engineering and State Key Laboratory of Electronic Thin Films and Integrated Devices, University of Electronic Science and Technology of China, Chengdu 610054, China}
\ead{yunshan.cao@uestc.edu.cn}

\author[3]{Akashdeep Kamra}[orcid=0000-0003-0743-1076]

\address[3]{Condensed Matter Physics  Center (IFIMAC) and Departamento de F\'{i}sica Te\'{o}rica de la Materia Condensada, Universidad Aut\'{o}noma de Madrid, E-28049 Madrid, Spain}
\ead{akashdeep.kamra@uam.es}

\author[1,4,5]{Rembert A. Duine}[]
\address[4]{Department of Applied Physics, Eindhoven University of Technology, P.O. Box 513, 5600 MB Eindhoven, The Netherlands}
\address[5]{Center for Quantum Spintronics, Department of Physics, Norwegian University of Science and Technology, NO-7491 Trondheim, Norway}

\ead{r.a.duine@uu.nl}

\author[2]{Peng Yan}[orcid=0000-0001-6369-2882]
\ead{yan@uestc.edu.cn}


\begin{abstract}
Spintronics and quantum information science are two promising candidates for innovating information processing technologies. The combination of these two fields enables us to build solid-state platforms for studying quantum phenomena and for realizing multi-functional quantum tasks. For a long time, however, the intersection of these two fields was limited due to the distinct properties of the classical magnetization, that is manipulated in spintronics, and quantum bits, that are utilized in quantum information science. This situation has changed significantly over the last few years because of the remarkable progress in coding and processing information using magnons. On the other hand, significant advances in understanding the entanglement of quasi-particles and in designing high-quality qubits and photonic cavities for quantum information processing provide physical platforms to integrate magnons with quantum systems. From these endeavours, the highly interdisciplinary field of quantum magnonics emerges, which combines spintronics, quantum optics and quantum information science. Here, we give an overview of the recent developments concerning the quantum states of magnons and their hybridization with mature quantum platforms. First, we review the basic concepts of magnons and quantum entanglement and discuss the generation and manipulation of quantum states of magnons, such as single-magnon states, squeezed states and quantum many-body states including Bose-Einstein condensation and the resulting spin superfluidity. We discuss how magnonic systems can be integrated and entangled with quantum platforms including cavity photons, superconducting qubits, nitrogen-vacancy centers, and phonons for coherent information transfer and collaborative information processing. The implications of these hybrid quantum systems for non-Hermitian physics and parity-time symmetry are highlighted, together with applications in quantum memories and high-precision measurements. Finally, we present an outlook on some of the challenges and opportunities in quantum magnonics.
\end{abstract}


\maketitle

\tableofcontents

\section{Introduction}
With the rapid accumulation of information and data in science, technology, economics and social activities, finding efficient means to store and process big data is becoming important and urgent. The traditional transistor technologies that make use of the charge degree of freedom of electrons in semiconductor devices are encountering a bottleneck, and suffer from problems such as high-energy consumption, Joule heating, and uncontrollable quantum effects, when the processing units scale down to the nanoscale. To overcome these limitations and promote further developments of information processing, two promising candidates, namely spintronics and quantum information science, stand out and have kept moving forward over the last few decades. Firstly, spintronics, emerged in the late 1980s, focuses on harnessing the spin of an electron, instead of its charge exploited in conventional electronics. Information is coded in the spin orientation of an ordered magnet as well as in the magnetic excitations of the ordered states, so-called magnons. Appealing features of spintronics include the stability of a large class of room temperature magnets, the high spin density of a magnet that allows for high density of information storage, and the tunability of spin orientations by both electric current and external fields \cite{DasmaRMP2004,FertRMP2008,HirohataReview2020}. The subfield of spintronics that focuses on the generation and manipulation of magnons is called magnonics or magnon spintronics \cite{ChumakNP2015}. In principle, information coded in magnons can propagate through a magnetic insulator over long distances due to the small dissipation, which may significantly reduce the energy consumption.

On the other hand, quantum computing and quantum information science manipulate the quantum nature of spins and utilize the basic principles of quantum mechanics, such as superpositions and entanglement, as resources to realize computing and communication tasks \cite{FeynmanIJTP1982,Nielsen2000,GisinNP2007,NAPbook2019}. For example, a spin-1/2 system has two states (up and down) and can thus represent a bit which records binary information. The simultaneous and coherent manipulation of $N$ spins would increase the storage capability in an exponential way as $2^N$, and thus increase the storage limit as well as the processing speed compared with classical computers. It has been reported that a programmable 53 qubit processor could complete a series of tasks in 200 seconds that would take a classical supercomputer approximately 1000 years \cite{AruteNature2019}. The worldwide focus on, and fast developments in, quantum science and technologies, have gradually put us in the age of the so-called second quantum revolution \cite{MacReview2003}, which may potentially shape our future lives.

In the early years, magnon spintronics and quantum information science developed independently. Recently, however, these two fields started to overlap, probably for the following reasons. (i) In modern quantum technologies, it is necessary and urgent to search for various hybrid platforms to achieve complementary functionalities and multitasking capabilities such as room temperature quantum computing and solid-state-based quantum simulations. With its long lifetime and tunability as information carrier, the magnon is an ideal choice to be integrated with the known quantum platforms, e.g., cavity photons and superconducting (SC) qubits; (ii) Spintronics mostly manipulates the classical magnetization orientation at room temperature and thermal stability of magnetic elements at the nanoscale is an important challenge. At this scale, quantum effects cannot be neglected any longer, and the interplay of spintronics and quantum mechanics has to be taken into account in a consistent manner; (iii) Currently, a young branch of quantum information, called continuous variable quantum information \cite{BraunRMP2005,ReidRMP2009,WeedRMP2012,AndersenNP2015,AlexRMP2017,BarzanjehNature2019,ShlomiScience2021,ArrazolaNature2021}, has emerged, which manipulates quasi-particles, for example photons and phonons which have continuous spectrum as opposed to traditional qubits. Magnons, falling perfectly into this category, can be strongly coupled to both photons and phonons, and thus provide a fertile platform to study both fundamental quantum physics and fruitful applications such as quantum transducers, quantum memories, high precision measurements and logic gates. The interplay of magnon spintronics with quantum information science shapes the new field \textit{quantum magnonics}, which covers the overlap between traditional spintronics, magnonics, quantum optics, quantum computing and quantum information science. In this review, we first introduce the language of quantum magnonics to make this field accessible to a general readership, and subsequently review its recent developments.

\begin{figure}
  \centering
  \includegraphics[width=1.0\textwidth]{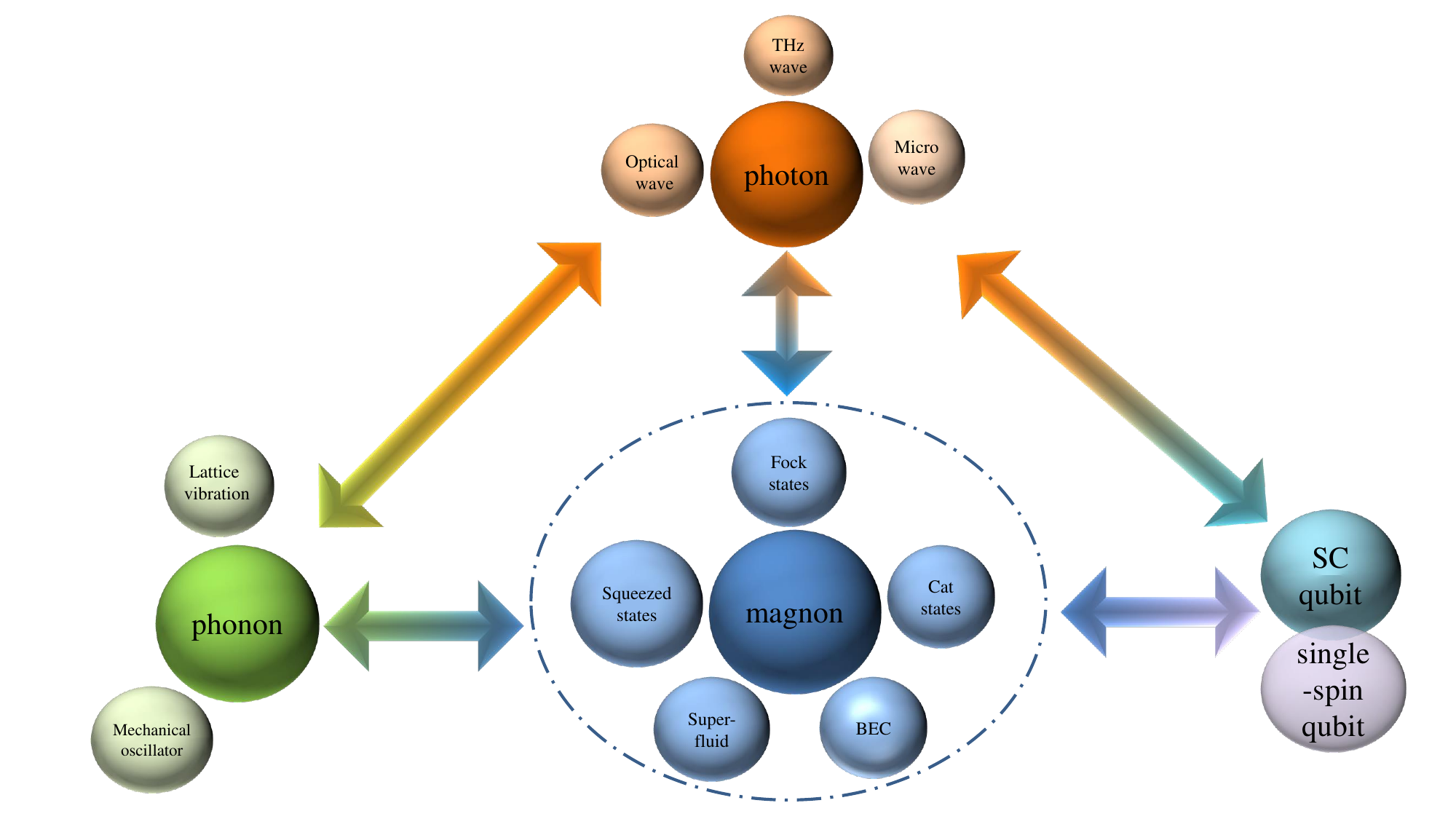}\\
  \caption{Framework of quantum magnonics. Centered around magnons, four routes are being developed. The central route concerns quantum states of magnons at low temperatures. The left route integrates magnons with phonons corresponding to the lattice vibrations within the magnet or mechanical vibrations of external oscillators. The top route entangles magnons with photons ranging from optical to microwave domains. The right route couples magnons to discrete variable systems including superconducting qubit and single-spin qubit (nitrogen-vacancy centers, silicon-vacancy centers etc). }
  \label{qmpicture}
\end{figure}

\textbf{Horizon of quantum magnonics.} The term ``quantum magnonics'' has been very frequently used in literature and at conference talks, even when reporting on classical spin-wave physics. At the outset of this review, we first have to clarify the horizon of quantum magnonics. Tabuchi \cite{TabuchCRP2016} restricted the field of quantum magnonics to the interplay of magnons with SC qubits. Later, Bunkov referred to quantum magnonics in the study of magnon Bose-Einstein condensation (BEC) and spin superfluidity \cite{BunkovJETP2020}. As we shall see, the field of quantum magnonics has developed explosively and has become more diverse in recent years. Kamra et al. \cite{KamraAPL2020} have emphasized the possibilities offered by magnon squeezing within quantum magnonics. Pirro et al. \cite{PirroNRM2021} discussed entanglement of magnons with photons in the context of quantum magnonics. Here, we use the following definition:
\begin{quote}
\textit{In general, any study of quantum states of magnons and the entanglement of magnons with other quantum platforms is part of quantum magnonics. In a broader sense, quantum many-body states of magnons, including Bose-Einstein condensation and its resulting spin superfluidity, are included in this field.}
\end{quote}

Following this definition, we schematically depict the interplay among the various branches of quantum magnonics in Fig. \ref{qmpicture}. The first direction concerns nonclassical states of magnons, including Fock states, squeezed states, Schr\"{o}dinger cat states and so on. Quantum many-body states also fall into this class. The second direction integrates magnonic systems with known quantum platforms, including photons, phonons, and qubits. Here the photonic systems range from optical frequencies down to the microwave domain. These photons can be coupled to magnons through off-resonant or resonant interactions. Phonons represent the quantization of the collective lattice vibration as well as the oscillations of mechanical oscillators. The qubit systems mainly include the SC qubits working under cryogenic conditions and nitrogen-vacancy (NV) centers at room temperature. Here the two directions are not isolated from each other, but have strong connections. For example, to detect the quantum states of magnons, one usually imprints the magnon state onto photons, which are measured with mature light detection methods. To probe the superfluid state of magnons, one may utilize quantum impurities placed on top of the magnetic system to detect the magnetic noise. Furthermore, hybrid magnonic systems provide a natural and controllable platform to study non-Hermitian physics and parity-time ($\mathcal{PT}$) symmetry, which will also be covered in the current review.

Our review is organized as follows. In Section \ref{sec_preliminary}, we give a preliminary introduction to magnons and their classical and quantum behaviours. Then, we introduce various measures to quantify the entanglement of discrete variables (qubit, qutrit) and continuous variables (magnon, photon, phonon). In Section \ref{sec_qmstate}, we discuss the developments concerning quantum states of magnons, including the single-magnon state, squeezed states, Schr\"{o}dinger cat states, as well as quantum many-body states. In Section \ref{sec_magnonX}, we overview the developments in the hybrid ``magnon+X'' systems, where X includes cavity photons, qubits, phonons and electrons. In Section \ref{sec_PT}, non-Hermitian physics and $\mathcal{PT}$ symmetry, particularly emerging in magnonic systems are highlighted. After briefly summarizing the applications of quantum magnonics in Section \ref{sec_application}, we come to the conclusion and outlook in Section \ref{sec_outlook}. If not stated otherwise, we decorate operators with a hat and set the reduced Planck constant $\hbar$ equal to one throughout the review.

\section{Preliminary knowledge} \label{sec_preliminary}
\subsection{Magnons}
\subsubsection{Magnetization dynamics and spin wave} \label{sec_llg}
\begin{figure}
  \centering
  \includegraphics[width=1.0\textwidth]{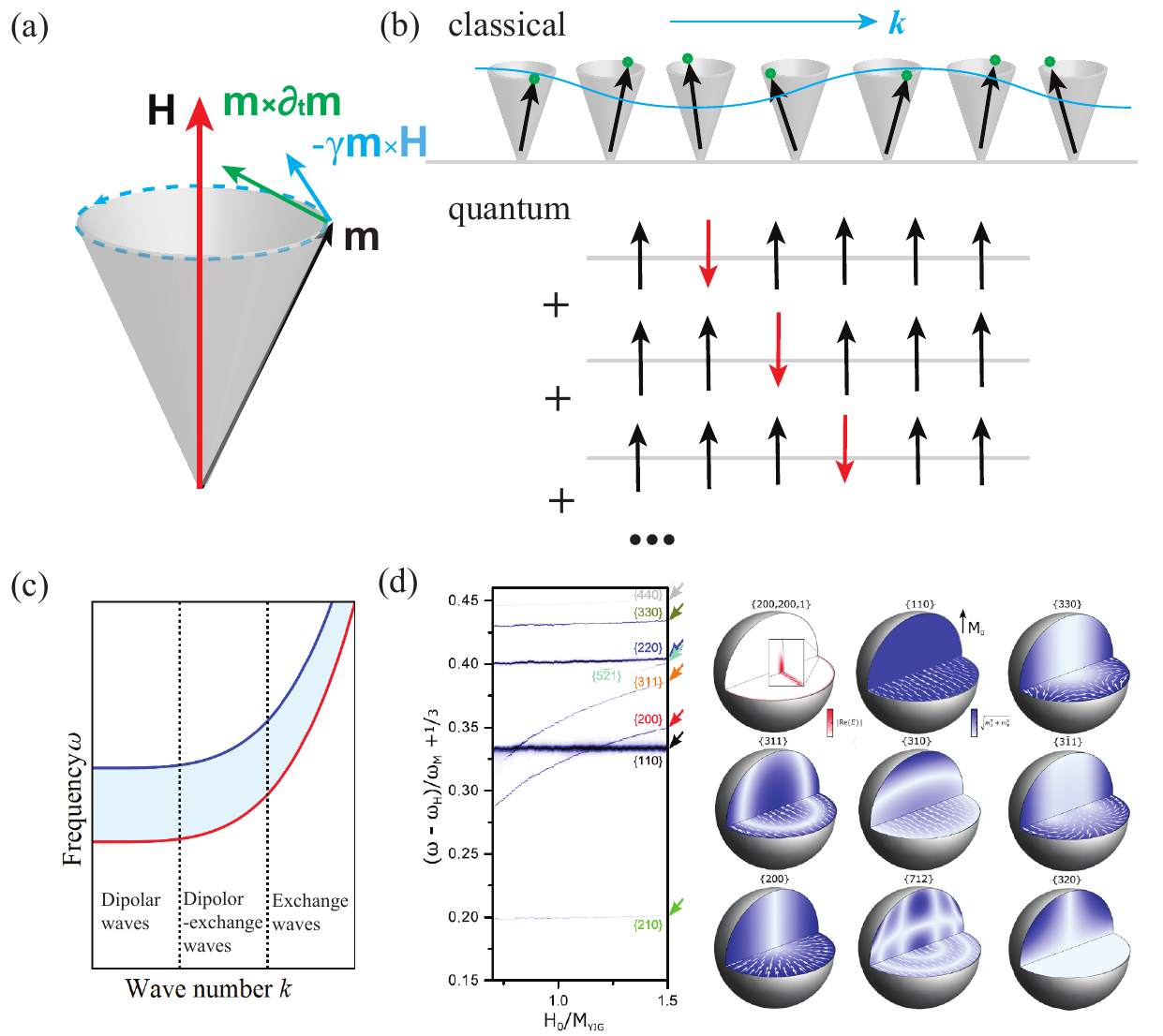}\\
  \caption{(a) Schematic of magnetic-moment precession and relaxation toward a magnetic field. (b) Spin-wave excitation in classical (wave-like) and quantum (localized) picture. (c) Typical spin-wave dispersion of a finite ferromagnetic body. With the increase of wave number $k$, dipolar, dipolar-exchange and exchange spin waves are classified.  (d) Field dependence of the dipolar spin-wave frequency of a magnetic sphere as a function of external field, and the wave profiles of typical modes involved. Here $(l,m,q)$ labels the number of radial ($q-1$), azimuthal ($m$) and polar modes ($l-m$), respectively. The (110) mode represents ferromagnetic resonance mode, where all spins precess in phase. Source: Figure (d) is  adapted from Ref. \cite{HaighPRB2018}.}\label{sw}
\end{figure}

Spin waves (SWs) are the collective excitations of magnetic moments in ordered magnets. Generation, propagation, manipulation and read-out of SW information have attracted significant attention over the past decade because of their promising applications in information technology. One major advantage of SWs over electrons as information carriers is that they can propagate in insulating systems, which can, in principle, weaken the effect of Joule heating that hampers traditional transistor technologies. Besides, there exist rich and mature knobs to manipulate the SWs, such as microwave, terahertz and attosecond excitations \cite{Pozarbook2011,KampfrathNP2011,SiegristNature2019,LiNature2020}, electric current through spin transfer \cite{SlonJMMM1996,BergerPRB1996,StilesPRB2002,ZhangPRL2004,TataraPRL2004,ThiavilleEPL2005}, spin-orbit effects \cite{MironNature2011,LiuPRL2012,NakayamaPRL2013,ManchonNM2015,LebrunNature2018,WadleyScience2016}, temperature gradients \cite{XiaoPRB2010,UchidaNature2008,BauerSSE2012,HinzkePRL2011,XiansiPRB2014}, ultrafast laser pulses, \cite{BeaurPRL1996,KimelNature2005,NishAPL2010,DebPRL2019}, and mechanical strains \cite{WeilerPRL2011,ScherPRL2010,SadovnPRL2018,XuSA2020}. To study and utillize SWs in spintronic devices, it is essential to first describe the dynamics of spins under various perturbations. Generally, the average spin angular momentum or magnetic moment in a unit volume is represented by a magnetization vector and the magnetization dynamics is governed by the phenomenological Landau-Lifshitz-Gilbert (LLG) equation \cite{Landau1935,Gilbert2004}
\begin{equation}
\centering
\frac{\partial \mathbf{m}}{\partial t}=-\gamma \mathbf{m} \times \mathbf{H}_{\text{eff}} + \alpha \mathbf{m} \times \frac{\partial \mathbf{m}}{\partial t} + \mathbf{\tau} (\mathbf{m}),
\end{equation}
where $\mathbf{m}$ is the magnetization direction that is spatiotemporally dependent in general, $\gamma>0$ is the magnitude of gyromagnetic ratio, $\mathbf{H}_{\text{eff}}$ is the effective field including exchange, anisotropy, and dipolar fields, and $\alpha$ is the Gilbert damping constant. The first and second terms respectively describe the precessional and damped motion of the magnetization around and towards its equilibrium orientation, as shown in Fig. \ref{sw}(a). The larger the damping, the faster the relaxation occurs. The term $\tau(\mathbf{m})$ is the spin torque caused by external drivings, for example, the thermal torque at finite temperature \cite{Brown1963}, spin-transfer torque from electric current \cite{SlonJMMM1996,BergerPRB1996,ZhangPRL2004}, and spin-orbit torque by spin current \cite{MironNature2011,LiuPRL2012}. The LLG equation was originally introduced by Landau and Lifshitz (LL) and extended by Gilbert on the dissipation term. In modern developments, this dynamic equation in both ferromagnets and antiferromagnets has been discussed based on the Lagrangian formalism by properly constructing the dissipation functional \cite{TataraPR2008,YuanEPL2019}. The microscopic mechanism of the Gilbert damping include the spin-orbit coupling \cite{KamberskyCJPB1976,HickeyPRL2009,LiuPRB2011,KhodPRL2020,AndreasPRB2014}, spin pumping \cite{YaroslavPRL2002,LiuPRL2014,ZhePRL2014}, two-magnon scattering \cite{HurbenJAP1998,AriasPRB1999}, etc.

Based on the LLG equation, the dynamics of physically-separated spins is coupled together and may form a wave-like pattern as a result of short-range exchange and long-range dipolar fields, as shown in Fig. \ref{sw}(b). If the exchange interactions dominate, the resulting SW has a short wavelength on the order of nanometer scale. It is sufficient to describe such exchange SW excitations and their spectrum by the LLG equation, as already presented in many textbooks and reviews \cite{Kittelbook1986, Gurevichbook1996,Coeybook2010}. If dipolar interactions dominate, the resulting SW has a longer wavelength, and is termed a magnetostatic SW. To fully describe the SW spectrum, one needs to consider the magnetic fields both inside and outside the magnetic sample, where Maxwell's equations are required to match the boundary conditions at the interface \cite{WalkerPR1957,EshbachPR1960,Kalinikos1986}. A special resonance mode with infinitely-long wavelength is the ferromagnetic resonance (FMR) \cite{KittelPR1948}, which corresponds to the uniform precession of all spins. This mode plays an important role when integrating magnonic systems with other quantum systems, such as photons, phonons and qubits. A schematic picture of the complete SW spectrum is shown in Fig. \ref{sw}(c). As an example, the SW spectrum of the widely used magnetic sphere is plotted in Fig. \ref{sw}(d). Here, the FMR mode is labeled as the $(1,1,0)$ mode \cite{WalkerJAP1958,HaighPRB2018}, where all spins inside the sphere precess in phase around the equilibrium magnetization direction defined by an external field.

\subsubsection{Quantization of spin wave: magnon}

In a quantum-mechanical picture, the SW corresponds to the quasi-particle called magnon. It is a boson obeying Bose-Einstein statistics, and, in the absence of fields and anisotropy, corresponds to a Goldstone boson. Magnons were originally proposed by Bloch in a spin-$1/2$ system \cite{Bloch1930} to explain the reduction of magnetization with increasing temperature, i.e., the celebrated $T^{3/2}$ law. The extension to the higher spin case is due to M\"{o}ller \cite{Muller1933}. Holstein and Primakoff \cite{HP1940} quantized the magnetic excitation based on the bosonic creation and annihilation operators and derived the Hamiltonian of magnons with both quadratic and higher-order terms, which give the non-interacting magnon behavior and interaction between magnons, respectively. This approach has become a standard technique to study the quantum behavior of magnons. Magnon theory was further developed by Dyson \cite{DysonPR1956} to consider the influence of magnon interactions on the low temperature behavior of thermodynamic quantities. Here we briefly review the  Holstein-Primakoff (HP) approach. To describe a spin-$S$ state in quantum mechanics \cite{SakuraiQM}, we consider the simultaneous eigenstates of the spin angular momentum operator $\hat{\mathbf{S}}^2$ and its component along the quantization axis ($\mathbf{e}_z$) $\hat{S}_z$ as
\begin{equation}
\hat{\mathbf{S}}^2|S,m\rangle = S(S+1)  |S,m\rangle, \quad \hat{S}_z|S,m\rangle=m|S,m\rangle,\quad m=-S,-S+1,...,S-1,S.
\end{equation}
However, with this approach it is difficult to treat the interaction between many spins and magnons. To overcome this difficulty, Holstein and Primakoff noticed that \cite{HP1940} the possible values of $m$ can be rewritten as $m\equiv S-n$, where the new quantum number $n$ takes only non-negative values $n=0,1,2,...2S$, regardless of whether one deals with fermions (half-integer $S$) or bosons (integer $S$). Transforming to this new basis $|S,n\rangle$, we have $\hat{S}_z|S,n\rangle =(S-n)|S,n\rangle$. In comparison with the quantization of the harmonic oscillator $\hat{a}^\dagger \hat{a} |n\rangle =n |n \rangle$, we immediately have the second-quantization form of $\hat{S}_z$, i.e., $\hat{S}_z=S-\hat{a}^\dagger \hat{a}$, where $\hat{a}$ ($\hat{a}^\dagger$) is the annihilation (creation) operator of a spin quantum or a magnon. To make the new representation complete, we need to find a representation of $\hat{S}_x$ and $\hat{S}_y$ by preserving the commutation relation for angular momentum $[\hat{S}_i,\hat{S}_j]=i\epsilon_{ijk} \hat{S}_k$, where $\epsilon_{ijk}$ is the Levi-Civita symbol. To this end, we recall the spin raising and lowering operators defined as, $\hat{S}^\pm=\hat{S}_x \pm i\hat{S}_y$. Based on the quantum theory of angular momentum, we have
\begin{equation}
\hat{S}^+ |S,n\rangle =\sqrt{2S-(n-1)}\sqrt{n} |S,n-1\rangle=\sqrt{2S-\hat{a}^\dagger \hat{a}} \hat{a}|S,n\rangle.
\end{equation}
We therefore obtain $\hat{S}^+ =\sqrt{2S-\hat{a}^\dagger \hat{a}} \hat{a}$ and, similarly, $\hat{S}^- =\hat{a}^\dagger \sqrt{2S-\hat{a}^\dagger \hat{a}}$, which allows us to express the excitation of a single spin $S$ in terms of magnon creation and annihilation operators.
A real magnet involves many interacting spins and is described by the general Hamiltonian \cite{Nowak2007,EvansJPCM2014}
\begin{equation}\label{generalHam}
\hat{\mathcal{H}}=-J\sum_{\langle i,j\rangle} \hat{\mathbf{S}}_i\cdot \hat{\mathbf{S}}_j + \hat{\mathcal{H}}_\mathrm{an}-  \mu_0 g_e \mu_B\sum_i \mathbf{H}_i \cdot \hat{\mathbf{S}}_i- \frac{\mu_0 (g_e\mu_B)^2}{4\pi}\sum_{i<j} \frac{3(\hat{\mathbf{S}}_i\cdot \mathbf{e}_{ij})(\hat{\mathbf{S}}_j\cdot \mathbf{e}_{ij})-\hat{\mathbf{S}}_i\cdot \hat{\mathbf{S}}_j}{|\mathbf{r}_{ij}|^3},
\end{equation}
where $J$ is the exchange coefficient and the first sum is over nearest neighbouring spins, $\mathbf{H}_i$ is the external magnetic field on the $i$-th spin, $\mu_0$ is vacuum magnetic permeability, $g_e$ is the Land\'{e} $g$-factor, $\mu_B$ is Bohr magneton, $\mathbf{r}_{ij}$ is the vector between the $i$-th and $j$-th spin, and $\mathbf{e}_{ij}$ is the unit direction vector connecting sites $i$ and $j$. On the right hand side of $\hat{\mathcal{H}}$, the first term is the exchange coupling, the second term is the magnetocrystalline anisotropy that prefers spins to align in particular directions, the third term is the Zeeman energy, and the last term is the long-range dipolar interaction. Depending on the competition of these interactions, the ground state of the system may be a collinear magnetic state, such as a homogeneous magnetic domain, or a noncollinear magnetic texture, such as a domain wall, vortex, skyrmion, etc \cite{Lilley1950,Hubertbook,BogdanovJMMM1994,RosserNature2006,WangCP2018}. To consider the magnon excitations above these magnetic states, one usually performs a HP transformation on the ground state of the system and derives the effective Hamiltonian for magnons as \cite{Akhiezerbook,Noltingbook,Kittelbook1963}
\begin{equation}
\hat{\mathcal{H}}=\hat{\mathcal{H}}_0 + \hat{\mathcal{H}}^{(2)} + \hat{\mathcal{H}}^{(3)} + \hat{\mathcal{H}}^{(4)} +...,
\end{equation}
where $\hat{\mathcal{H}}_0$ is ground state energy, $\hat{\mathcal{H}}^{(2)},\hat{\mathcal{H}}^{(3)}, \hat{\mathcal{H}}^{(4)}$ denote the two-magnon processes contributed by exchange field, external field, anisotropy field and dipolar interactions, three-magnon processes from dipolar interactions, and four magnon processes from exchange and dipolar interactions, respectively. If the magnon density of the system is very low, which typically holds for weak perturbations well below the Curie temperature, the three- and four-magnon processes can be neglected. It is then sufficient to diagonalize the quadratic component $\hat{\mathcal{H}}^{(2)}$ to derive the eigenspectrum. Note that this approach works for both collinear and noncollinear magnetic structures \cite{SwanPRB1983,RoldPRB2015,ZhangPRB2018,SantosPRB2018,CamiloPRL2018}.

For sufficiently large $S$, the quantum approach should give an identical prediction of the low energy SW dispersion as starting from the LLG equation, because the diagonalization procedure of the Hamiltonian corresponds to solving the eigenvalues of the dynamic matrix of the quantum Heisenberg equation, which is equivalent to solving the linearized LLG equation in the absence of damping. As an example, let us consider a soft magnetic film with magnetization aligned along the in-plane field $\mathbf{H}=H\mathbf{e}_z$. The Hamiltonian of the system is
\begin{equation}\label{biaxialHam}
\hat{\mathcal{H}}=-J\sum_{\langle i,j\rangle} \hat{\mathbf{S}}_i\cdot \hat{\mathbf{S}}_j + K_x \sum_i \left (\hat{S}_{i,x} \right )^2-K_z \sum_i\left (\hat{S}_{i,z} \right )^2-\mu_0 g_e \mu_B H\sum_i \hat{S}_{i,z},
\end{equation}
where $K_x$ and $K_z$ are anisotropies along the $x$ and $z$ directions, respectively, with $K_x>0, K_z>0$. The classical ground state of the system is $\langle \hat{\mathbf{S}} \rangle=S\mathbf{e}_z$. By keeping the linear terms of the HP transformation of the spin operators along this ground state, i.e.,
\begin{equation}\label{HPtransform}
\hat{S}_{i,z}=S-\hat{a}_i^\dagger \hat{a}_i, \quad \hat{S}_{i,x}=\frac{\sqrt{2S}}{2}(\hat{a}_i + \hat{a}^\dagger_i), \quad \hat{S}_{i,y}=\frac{\sqrt{2S}}{2i}(\hat{a}_i-\hat{a}_i^\dagger),
\end{equation}
where $\hat{a}_i$ ($\hat{a}_i^\dagger$) is the magnon annihilation (creation) operator on the $i-$th spin site, the effective Hamiltonian of magnon excitations to the quadratic terms can be written as
\begin{equation}
\mathcal{\hat{H}}^{(2)}=-JS\sum_{\langle i,j\rangle} (\hat{a}_i \hat{a}^\dagger_j+h.c.) +  \sum_i \mu_i \hat{a}_i^\dagger \hat{a}_i + \frac{K_xS}{2} \sum_i (\hat{a}_i \hat{a}_i + h.c.),
\label{hamsse}
\end{equation}
where $\mu_i=2ZJS + \mu_0 g_e \mu_B H + 2K_zS + K_xS$ with $Z$ being the coordination number. By switching to Fourier space and diagonalizing the Hamiltonian, we obtain the bosonic Hamiltonian of the system as
\begin{equation}\label{sq:eq:Hmag}
\mathcal{\hat{H}}^{(2)}=const. + \sum_k \omega_k \hat{a}^{\dagger}_\mathbf{k} \hat{a}_\mathbf{k}.
\end{equation}
Here the eigenfrequency is
\begin{equation}
\omega_k = \sqrt{[2JS(kd)^2 + \mu_0 g_e \mu_B H+2K_xS][2JS(kd)^2 + \mu_0 g_e \mu_B H+2K_zS+2K_xS]},
\end{equation}
in the long wavelength limit with $d$ the lattice constant. The same dispersion is obtained employing the LLG phenomenology.

The anisotropy may substantially alter the nature of the magnonic eigenmodes in a ferromagnet leading to squeezing, as discussed in Section~\ref{sec_squeezedmagnon} below. Here and in the next section, we examine magnons in an isotropic ferromagnet \ref{sec:cohstate}. The corresponding considerations in an anisotropic ferromagnet will be addressed in Section~\ref{sec_squeezedmagnon}. For a magnetic sphere with negligible magnetocrystalline anisotropy ($K_x=K_z=0$), i.e., an istrotropic magnet, the resonance frequency $\omega_k= \mu_0 g_e \mu_B H+2JS(kd)^2$. Employing $\hat{S}_{i,z}=S-\hat{a}_i^\dagger \hat{a}_i$ from the HP approach, the total spin $z$-component in an eigenstate is evaluated as
\begin{align}\label{sq:eq:spin1}
\left \langle  \hat{\mathcal{S}}_{z} \right \rangle \equiv \left \langle \sum_i \hat{S}_{i,z} \right \rangle & =  N S - \sum_\mathbf{k} \left \langle \hat{a}_\mathbf{k}^\dagger \hat{a}_\mathbf{k} \right \rangle = N S - \sum_{\mathbf{k}} n_\mathbf{k},
\end{align}
where $N$ is the total number of lattice sites and $n_\mathbf{k}$ is the number of $\mathbf{k}$ magnons. Equation \eqref{sq:eq:spin1} shows that excitation of each magnon alters the total spin by unity, which is why each magnon carries one unit ($\hbar$) of spin angular momentum along $\mathbf{e}_z$~\cite{Kittelbook1963}. Within this linear approximation [Eq.~\eqref{HPtransform}], magnons are non-interacting bosonic modes. Hence, without loss of generality, we may focus on the $\mathbf{k} = \pmb{0}$ mode addressed in typical ferromagnetic resonance experiments
\begin{align}\label{sq:eq:H1}
\hat{\mathcal{H}} & = \omega_r \hat{a}^\dagger \hat{a},
\end{align}
where $\omega_r \equiv \omega_{\mathbf{k}\rightarrow \pmb{0}}$ and we have further dropped the subscript $\pmb{0}$ of the magnon ladder operators for convenience.

\subsubsection{Coherent states and classical dynamics}\label{sec:cohstate}
Now we wish to relate the magnonic quantum picture to the phenomenological LL description, considering the $\mathbf{k} = \pmb{0}$ mode in an isotropic ferromagnet [Eq.~\eqref{sq:eq:H1}]. Employing Eq.~\eqref{HPtransform} and the properties of the number states, we evaluate the expectation value
\begin{align}\label{sq:eq:expect1}
\left \langle \hat{\mathcal{S}}_{x} \right \rangle & \equiv  \left \langle \sum_{i} \hat{S}_{i,x} \right \rangle = \sqrt{\frac{N S}{2}} \left \langle \hat{a} + \hat{a}^\dagger \right \rangle = 0,
\end{align}
in any eigenstate of the Hamiltonian Eq.~\eqref{sq:eq:H1}. The expectation value of $\hat{\mathcal{S}}_y\equiv  \left \langle \sum_{i} \hat{S}_{i,y} \right \rangle$ vanishes similarly. This is in contrast with typical experiments which find a finite value for the spin's $x$ and $y$ components. What has gone wrong? The equivalent question for harmonic oscillators bothered Schr{\"o}dinger and led him to find the states that reproduce the classically observed dynamics~\cite{NietoPLA1997}. These are called coherent states~\cite{GlauberPR1963,SudarshanPRL1963,RezendePLA1969} and are eigenstates of the annihilation operator
\begin{align}
\hat{a} \ket{\beta} & = \beta \ket{\beta},
\end{align}
where $\beta$ is a complex number. Using this property and Eq.~\eqref{sq:eq:expect1}, it is seen directly that the expectation values of $\hat{\mathcal{S}}_{x,y}$ in a coherent state do not vanish~\cite{RezendePLA1969}. The coherent state can further be expressed in terms of Fock states $\ket{n}$~\cite{Gerry2004}
\begin{align}\label{sq:eq:alphanum}
\ket{\beta} & = \sum_n \exp \left( -\frac{|\beta|^2}{2} \right) \frac{\beta^n}{\sqrt{n!}} \ket{n},
\end{align}
from which we see that a coherent state has a Poissonian distribution of the magnon number $n$. We further evaluate the mean $\bar{n}$ and uncertainty $\Delta n$ of magnon number
\begin{align}\label{sq:eq:cohprop}
\bar{n} \equiv \bra{\beta} \hat{n} \ket{\beta} \equiv \expect{\hat{n}} =  |\beta|^2, \quad \left( \Delta n \right)^2 \equiv \expect{\hat{n}^2} - \left(\expect{\hat{n}} \right)^2 = |\beta|^2, \quad \implies \frac{\Delta n}{\bar{n}} = \frac{1}{|\beta|} = \frac{1}{\sqrt{\bar{n}}}.
\end{align}
The properties above clarify the classical-quantum correspondence that we have sought. It is not sufficient to consider a high magnon number alone, as one might guess at first, to recover classical dynamics from the quantum picture. This is because the expectation values of $\hat{\mathcal{S}}_{x,y}$ in any number state vanish [Eq.~\eqref{sq:eq:expect1}]. On the other hand, the coherent state which is a superposition of number states yields a finite expectation value~\cite{GlauberPR1963,SudarshanPRL1963,RezendePLA1969}. Furthermore, in the limit of a large average magnon number $\bar{n}$, the relative fluctuations become small [Eq.~\eqref{sq:eq:cohprop}] giving rise to the classical picture of a spin precession with (nearly) constant amplitude~\cite{RezendePLA1969,BenderPRL2019}.

Let us take one step further and establish the method to determine $\beta$. From a theoretical perspective, a coherent state is generated by the ``displacement operator'' $\hat{D}(\beta)$
\begin{align}\label{sq:eq:disp1}
\ket{\beta} & = \hat{D}(\beta) \ket{0}, \quad \mathrm{where}~ \hat{D}(\beta) \equiv \exp \left( \beta \hat{a}^\dagger - \beta^* \hat{a}\right).
\end{align}
In certain ideal situations, the system's time evolution under an external drive can be cast in the form of Eq.~\eqref{sq:eq:disp1}, which determines the generated coherent state~\cite{Gerry2004} (see also Section~\ref{sec_qm_equation}). Furthermore, Equation~\eqref{sq:eq:disp1} provides a powerful theoretical tool for deriving various properties of coherent states. We discuss an alternative method~\cite{GardinerPRA1985,KamraPRL2016,KamraPRB2016} which is especially useful for studying spin dynamics. A more detailed analysis is presented in the next section \ref{sec_qm_equation}.

Accounting for a small microwave magnetic field $h \cos (\omega_d t) \hat{\pmb{x}}$, the drive Hamiltonian becomes
\begin{align} \label{sq:eq:Hd}
\hat{\mathcal{H}}_{d} & = \mu_0 h \cos (\omega_d t) g_e \mu_B \hat{\mathcal{S}}_x = \mu_0 h \cos (\omega_d t) g_e \mu_B \sqrt{\frac{NS}{2}} \left( \hat{a} + \hat{a}^\dagger \right) \equiv \epsilon_{\text{rf}}(t) \left( \hat{a} + \hat{a}^\dagger \right).
\end{align}
Working in the Heisenberg picture, the equation of motion $\dot{\hat{a}} = i [\hat{\mathcal{H}} + \hat{\mathcal{H}_d}, \hat{a}]$ yields [Eqs.~\eqref{sq:eq:H1} and \eqref{sq:eq:Hd}]
\begin{align}
\frac{d\hat{a}}{dt} & = - i \omega_r \hat{a} - i \epsilon_{\text{rf}}(t).
\end{align}
In steadily oscillating state, we may take the expectation value of the above equation defining $\expect{\hat{a}} \equiv \beta$
\begin{align}\label{sq:eq:dyn1}
\frac{d\beta}{dt} & = - i \omega_r \beta - i \epsilon_{\text{rf}}(t) - \alpha \omega_r \beta,
\end{align}
where the last term on the right has been added to account for dissipation. This can be derived rigorously considering a specific bath~\cite{KamraPRL2016,KamraPRB2016}, but can also be seen as a phenomenological dissipation term with $\alpha$ the Gilbert damping parameter that also enters the LLG phenomenology~\cite{Gilbert2004}.

Equation~\eqref{sq:eq:dyn1} is a powerful result for multiple reasons. First, its solution expressed in terms of the spin component expectation values is nothing but the linearized LLG equation~\cite{Gilbert2004,KamraPRL2016,KamraPRB2016}. Hence, it serves as a derivation of the linearized classical spin dynamics from a microscopic quantum Hamiltonian. Second, it actually goes beyond the simplified understanding that classical dynamics corresponds to a coherent state. A realistic system, which includes dissipation and nonideal fluctuating drives, is not described by a pure state and thus cannot be a simple coherent state. Nevertheless, Equation~\eqref{sq:eq:dyn1} accounts for such nonidealities. The explicit example of a bosonic dissipative bath is presented in the following Section \ref{sec_qm_equation}.

Finally, assuming that the influence of dissipation and drive nonideality is small, we may approximate the state of our system by a coherent state
\begin{align}\label{sq:eq:psit}
\ket{\psi(t)} & = \hat{D}(\beta(t)) \ket{0},
\end{align}
where $\beta(t)$ is determined by Eq.~\eqref{sq:eq:dyn1}. Furthermore, the quantum fluctuations in the coherent state are the same as in vacuum
\begin{align}\label{sq:eq:fluc1}
\Delta \mathcal{S}_x \equiv \sqrt{ \bra{\beta} \hat{\mathcal{S}}_x^2 \ket{\beta} - \left( \bra{\beta} \hat{\mathcal{S}}_x \ket{\beta} \right)^2 } = \sqrt{ \bra{0} \hat{\mathcal{S}}_x^2 \ket{0} - \left( \bra{0} \hat{\mathcal{S}}_x \ket{0} \right)^2 } = \sqrt{\frac{NS}{2}} = \Delta \mathcal{S}_y.
\end{align}
Thus, coherent states correspond to the minimum Heisenberg uncertainty, which is not the case for general states~\footnote{For example, the squeezed states with $\theta \neq 0,\pi$ [Eqs.~\eqref{sq:eq:deltasx} and \eqref{sq:eq:deltasy}] bear quantum fluctuations in $\mathcal{S}_{x,y}$ larger than the Heisenberg minimum requirement.}. Moreover, they have an equal magnitude of quantum fluctuations in the two spin components.

\begin{figure}
	\begin{center}
		\includegraphics[width=1.0\textwidth]{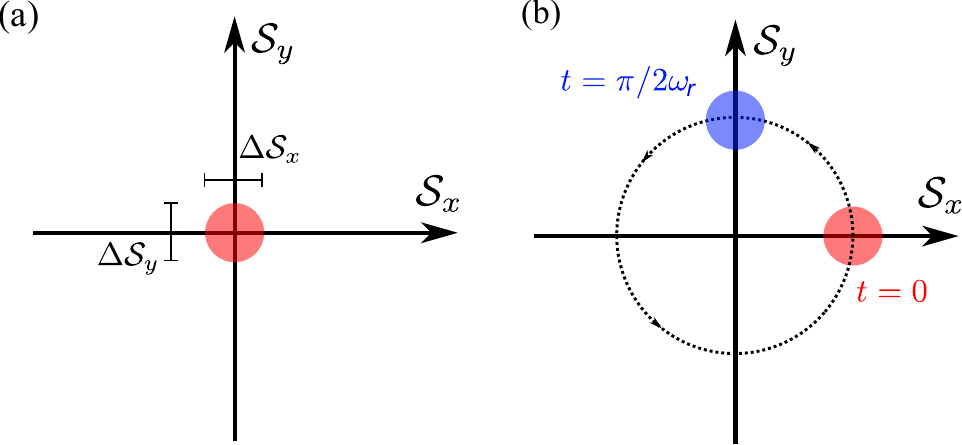}
		\caption{Phase space depiction of (a) isotropic ferromagnetic ground state, corresponding to the magnon vacuum, and (b) a magnon coherent state. The shaded region represents the quantum fluctuations due to the Heisenberg uncertainty principle with its area $\propto \hbar$. In the coherent state, depicted in (b), the uncertainty region center undergoes precession along a circular trajectory, shown as a dotted curve. For macroscopically large amplitude of the precession, the Heisenberg uncertainty region appears as an infinitesimal point and the classical spin precession picture is recovered.}
		\label{sq:fig:mag1}
	\end{center}
\end{figure}

Equipped with the analysis above culminating in Eqs.~\eqref{sq:eq:psit} and \eqref{sq:eq:fluc1}, we portray the spin dynamics, including quantum fluctuations, in its phase space comprised of $\mathcal{S}_x \propto \Re (\beta)$ and $\mathcal{S}_y \propto \Im (\beta)$. This has been depicted for the magnon vacuum (ground state of an isotropic ferromagnet) in Fig.~\ref{sq:fig:mag1} (a) and ``classical'' spin precession in Fig.~\ref{sq:fig:mag1} (b). To summarize, the quantum spin dynamics follows the classical precession expected from the LLG description with a circular minimum Heisenberg uncertainty region around the classical trajectory~\cite{KamraPRL2016,KamraPRB2016}. This picture is an approximation valid when the microwave drive is in an ideal coherent state and the role of dissipation is negligible.

\subsubsection{Quantum equation of motion} \label{sec_qm_equation}
In this section, we discuss the quantum dynamics of magnons, now also accounting for anisotropies in our description. The Heisenberg equation corresponding to the Hamiltonian \eqref{hamsse} reads
\begin{equation} \label{HeisenEq}
i\frac{d\hat{a}_i}{dt}=[\hat{a}_i,\hat{\mathcal{H}}^{(2)}]=-2JS\sum_j \hat{a}_j + \mu_i \hat{a}_i + K_x S \hat{a}^\dagger_i.
\end{equation}
where the first sum is over neighboring spins of the $i$-th spin. By reformulating these dynamic equations into the form of spin operators, we obtain, $\partial_t \hat{\mathbf{S}}_i = - \gamma (\hat{\mathbf{S}}_i \times \hat{\mathbf{H}}_{i,\text{eff}})$, where the effective field is equal to $\hat{\mathbf{H}}_{i,\text{eff}} \equiv (2J\sum_j\hat{S}_{j,x}-2K_x\hat{S}_{i,x})\mathbf{e}_x + J\sum_j\hat{S}_{j,y}\mathbf{e}_y + (\mu_0 g_e \mu_B H+2K_zS + 2ZJS)\mathbf{e}_z$ and $\gamma = g_e \mu_B$ appears naturally. After taking the mean-field approximation, this equation becomes the same as the classical LLG equation.

When the dissipation of spins is considered, recovering the classical dynamic equations from the quantum mechanical evolution is an ongoing research. The difference between various approaches lies in the way how dissipation is introduced in the quantum models and how the damping of the classical magnetization is recovered \cite{KamraPRL2016,KamraPRB2016,WieserPRL2013,MondalPRB2016,NorambuenaNJP2020,YuanSqueezed2021}. In principle, all these methods, no matter quantum or classical, should give identical predictions for the low-energy spectrum of SWs. Two known methods to introduce dissipation in a quantum system are the master equation approach based on the Lindblad formalism \cite{LindbladCMP1976,ManzanoAIP2020}, and the Heisenberg-Langevin approach based on the fluctuation-dissipation theorem \cite{DFWalls}. Since the latter will be frequently used in this review, we will briefly introduce it here.

The Hamiltonian of the total system including the magnon mode, a bath and their interactions is written as
\begin{equation}
\hat{\mathcal{H}}= \hat{\mathcal{H}}_{s} + \hat{\mathcal{H}}_b + \hat{\mathcal{H}}_{\mathrm{int}},
\label{hambath}
\end{equation}
where $\hat{\mathcal{H}}_{s},\hat{\mathcal{H}}_{b}, \hat{\mathcal{H}}_{\mathrm{int}}$ are respectively the Hamiltonian for the magnon mode, bath and the mutual interaction. Without loss of generality, we shall focus on the FMR mode, and the Hamiltonian for the magnons is $\hat{\mathcal{H}}_s=\omega_r \hat{a}^\dagger \hat{a}$, where $\omega_r = \omega (\mathbf{k}\rightarrow 0)$. The general formalism involving many interacting bosonic modes can be found in Ref. \cite{YuanPRA2020}. Next, $\hat{\mathcal{H}}_b$ is the free Hamiltonian of the bath, which may be formulated as a collection of harmonic oscillators, i.e.,
\begin{equation}
\hat{\mathcal{H}_b} = \sum_i \omega_i \hat{b}^\dagger_i \hat{b}_i,
\end{equation}
where $\hat{b}_i^\dagger$ ($\hat{b}_i$) is the creation (annihilation) operator for the oscillators which
satisfy the bosonic commutation relations $[\hat{b}_i,\hat{b}_j^\dagger]=\delta_{ij}$.
Since the bath usually has a large number of degrees of freedom, we recast the Hamiltonian in the continuum limit as
\begin{equation}
\hat{\mathcal{H}}_b = \int \omega  \hat{b}^\dagger (\omega) \hat{b} (\omega) d\omega,
\end{equation}
with the commutations $[\hat{b}(\omega),\hat{b}^\dagger(\omega')]=\delta(\omega-\omega')$.
$\hat{\mathcal{H}}_{\mathrm{int}}$ is the interaction between the magnon modes
and the bath
\begin{equation}
\hat{\mathcal{H}}_{\mathrm{int}} = \int d\omega \left[g(\omega) \hat{a}^\dagger \hat{b}(\omega) + g^*(\omega)\hat{a} \hat{b}^\dagger(\omega) \right].
\label{couple}
\end{equation}
where $g(\omega)$ is the magnon-bath coupling strength.

The Hamiltonian (\ref{hambath}) leads to the following Heisenberg equation of motion for the magnon mode and bath mode
\begin{subequations}
\begin{align}
&\frac{d\hat{a}(t)}{dt}=-i[\hat{a},\hat{\mathcal{H}}_s]-i\int d\omega g(\omega)\hat{b}(\omega,t) \label{hl1},\\
&\frac{d\hat{b}(\omega,t)}{dt}=-i\omega \hat{b}(\omega,t) - i g^*(\omega) \hat{a}(t)\label{hl2}.
\end{align}
\end{subequations}
According to Eq. (\ref{hl2}), the equation for the bath mode can be explicitly solved as
\begin{equation}
\hat{b}(\omega,t) = \hat{b}(\omega,t_0)e^{-i\omega(t-t_0)}-i g^*(\omega) \int_{t_0}^t dt' \hat{a}(t')e^{-i\omega(t-t')}.
\end{equation}
By substituting this solution into Eq. (\ref{hl1}), we obtain
\begin{equation}
\frac{d\hat{a}(t)}{dt}=-i[\hat{a},\hat{\mathcal{H}}_s]-i\int d\omega g(\omega)\hat{b}(\omega,t_0)e^{-i\omega(t-t_0)}
-\int d\omega g(\omega)g^*(\omega) \int_{t_0}^t dt' \hat{a}(t')e^{-i\omega(t-t')}.\\
\label{aeq}
\end{equation}

Equation \eqref{aeq} shows that the state of the magnon at time $t$ depends on its evolution at earlier times, which makes it difficult to handle. The Markov approximation is usually adopted to further simplify the dynamic equations. If the bath is a large system maintained in thermal equilibrium, we expect that it quickly dissipates the information taken from the magnon system and yields no feedback \cite{Carmichaelbook}. Mathematically, we now assume that $g(\omega)$ is
independent of frequency such that we define $|g(\omega)|^2 \equiv \kappa_m/\pi$,
where the frequency-independent coefficient $\kappa_m$ denotes the coupling strength between magnons and the bath. Then, Equation (\ref{aeq}) is reduced to
\begin{equation}
\frac{d\hat{a}(t)}{dt}=-i[\hat{a},\hat{\mathcal{H}}_s]-\kappa_m \hat{a}(t)+\sqrt{2\kappa_m} \hat{\zeta}(t),
\label{ain}
\end{equation}
where the noise operator is defined as $\hat{\zeta}(t)=-i/\sqrt{2\pi}\int d\omega \hat{b}(\omega)e^{-i\omega t}$. The damping rate is related to the Gilbert damping parameter as $\kappa_m = \alpha \omega_r$ as used in Eq. \eqref{sq:eq:dyn1}. It is straightforward to verify that the noise has zero mean and correlations given by \cite{GardinerPRA1985}
\begin{equation}
\langle \hat{\zeta}(t)\hat{\zeta}^{\dagger} (t') \rangle = (n_{\mathrm{th}}+1) \delta(t-t'), \quad \langle \hat{\zeta}^{\dagger}(t)\hat{\zeta} (t') \rangle = n_{\mathrm{th}} \delta(t-t'),
\end{equation}
where $n_{\mathrm{th}}=1/[\text{exp}(\omega_r/k_BT)-1]$ with $k_B$ Boltzmann constant and $T$ being the temperature. Equation (\ref{ain}) is also called Heisenberg-Langevin equation. It shows that the evolution of magnons at time $t$ depends only on the information of modes at $t$, and is thus Markovian. The linearity of this equation together with the Gaussian nature of the noise implies that the magnon mode will decay to a stationary Gaussian state.

Let us consider magnon dynamics under a periodic driving, $\hat{\mathcal{H}}_s= \omega_r \hat{a}^\dagger \hat{a}+ \xi (\hat{a} e^{i \omega_d t}+ \hat{a}^\dagger e^{-i \omega_d t})$. In the rotating frame with $\mathcal{R}=\exp(-i\omega_d \hat{a}^\dagger \hat{a} t)$, the Hamiltonian becomes $\hat{\mathcal{H}}_s= \Delta \hat{a}^\dagger \hat{a}+ \xi (\hat{a} + \hat{a}^\dagger )$, where $\Delta=\omega_r-\omega_d$ is the detuning of the magnon mode with respect to the driving. Then the Heisenberg-Langevin equation for the system is
\begin{equation}
\frac{d\hat{a}}{dt} = - i (\Delta-i \kappa_m) \hat{a} - i \xi  + \sqrt{2\kappa_m} \hat{\zeta}(t).
\end{equation}
By solving this equation for the steady value of $\langle a \rangle$ and $\langle a^\dagger a \rangle$, we have
\begin{equation}
\langle \hat{a} \rangle=-\frac{\xi}{\Delta - i\kappa_m}, \quad \langle \hat{a}^\dagger \hat{a} \rangle = \frac{\xi^2}{\Delta^2 + \kappa_m^2} + n_{\mathrm{th}},\quad \langle \hat{a} \hat{a} \rangle=\frac{\xi^2}{(\Delta - i\kappa_m)^2}.
\end{equation}
Based on the HP transformation shown above, $\hat{\mathcal{S}}_x =\sqrt{2NS}(\hat{a}+\hat{a}^\dagger)/2,\hat{\mathcal{S}}_y =\sqrt{2NS}(\hat{a}-\hat{a}^\dagger)/2i$, we can readily evaluate the uncertainty of the spin components
\begin{equation}
\begin{aligned}
&\Delta \mathcal{S}_x=\Delta \mathcal{S}_y=\sqrt{\frac{NS}{2} \left ( 2n_\mathrm{th}+ 1 \right )}.
\end{aligned}
\end{equation}
At zero temperature ($n_\mathrm{th}=0$), one immediately sees that
\begin{equation}
\Delta S_x =\Delta S_y=\sqrt{\frac{NS}{2}},
\end{equation}
which recovers the uncertainty relation in Eq. \eqref{sq:eq:fluc1}. This implies that the quantum fluctuations of $\hat{\mathcal{S}}_x$ and $\hat{\mathcal{S}}_y$ are equal and further satisfy the minimum uncertainty relation, which is a signature of a coherent state ~\cite{GlauberPR1963,SudarshanPRL1963,RezendePLA1969}. This indicates that the magnons under strong driving will reproduce the classically observed dynamics ~\cite{NietoPLA1997}, and justify the arguments in the discussion of Eq. \eqref{sq:eq:fluc1}.
At finite temperature, the thermal magnon will play its role, leading to a larger uncertainty of the spin components.

\subsection{Quantum entanglement}\label{sec_en_measure}
The quantum state of a physical system is represented by a state vector in Hilbert space. Depending on the dimension of the Hilbert space, two types of states are classified. (i) If the dimension of the Hilbert space is finite, it is a discrete variable system. The best-known example is the spin-1/2 system or qubit, with only two elements in the Hilbert space ($\pm 1/2$, or up and down, or 0 and 1). Effective qubits have been realized in various solid-state platforms, including superconducting Josephson junctions \cite{MartinisPRL2002,MakRMP2001}, photonic systems \cite{KokRMP2007,SergAPR2019}, NV centers \cite{ChildressNV2006}, silicon-vacancy (SV) centers \cite{LachPRL2014,PingPRL2014}, nuclear spins \cite{KaneNature1998}, quantum dots \cite{LossPRA1998}, trapped ions \cite{CiracPRL1995} and topological states of matter \cite{NayakRMP2008}. Coherent control of qubits has become the cornerstone of quantum information processing. (ii) If the dimension of the Hilbert space is infinite, this is a continuous variable system. Examples are the position and momentum degrees of freedom of a particle. Besides photons, quasi-particle excitations in solid-state physics are also classified as a continuous variable system, for example, phonons and magnons, because the spectrum of these quasi-particles is continuous. The quantum states of quasi-particles, such as the antibunched state, squeezed state, cat state and so on, provide a promising alternative for quantum computing, quantum communication and quantum key distribution \cite{BraunRMP2005,WeedRMP2012} as we shall see below.

Entanglement corresponds to a nonclassical relation between two distant quantum systems and was first proposed by Einstein, Podolsky, Rosen and Schr\"{o}dinger in the 1930s \cite{EPR1935,Schrodinger1935,Schrodinger1936}. Later, such non-local correlations were shown to yield Bell inequalities \cite{Bell1964} that have been experimentally tested since 1972 \cite{Freedman1972}. Nowadays, entanglement of two or more quantum systems has become an indispensable resource for quantum cryptography, quantum teleportation and quantum computation \cite{Nielsen2000}. A fundamental question in this field is how to quantify and manipulate the entanglement of two or more systems. A single answer to this question does not exist. In general, the problem of deciding whether a state is entangled or not is difficult and is considered NP-hard under some reductions \cite{Gurvits2003,Ghar2010}. Only in low dimensions and for some special cases, one can quantify the entanglement based on a given density matrix of the joint system. Furthermore, the existing criteria for entanglement for discrete and continuous quantum systems are quite different \cite{GuhneReview2009}, as briefly summarized below. Because it is not our aim to extensively discuss entanglement measures in this review, we will focus on the most important and widely used measures for quantum magnonics.

\subsubsection{Discrete variable system: qubit} \label{sec_en_qubit}
{\it Entanglement entropy.} The most popular measure of entanglement for pure states is the von Neumann entanglement entropy \cite{BZbook2006}. Let us consider a physical system, that has been partitioned into two parts labelled 1 and 2, described by the density matrix $\rho_{12}$. We wish to examine the entanglement between these two parts. The von Neumann entanglement entropy is defined as $E \equiv -\text{tr}(\hat{\rho}_1 \ln \hat{\rho}_1)$, where $\hat{\rho}_1$ is the reduced density matrix of the subsystem, obtained by tracing out the degrees of freedom of the other subsystem, i.e., $\hat{\rho}_1=\text{tr}_2(\hat{\rho}_{12})$. If the two systems are separable, $\hat{\rho}_1$ is the density matrix of a pure state satisfying $\hat{\rho}_1^2=\hat{\rho}_1$, thus the entropy $E(\hat{\rho}_1)$ vanishes. On the other hand, if the two subsystems are entangled, $\hat{\rho}_1$ will be a mixed state with finite entropy. The magnitude of $E$ directly quantifies the strength of entanglement. Considering two entangled qubits, the maximally entangled states are the four Bell states
\begin{equation}
\frac{1}{\sqrt{2}} \left(| 01\rangle + | 10\rangle \right ), \frac{1}{\sqrt{2}}\left(| 00\rangle + | 11\rangle \right),
\frac{1}{\sqrt{2}} \left(| 01\rangle - | 10\rangle \right ), \frac{1}{\sqrt{2}}\left(| 00\rangle - | 11\rangle \right)
\end{equation}
with entropy $\ln2$, which form a maximally entangled basis. For the two-qutrit case, one maximally entangled state is $|\varphi \rangle= (| 11\rangle + | 22\rangle + | 33\rangle)/\sqrt{3}$  with entanglement $\ln3$ \cite{Caves2000}.
The von Neumann entanglement entropy has become a standard theoretical tool for characterizing, for example, the topological ground states and excitations of quantum spin liquids \cite{BalentsNature2010,SavaryIOP2016}. Its strength as an entanglement measure comes from its broad applicability to pure quantum states - discrete or continuum. Furthermore, by partitioning the system of interest in different ways, it allows for revealing new information about the quantum states \cite{EisertRMP2010,AmicoRMP2008}.

{\it Concurrence.} In most cases, the von Neumann entanglement entropy is a good metric for pure states only. When the two qubits are subject to relaxation and dephasing, the joint density matrix $\hat{\rho}_{12}$ represents a mixed state. In general, a mixed state is called separable, if there exist convex weights $p_i$ and product states $\hat{\rho}_{1,i}\otimes \hat{\rho}_{2,i}$ such that
$\hat{\rho}_{12}=\sum_i p_i \hat{\rho}_{1,i}\otimes \hat{\rho}_{2,i}$ holds. Otherwise the state is called an entangled state \cite{GuhneReview2009}.

Now the entropy cannot distinguish between classical and quantum correlations and it is no longer a good entanglement measure, because the von Neumann entropy of a subsystem is not zero when the two subsystems are separable \cite{VedralPRL1997}. Other metrics for entanglement quantification are now more useful, for example, the entanglement as measured by Wooters' concurrence \cite{WootersPRL1998}
\begin{equation}
C_{12}=\max (0,\lambda_1-\lambda_2-\lambda_3-\lambda_4),
\end{equation}
where $\lambda_1,\lambda_2,\lambda_3,\lambda_4$ are the square roots
of the eigenvalues of $\hat{\rho}_{12}[ (\hat{\sigma}^y \otimes \hat{\sigma}^y)
\hat{\rho}^*_{12} (\hat{\sigma}^y \otimes \hat{\sigma}^y)]$ in the decreasing order, and where $\hat{\sigma}^\nu~(\nu=x,y,z)$ are Pauli matrices and $\hat{\rho}^*_{12}$ is the complex conjugate of the joint density matrix $\hat{\rho}_{12}$.
The concurrence is zero if and only if the two qubits are in a separable state. Concurrence has the merit of universal applicability to arbitrary two-qubit states and a larger value of $C_{12}$ indicates a stronger entanglement. Its generalization to high dimensional pure states was reported in \cite{RungtaPRA2001}.

\begin{figure}
  \centering
  \includegraphics[width=1.0\textwidth]{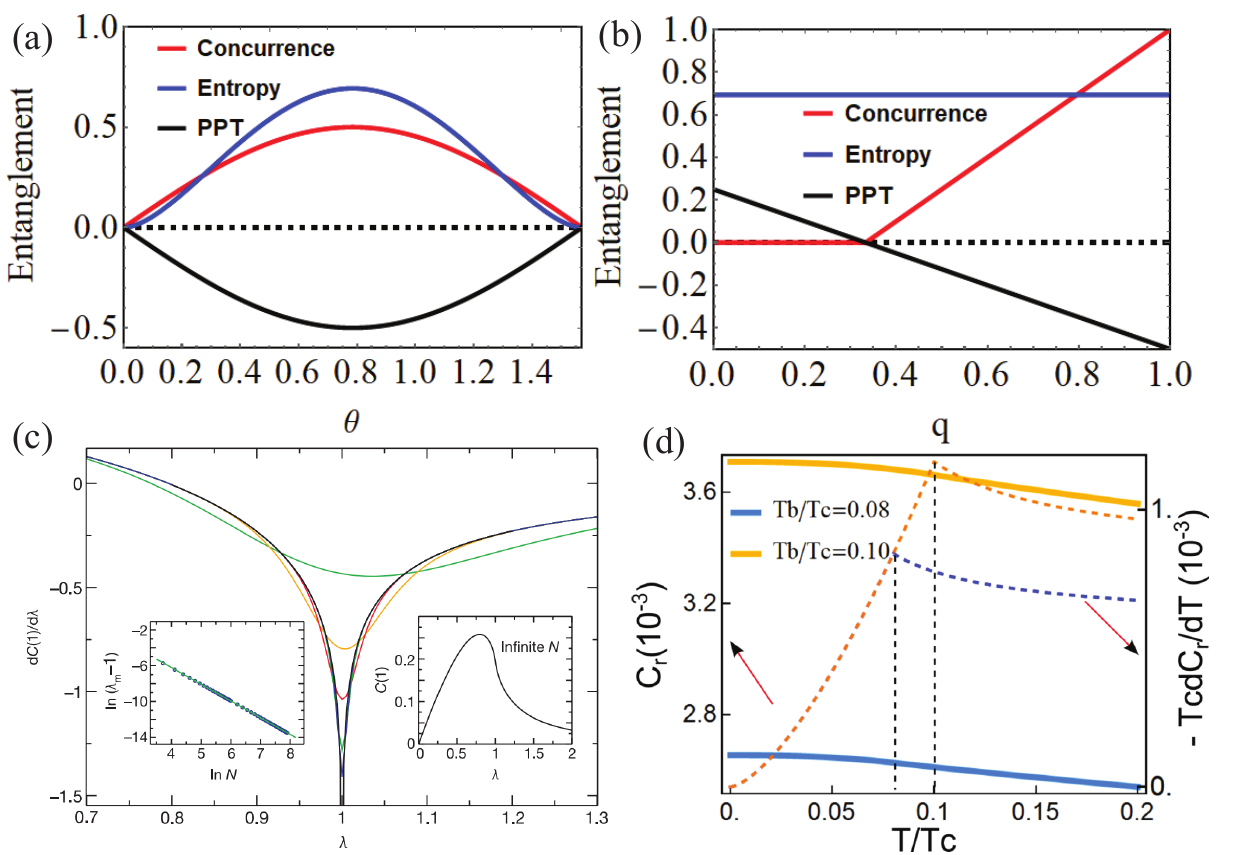}\\
  \caption{Entanglement measures for a pure state (a) and mixed state (b). For PPT criteria, the minimum eigenvalue of $\hat{\rho}^{T_B}$ is plotted and its negativeness indicates the existence of entanglement. (c) Concurrence of neighboring spins in a ferromagnetic spin-1/2 chain as a function of the reduced external field $\lambda$. $\lambda \equiv 2J/h$. As the system size increases, the divergence of $dC/d\lambda$ becomes more pronounced. (d) Concurrence of neighboring spins as a function of temperature in a magnonic system supporting Bose-Einstein condensation. $T_B$ and $T_C$ are respectively the condensation temperature and Curie temperature of the system. Source: Figures (c) and (d) are adapted from Refs. \cite{OsterlohNature2002,YuanBEC2018}. }\label{measure}
\end{figure}

{\it Peres-Herodecki criteria.} Another important measure for entanglement is the Peres-Herodecki criterion or positive partial transpose (PPT) \cite{PPT1996,Horodecki1996}. It states that the partial transpose of the joint density matrix of a bipartite system being positive definite is necessary for the joint states being separable. In systems with $2\times 2$ and $2\times3$ dimensional Hilbert spaces, this condition is also sufficient. This criteria can be extended to the infinite dimensional case as we shall see below.
As an example to illustrate the usefulness and relation between these measures, we first consider a two-qubit pure state, $|\varphi \rangle = \cos \theta |01\rangle - \sin \theta |10\rangle$ and plot the entanglement between the two qubits using the measures that were introduced, in Fig. \ref{measure}(a). Here both concurrence and von Neumann entropy, though different in magnitude, consistently capture the entanglement as a function of $\theta$. For a mixed state, we consider the Werner state \cite{WernerPRA1989,BarbPRL2004} $\hat{\rho}_{12}=(1-q)\hat{I}/4 + q|\varphi_s \rangle \langle \varphi_s|$, where $\hat{I}$ is a $4\times 4$ identity matrix and $|\varphi_s \rangle =( |01\rangle -  |10\rangle )/\sqrt{2}$ is the spin singlet-state. The entanglement of such a state is plotted in Fig. \ref{measure}(b). Now the concurrence and PPT criteria give mutually consistent predictions of the amount of entanglement, but the von Neumann entropy fails, as expected.

In a many-spin system, the entanglement between spins can behave as a measure for a quantum phase transition \cite{OsterlohNature2002,VidalPRL2003,VerstPRL2004,OsterPRL2006,PollPRL2009,DePRL2012,LeePRL2014,PezzePRL2017,VidmarPRL2018,DeReview2018}. Osterloh et al. \cite{OsterlohNature2002} first studied a ferromagnetic spin-1/2 chain with an exchange coupling $J$ and transverse field $h$, described by the Hamiltonian
\begin{equation}
\hat{\mathcal{H}} = -\frac{J}{2} (1+\delta ) \sum_{i=1}^N \hat{\sigma}_i^x\hat{\sigma}_{i+1}^x -\frac{J}{2} (1-\delta ) \sum_{i=1}^N \hat{\sigma}_i^y\hat{\sigma}_{i+1}^y-h\sum_{i=1}^N \hat{\sigma}_z.
\end{equation}
When $0<\delta \le 1$, the system undergoes a quantum phase transition for $N= \infty$ from ordered to disordered phase at the critical value $\lambda \equiv J/2h=1$. The authors calculated the entanglement between neighboring spins as quantified by concurrence $C$ as a function of $\lambda$, as shown in Fig. \ref{measure}(c) and found a scaling behavior of the first-order derivative of $C$ at the transition point. This suggests that the entanglement can be a good indicator of the quantum phase transition, and thus provides a way to connect the theory of critical phenomena with quantum information. Yuan et al. \cite{YuanBEC2018} studied the entanglement of spins in a magnetic systems allowing for magnon BEC [see Section \ref{sec:quantummanybodystatesofmagnons}] and found that the entanglement exhibits a distinct kink behavior near the condensation temperature, as shown Fig. \ref{measure}(d). Below the condensation temperature, there exists a finite entanglement between distant spins even when they are well separated in space, which is due to the off-diagonal long-range order in the system \cite{YangRMP1962}. Similar results were also reported by Wong et al. \cite{WongPRB2017}. Besides, the entanglement of spins can also serve as a measure of the transition between collinear and noncollinear magnetic structures \cite{YuanQDW2018}. Recently, Zou et al. \cite{ZouPRB2020} showed that the amount of entanglement of arbitrary spins in a magnetic system is tunable by external magnetic fields.

\subsubsection{Continuous variable system: magnon, photon, phonon} \label{sec_cvs}
A continuous variable system has an infinite dimensional Hilbert space, and therefore the dimension of the density matrix to describe such a system is also infinite.
Now, the entanglement measures that are obtained by directly evaluating the properties of the density matrix become difficult to implement. It is under such circumstances that the von Neumann entanglement entropy becomes a powerful metric for characterizing entanglement of pure states~\cite{SavaryIOP2016,EisertRMP2010,AmicoRMP2008,KamraPRB2019,HartmannPRB2021}. We refer the readers to other reviews~\cite{SavaryIOP2016,EisertRMP2010,AmicoRMP2008} for further examples. Here, we confine our discussion to other entanglement metrics that are more useful in the presence of dissipation. First, one has to find manageable representations to reduce the amount of degrees of freedom in such systems \cite{AdessoReview2007}.  A widely-used representation is based on the position and momentum space of a quasi-particle, defined as $\hat{q}\equiv (\hat{a}+\hat{a}^\dagger)/\sqrt{2},\hat{p}\equiv(\hat{a}-\hat{a}^\dagger)/\sqrt{2}i$, where $\hat{a}$ ($\hat{a}^\dagger$) is the annihilation (creation) operator of the quasi-particle excitation. Here $\hat{a}$ and $\hat{a}^\dagger$ satisfy the bosonic commutation relation $[\hat{a},\hat{a}^\dagger]=1$, which is a natural implication of the commutation relation between the position and momentum operators  $[\hat{q},\hat{p}]=i$. In general, for a continuous variable system containing $N$ types of excitations, the quantum state of the system is described by a vector $(\hat{q}_1,\hat{p}_1,\hat{q}_2,\hat{p}_2,...\hat{q}_N,\hat{p}_N)^T$ with a reduced dimension of $2N$. Here we focus on a bipartite system described by a four dimensional state vector $\hat{\mathbf{R}}=(\hat{q}_1,\hat{p}_1,\hat{q}_2,\hat{p}_2)^T$, which satisfies the commutation relations
\begin{equation}\label{commut}
[\hat{R}_i,\hat{R}_j]=i\Omega_{ij},\mathbf{\Omega} = \left (\begin{array}{cc}
                                                                                       \mathbf{J} & 0 \\
                                                                                       0 & \mathbf{J}
                                                                                     \end{array}\right ),
\mathbf{J} = \left (\begin{array}{cc}
                                                                                       0 & 1 \\
                                                                                       -1 & 0
                                                                                     \end{array}\right ).
\end{equation}
In this representation, the density matrix can be expressed via the Wigner function of the system
\begin{equation}
W(\mathbf{q},\mathbf{p})=\frac{1}{\pi^2} \int d^2 \mathbf{q}' \langle \mathbf{q-q}'|\hat{\rho} |\mathbf{q+q}'\rangle e^{2i\mathbf{q}'\cdot \mathbf{p}},
\end{equation}
where $\mathbf{q}=(q_1,q_2)$ and $\mathbf{p}=(p_1,p_2)$. In 2000, Simon \cite{SimonPRL2000} generalized the PPT criterion presented in Section \ref{sec_en_qubit} to bipartite continuous variable states as,

\textit{Theorem: If $\hat{\rho}$ is separable, its Wigner distribution necessarily goes over into a Wigner distribution under the phase space mirror reflection $\hat{\mathbf{R}} \rightarrow \Lambda \hat{\mathbf{R}}$ with $\Lambda = diag(1,1,1,-1)$.}

On the other hand, the commutation relations Eq. (\ref{commut}) and non-negativeness of the density matrix leads to the following uncertainty inequality \cite{SimonPRA1994}
\begin{equation} \label{uncertainty}
\mathbf{V}+\frac{i}{2} \mathbf{\Omega} \ge 0,
\end{equation}
where $\mathbf{V}$ is the bipartite covariance matrix defined as $V_{ij}\equiv \langle (\hat{R}_i- \langle \hat{R}_i \rangle) (\hat{R}_j- \langle \hat{R}_j \rangle) \rangle$. Note that the first moments $\langle \hat{R}_i \rangle$ can be arbitrarily adjusted by displacement operations in the phase space, which do not alter the entanglement properties. Therefore, it is usually adjusted to be zero without loss of generality \cite{AdessoReview2007}.

By combining the PPT criteria and the uncertainty inequality (\ref{uncertainty}), Simon found a necessary condition for separability of a bipartite quantum state
\begin{equation}
\det \mathbf{A} \det \mathbf{B} + \left(\frac{1}{4} -\det \mathbf{C}\right )^2 - \text{tr} (\mathbf{AJCJBJC}^T\mathbf{J}) \ge \frac{1}{4} (\det \mathbf{A} + \det \mathbf{B}),
\end{equation}
where $\det \mathbf{A}(\mathbf{B})$ is the determinant of $\mathbf{A}(\mathbf{B})$ and the matrices $\mathbf{A},\mathbf{B}$ and $\mathbf{C}$ are defined by expressing the covariance matrix in the following $2\times2$ block matrix
\begin{equation}\label{cv_twomode}
\mathbf{V}= \left (\begin{array}{cc}
                                                                                       \mathbf{A} & \mathbf{C} \\
                                                                                       \mathbf{C}^T & \mathbf{B}
                                                                                     \end{array}\right ).
\end{equation}
The violation of this inequality implies a finite entanglement between the two particles. It was further proven that this inequality is a necessary and sufficient condition for separability, for all bipartite Gaussian states. Here, a Gaussian state is a class of states whose Wigner distribution is a Gaussian function in phase space, such as thermal states, squeezed states, and coherent states \cite{BraunRMP2005}. An equivalent separability criterion for bipartite Gaussian state was independently derived by Duan, Giedke, Cirac and Zoller (DGCZ) \cite{DuanPRL2020}, which is called the DGCZ inequality. Generalizations of the Simon criterion to a wider class of multi-mode states without limiting  oneself to two-mode Gaussian states were attempted in \cite{HilleryPRL2006,SerafiniPRL2006,ShchukinPRA2006,AgarwalNJP2005}.

{\it Logarithmic negativity.}  Besides checking the inequality, another widely-used measure to quantity the entanglement of two-mode Gaussian states is logarithmic negativity \cite{VidalPRA2002}. It is defined as
\begin{equation}
E_N=\max(0,-\ln(2\eta^-)), \mathrm{with} ~\eta^-\equiv\frac{1}{\sqrt{2}}\sqrt{\left (\sum \mathbf{V} \right)^2 - \sqrt{\sum \mathbf{V}-4\det \mathbf{V}}},
\end{equation}
where $\sum \mathbf{V}= \det \mathbf{A} + \det \mathbf{B}-2\det \mathbf{C}$. Note that this measure is valid for both pure and mixed states, and it will be repeatedly used below to quantify the entanglement among magnons, photons and phonons.

As an example to illustrate the meaning of continuous variable entanglement, let us consider a two-mode squeezed state defined as
\begin{equation}
|r\rangle =\exp \left [r (\hat{a}_i^\dagger \hat{a}_j^\dagger - \hat{a}_i \hat{a}_j) \right ] |0\rangle_i \otimes |0\rangle_j,r\in \mathrm{Reals},
\end{equation}
whose covariance matrix is
\begin{equation}
\begin{aligned}
\mathbf{V}&=\left ( \begin{array}{cc}
            \mathbf{A} & \mathbf{C} \\
            \mathbf{C}^T & \mathbf{B}
          \end{array} \right )=\frac{1}{2} \left ( \begin{array}{cccc}
            \cosh 2r & 0& \sinh 2r & 0 \\
            0& \cosh 2r & 0 & -\sinh 2r \\
            \sinh 2r& 0 & \cosh 2r & 0 \\
            0& -\sinh 2r & 0 & \cosh 2r \\
          \end{array}
\right ).
\end{aligned}
\end{equation}
We readily derive
\begin{align}
\sum \mathbf{V} =\det \mathbf{A} +\det \mathbf{B}-2\det \mathbf{C}=\frac{1}{2}\cosh 4r,\mathrm{Det}\mathbf{V} &=\frac{1}{16},
\end{align}
and the smallest symplectic eigenvalue
\begin{align}
\eta^- =\frac{1}{2}\left [ \cosh 4r - (\cosh^2 4r -1)\right ]^{1/2}=\frac{1}{2}e^{-2|r|}.
\end{align}
Therefore the entanglement is quantified as $E_N = -\ln(2\eta^-)=2|r|$. On the other hand, by defining the two-mode position and momentum quadratures as $\hat{X} =(\hat{a}_i + \hat{a}_i^\dagger +\hat{a}_j + \hat{a}_j^\dagger )/2^{3/2}$ and $\hat{P}=(\hat{a}_i - \hat{a}_i^\dagger +\hat{a}_j - \hat{a}_j^\dagger )/(2^{3/2}i)$, the uncertainties of the quadratures are evaluated as $\Delta X= e^{r}/2$ and $\Delta P=e^{-r}/2$. Given $r>0$, we see that the larger the squeezing of the momentum uncertainty is, the stronger the two modes are entangled. Hence the strength of entanglement for a two-mode continuous variable system may be understood as how much the collective quadrature is squeezed below the quantum limit.

{\it EPR steering.} Further, when two subsystems are entangled, one intriguing property is that one subsystem can steer the state of the other by local operations within its own Hilbert space and by classical communication (LOCC). This concept is known as EPR steering and was originally proposed by Schr\"{o}dinger \cite{Schrodinger1935,Schrodinger1936}. The detection of the steering was first proposed by Reid \cite{ReidPRA1989} and experimentally verified later \cite{OuPRL1992,BowenPRL2003,HowellPRL2004}. Unlike entanglement, EPR steering is an asymmetric property between two subsystems, i.e., one subsystem may steer the quantum state of the other, but not vice versa \cite{ReidRMP2009}. This makes EPR steering useful in realizing one-sided quantum key distribution \cite{BranPRA2012,KogiasPRL2015}. More subtle differences between entanglement and EPR steering can be found in Refs. \cite{WisemanPRL2007,HePRA2011,HePRL2015}. To quantify the strength of EPR steering, Kogias et al. \cite{KogiasPRL2015} proposed a computable measure valid for arbitrary bipartite Gaussian states based on their covariance matrix (\ref{cv_twomode}) as
\begin{equation}
\mathcal{G}^{a\rightarrow b} = \max \left \{0,\frac{1}{2} \ln \frac{\det 2\mathbf{A}}{\det 2\mathbf{V}} \right \},\mathcal{G}^{b\rightarrow a} = \max \left \{0,\frac{1}{2} \ln \frac{\det 2\mathbf{B}}{\det 2\mathbf{V}} \right \}.
\end{equation}
The magnitude of $\mathcal{G}^{a\rightarrow b}$ ($\mathcal{G}^{b\rightarrow a}$ ) quantifies how much the joint state is steerable, by local measurements on the $a$ ($b$) side. The operational interpretation of this quantifier can be reduced to the seminal proposal by Reid \cite{ReidPRA1989}.

So far, we have introduced magnons, the dynamic equations that describe the quantum behavior of magnons, and the language to describe the entanglement between two subsystems for both discrete and continuous variable systems. Equipped with these preliminary knowledge, we review the recent developments concerning the generation and manipulation of quantum states of magnons in the next section.

\section{Quantum states of magnons}\label{sec_qmstate}
It is known that the particle-like nature of light gave birth to the concept of photons, possessing many intriguing quantum properties, such as the photoelectric effect, BEC, squeezing, antibunching, and entanglement. As bosonic particles, magnons share many similarities to photons, but their quantum nature was not thoroughly studied until very recently. In such an infant stage, there still exist some controversial understandings in whether a phenomenon is quantum or classical, such as magnon BEC. In this section, we would like to review the recent developments in this direction including single-magnon state in Section \ref{sec_singlemagnon}, squeezed states in Section \ref{sec_squeezedmagnon}, magnonic cat states in Section \ref{sec_cat_magnon}, and quantum many-body states of magnons in Section \ref{sec:quantummanybodystatesofmagnons}.

\subsection{Single-magnon state} \label{sec_singlemagnon}
\begin{figure}
  \centering
  \includegraphics[width=1.0\textwidth]{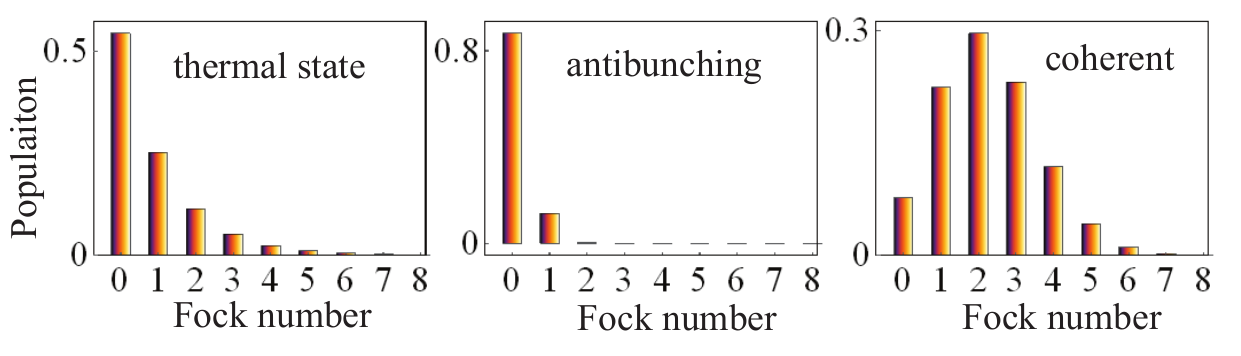}\\
  \caption{Illustration of the thermal state, antibunched state and coherent state of bosons. They are respectively characterized by super-Poissonian ($g^{(2)}(0)>1$), sub-Poissonian ($g^{(2)}(0)<1$) and Poissonian statistics ($g^{(2)}(0)=1$). Source: The figures are adapted from Ref. \cite{Yuanantibunch2020}.}\label{antibunch}
\end{figure}

Magnons that are excited in a magnet will reach equilibrium with the environment in the absence of external driving and form a thermal magnon gas.
Such a thermal state is usually described by a super-Poisson distribution in Fock space, where the mean magnon occupation is smaller than the variance of the magnon number. Under a strong driving, for example by microwaves, the magnons will reach a coherent state, with mean magnon number equal to its variance [see Eq. \eqref{sq:eq:cohprop}]. As discussed above in Section \ref{sec:cohstate}, Coherent magnons behave classically as a wavepacket and can serve as carrier for information processing. In quantum science and technology, it is meaningful to study nonclassical states of magnons for the integration of magnons with other quantum platforms.
In line with the single-photon state in quantum optics \cite{LounisRev2005}, a single-magnon state is a quantum state with large occupation probability of the first-excited state corresponding to a single magnon. The excitation of two or more magnons should be negligible. The magnon population in Fock space then obeys sub-Poisson statistics, which is different from a thermal magnon gas and a coherently driven magnonic system. Mathematically, these states can be characterized by a zero-delay second-order correlation function \cite{DFWalls} defined as
\begin{equation}
g^{(2)}(0)\equiv\frac{\langle \hat{a}^\dagger \hat{a}^\dagger \hat{a} \hat{a} \rangle}{\langle \hat{a}^\dagger \hat{a} \rangle^2} =1+\frac{\langle (\Delta \hat{n})^2 \rangle -\langle \hat{n} \rangle}{\langle \hat{n} \rangle ^2 },
\end{equation}
where $\hat{a}$ ($\hat{a}^\dagger$) is the magnon annihilation (creation) operator, $\hat{n}= \hat{a}^\dagger \hat{a}$ is number operator, $\Delta \hat{n}= \hat{n}- \langle \hat{n} \rangle$ and  $\langle \hat{A} \rangle\equiv \text{tr}(\hat{\rho} \hat{A})$ is the ensemble average of the observable $\hat{A}$. According to the value of $g^{(2)}(0)$ or statistics of the magnon gas, three regimes are identified \cite{PlenioRMP1998}: (i) When the magnon distribution is super-Poissonian, $\langle (\Delta \hat{n})^2 \rangle > \langle \hat{n} \rangle$, then $g^{(2)}(0)>1$. For example, $g^{(2)}(0)=2$ is achieved by a thermal magnon gas, which is bunched due to its bosonic statistics \cite{BenderPRL2019,KimPRA1989}. (ii) When the magnon distribution is Poissonian, for example, a coherent state, one has $(\Delta \hat{n})^2 \rangle = \langle \hat{n} \rangle$ [see Eq. \eqref{sq:eq:cohprop}] and thus $g^{(2)}(0)=1$. The magnonic system under strong driving falls into this class, as already discussed in Sections \ref{sec:cohstate} and \ref{sec_qm_equation}. (iii) When the magnon distribution is sub-Poissonion, $\langle (\Delta \hat{n})^2 \rangle < \langle \hat{n} \rangle$ and $g^{(2)}(0)<1$ implying antibunched magnons. This is a purely quantum-mechanical state that cannot be captured by classical statistics \cite{ReidPRA1986}. The limiting case $g^{(2)}(0)=0$ indicates a perfect single-magnon source. Such case leads to a magnon blockade, i.e., the excitation of a second magnon is blocked. The Fock distributions of a typical thermal state, antibunched state and coherent state are shown in Fig. \ref{antibunch}. Bender et al. \cite{BenderPRL2019} introduced a normalized spin current cross-correlation function $c^{(2)}(0)$ to quantify the quantum correlations of spin currents injected into two metallic leads attached to a ferromagnetic insulator. This function approaches 1 and 2 when magnons are in a coherent and thermal states, respectively, resembling the behavior of $g^{(2)}(0)$.

\begin{figure}
  \centering
  \includegraphics[width=1.0\textwidth]{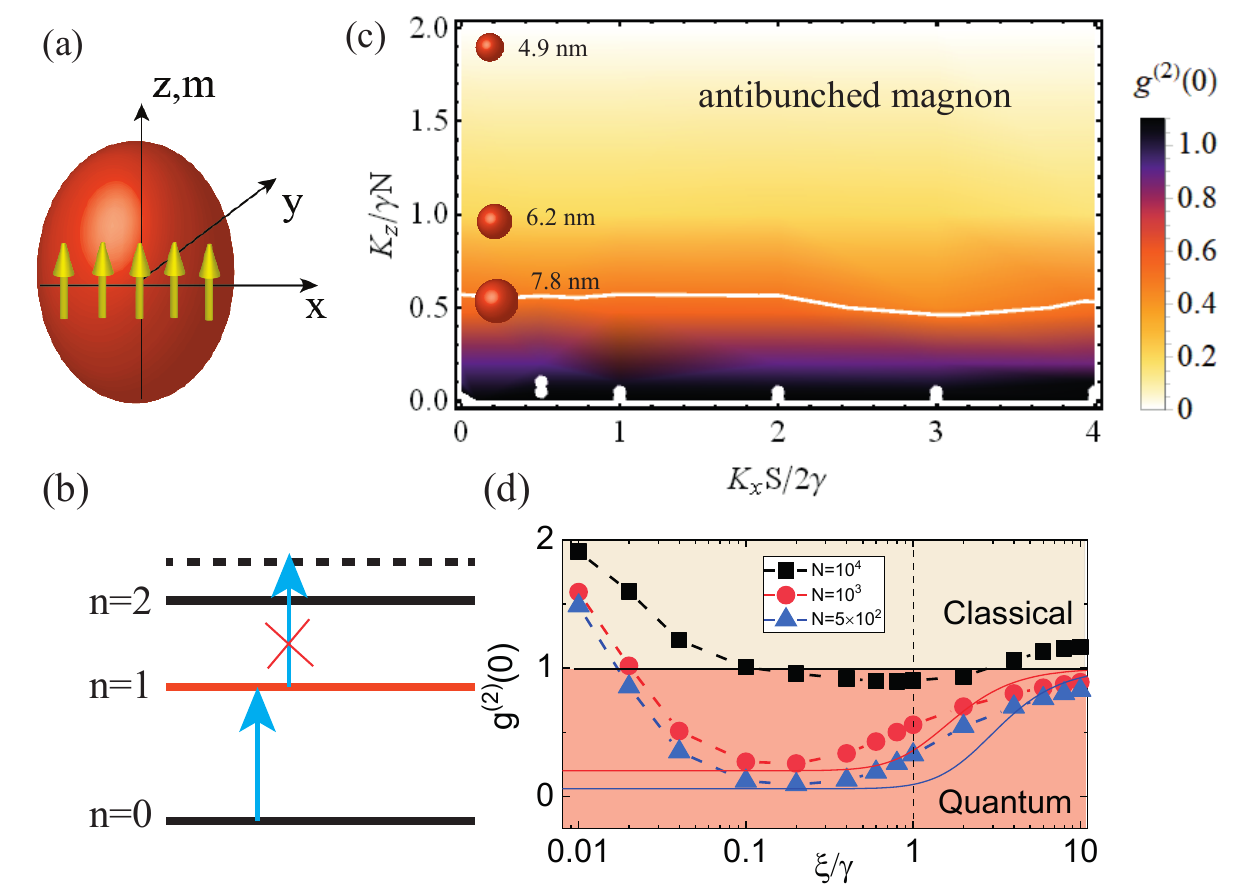}\\
  \caption{(a) Schematic of a nanomagnet. (b) Anharmonic energy-level distribution of the system with strong nonlinearity. The excitation of a magnon from $n=1$ to 2 by a microwave with frequency $\omega_1-\omega_0$ will be suppressed due to frequency mismatch. (c) Second-order correlation function $g^{(2)} (0)$ of magnons as a function of $K_xS/2\gamma$ and $K_z/\gamma N$. The orange balls sketch typical magnetic spheres with bunched and antibunched behaviors. (d) Correlation function $g^{(2)} (0)$ as a function of driving strength for $N=10^4$ (black squares), $10^3$ (red dots) and $5 \times 10^2$ (blue triangles), respectively. Source: The figures are adapted from Ref. \cite{Yuanantibunch2020}.}\label{antibunch_yuan}
\end{figure}

How to generate and manipulate antibunched magnons has started to attract attention recently. The conventional method to generate a single-magnon state is to enhance the nonlinearities in the magnonic system, which induces anharmonicity of the energy levels of the system. Kerr-type nonlinearities $(\hat{a}^\dagger \hat{a})^2$ have been widely used to study photon blockade in quantum optics \cite{PaulRMP1982,LiewPRL2010,HoffmanPRL2011,RablPRL2011}. Wang et al. \cite{WangPRB2016} experimentally demonstrated the magnon Kerr effect in a small yttrium-iron-garnet (YIG) sphere, which induced a perceptible shift of the magnon frequency. Yuan et al. \cite{Yuanantibunch2020} studied the magnon excitation in a nanometer-sized magnet as shown in Fig. \ref{antibunch_yuan}(a) and investigated the role of Kerr-type nonlinearities on the magnon correlations. In particular, they considered a biaxial magnet described by the Hamiltonian (\ref{biaxialHam}) and considered the Hamiltonian up to the fourth order
\begin{equation}
	\hat{\mathcal{H}}=\omega_a \hat{a}^\dagger \hat{a} + w(\hat{a}^\dagger \hat{a}^\dagger + \hat{a}\hat{a}) + v(\hat{a}^\dagger \hat{a})^2+u(\hat{a}^\dagger \hat{a}\hat{a}\hat{a} +\hat{a}^\dagger \hat{a}^\dagger \hat{a}^\dagger \hat{a}) +
	\xi (\hat{a} e^{i\omega t} + \hat{a}^\dagger e^{-i\omega t}),
\end{equation}
where $\omega_a=2K_z+K_x+\mu_0 g_e \mu_B H$, $w=K_xS/2$, $v=-K_z/N-K_x/2N$, $u =-K_x/4N$, with $N$ the number of spins in the magnet. Here, $\xi$ is driving term produced by an external microwave. The nonlinear terms caused by four-magnon interactions may come from either shape or magnetocrystalline anisotropy and are inversely proportional to the number of spins \cite{Yuanantibunch2020,HaghJAP2020}. Therefore, as the magnet becomes smaller, the interaction causes significant nonuniform spacing of the energy levels as shown in Fig. \ref{antibunch_yuan}(b). When the input microwave driving matches the frequency of the first excited level ($n=0 \rightarrow 1$), it will efficiently pump the magnons to the level $n=1$, but cannot excite them further due to the frequency mismatch. Therefore, magnon antibunching may be realized in a small nanomagnet and the required size of magnet is estimated to be below 10 nm, which is accessible within current fabrication techniques \cite{OyaSR2015,OroJPC2016}. A full phase diagram in the $(K_xS/2\gamma,K_z/\gamma N)$ plane is show in Fig. \ref{antibunch_yuan}(c). To study the classical-quantum transition in the presence of thermal fluctuations, nonlinearities and external drivings, the authors further simulated the steady-state correlations of magnons, the results of which are shown in Fig. \ref{antibunch_yuan}(d). In the weak driving regime, thermal effects dominate the steady state such that the steady-state correlations are close to 2. In the opposite limit of strong driving, the steady state approaches a coherent state with $g^{(2)}(0)=1$. Only in the regime of intermediate driving, the nonlinearity generates an antibunched state. At temperatures below 1 K, this magnonic quantum state survives. To detect the antibunched magnons, it was proposed to couple the magnonic system to photons and encode the magnon correlations onto the photons. By measuring the photon intensity correlations through a Hanbury Brown-Twiss interferometer \cite{HanburyNature1956}, one could probe the magnon antibunching.

\begin{figure}
  \centering
  \includegraphics[width=1.0\textwidth]{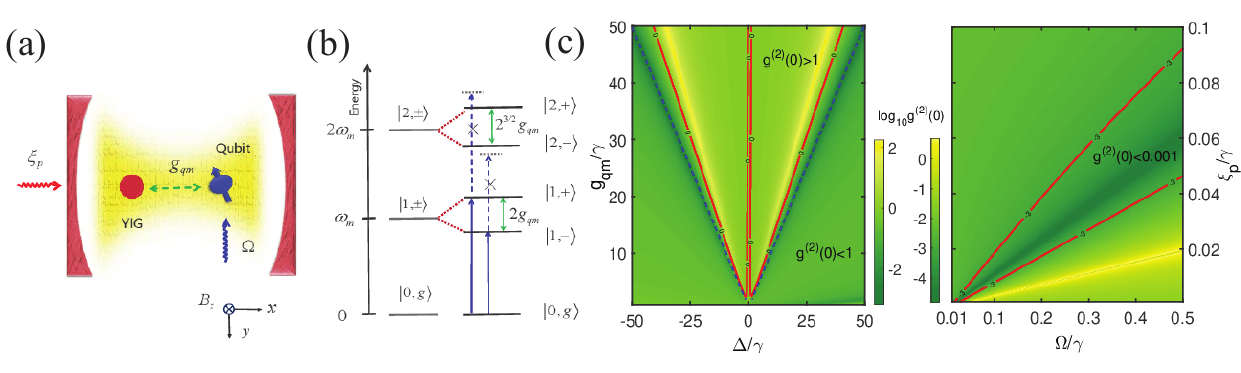}\\
  \caption{(a) Schematic illustration of a hybrid magnet-qubit system inside a cavity. (b) Anharmonic energy level diagram of the system induced by the qubit-magnon coupling $g_{qm}$. (c) Steady-state second-order correlation as a function of the frequency detuning ($\Delta$) between intrinsic magnon frequency and driving frequency, the qubit-magnon coupling strength ($g_{qm}$), the driving amplitude of the qubit ($\Omega$) and magnon ($\zeta_p$).  Source: The figures are adapted from Ref. \cite{LiuPRB2019}.}\label{antibunch_Liu}
\end{figure}

Another way to induce anharmonicities for magnons is to couple the magnon mode to an ancillary mode. Liu et al. \cite{LiuPRB2019} considered a hybrid magnet-qubit system inside a cavity [Fig. \ref{antibunch_Liu}(a)], where the magnet and cavity are coupled in a coherent way and are driven by microwaves with strengths $\xi_p$ and $\Omega$, respectively.  They found that the qubit mode can change the energy level diagram of the magnons to an anharmonic form, as shown in Fig. \ref{antibunch_Liu}(b), where $|n,g\rangle$ are the dressed states of the hybrid system with magnon number $n$ and the qubit in its ground state. Here, the transitions from the ground state $|0,g\rangle$ to $|1,\pm \rangle$ are allowed when the driving matches the frequency difference, but the transitions from $|1,\pm \rangle$ to $|2,\pm \rangle$ are suppressed due to a frequency mismatch, where $\pm$ denote the splitting of the qubit's energy level. By carefully manipulating the detuning between the driving and the intrinsic magnon frequency ($\Delta$), and the driving amplitudes of the qubit ($\Omega$) and magnon ($\zeta_p$), a nearly perfect magnon blockade with $g^{(2)}(0)<0.001$ is realized, as shown in Fig. \ref{antibunch_Liu}(c). Similar results were independently obtained by Xie et al. \cite{XiePRA2020}.

Besides the method to generate magnon blockade by modifying their energies, Wang et al. reported unconventional blockade in the absence of nonlinearities, by playing with the magnon-photon coupling strength, the gain of the cavity and loss of the magnet to generate a $\mathcal{PT}$ symmetric system \cite{WangAP2020}. In such a case, different channels interfere which leads to a significant occupation of the single-magnon energy level.

Lachance-Quirion et al. \cite{LachScience2020} have experimentally verified that one can detect the single-magnon state by coupling the magnonic system to a superconducting qubit. The details shall be discussed in Section \ref{sec_magnon_qubit}.   Bittencourt et al. \cite{BittenPRA2019} proposed a magnon-heralding protocol in a cavity optomagnonic setup. They first prepared the magnon state in vacuum. Then they sent a write pulse to generate an entangled photon-magnon state, and subsequently sent a second pulse to read the information of the photons. By measuring the statistical properties of the output photons, the magnon state collapses into a single-magnon Fock state. Hence the read photon is a witness of the heralding. This heralding protocol is sensitive to the initial state of the magnons and cooling of the magnons before the write pulses is preferred to reduce the initial mean number of magnons. The phononic counterpart of this proposal was reported by Galland et al. \cite{GallandPRL2014}. On the application side, the single-magnon excitation may be entangled to photons for Bell tests \cite{YuanBell2020} and it can also be used as a quantum memory to store information from an incoming photon \cite{TanjiPRL2009}. The realizations of single-magnon detectors is important to make use of magnetic states for quantum information processing \cite{PsaPRL2021}.

\subsection{Squeezed states}\label{sec_squeezedmagnon}
The Heisenberg uncertainty principle states that the product of quantum fluctuations in two noncommuting observables has a minimum value determined by $\hbar$. Nevertheless, it is possible to realize states with reduced quantum noise in one observable at the expense of increased fluctuations in the other~\cite{WallsNature1983}. Such states, generally called ``squeezed states'', have been the subject of intense research in the field of quantum optics~\cite{Gerry2004,WallsNature1983,SchnabelPR2017,GrotePRL2013,LIGONP2011,LIGONP2013,LIGOPRL2016,Walls2008}. Considering a single electromagnetic or photon mode, its associated electric and magnetic fields become two such noncommuting observables. These are proportional to the two ``quadratures'' $\sim (\hat{c}^\dagger \pm \hat{c})$, where $\hat{c}$ is the annihilation operator for the photon mode~\cite{Gerry2004,Walls2008}. While a graphic and intuitive picture is provided by this squeezing of the quantum fluctuations, the wide range of quantum phenomena (such as entanglement) embodied by these states can be appreciated best via their wave functions.

Such squeezed states and related phenomena carry over to other bosonic systems, such as mechanical resonators or phonons~\cite{MeekhofPRL1996,GarrettScience1997,AspelmeyerRMP2014,WollmanScience2015}, in a natural way. In optical and mechanical systems, squeezed states are nonequilibrium in nature and are generated via external drives, for example. Magnets, in contrast, host robust squeezed magnonic states in equilibrium, as shown by Kamra and Belzig in 2016~\cite{KamraPRL2016,KamraPRB2016,KamraPRB2017}. This equilibrium squeezing results from energy minimization and forms the primary focus of the present subsection. A physically intuitive discussion, avoiding formalism, of this niche admitted by magnons and an overview have been presented in Ref.~\cite{KamraAPL2020}. Here, we follow a complementary approach and clarify some key aspects of these phenomena by introducing formalism and analyzing single-mode squeezing of the $\mathbf{k} = \pmb{0}$ mode in ferromagets~\cite{KamraPRL2016,KamraPRB2017}. This also provides us tools to appreciate the key ideas in a wide range of recent studies, which are briefly discussed below. In this process, we also review some important insights from quantum optics~\cite{Gerry2004,Walls2008}, but using our magnon language. The latter turns out to be particularly convenient as the two noncommuting observables (or quadratures) simply become the spin components transverse to the equilibrium order direction. The discussion below borrows heavily from Refs.~~\cite{Gerry2004,KamraPRL2016,KamraPRB2016,Skogvoll2021} which should be consulted for further details. Further, Section~\ref{sec:cohstate} serves as a prelude to the following discussion.

\begin{figure}
	\begin{center}
        \includegraphics[width=1.0\textwidth]{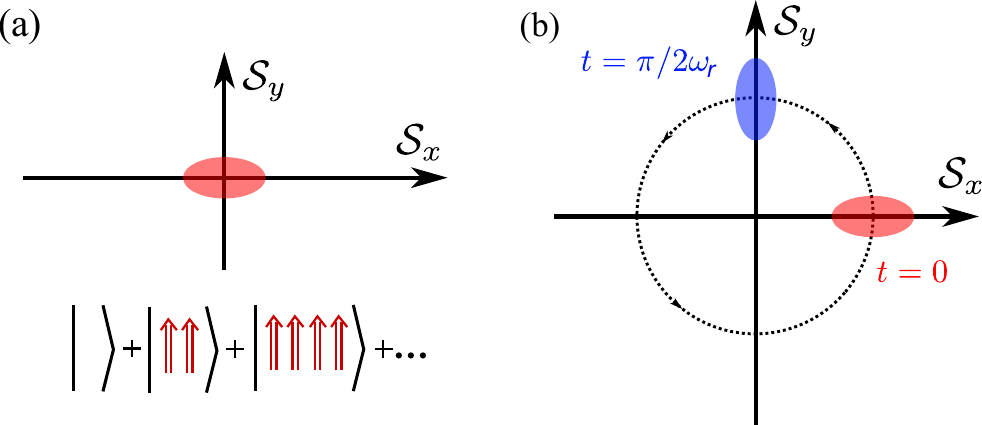}
		\caption{Phase space depiction of (a) squeezed vacuum state and (b) squeezed coherent state. (a) The squeezed vacuum is constituted by superposition of even magnon number states, schematically depicted in the lower panel~\cite{KamraAPL2020}. (b) Time evolution of the squeezed coherent state prepared at $t=0$ with squeezing in $\mathcal{S}_y$~\cite{Gerry2004}. Circular trajectory of the uncertainty region center represents a classical property. Phase-dependent quantum fluctuations due to squeezing of the uncertainty region is a quantum feature.}
		\label{sq:fig:sqofmag}
	\end{center}
\end{figure}

{\it Squeezed coherent state of magnon.} We first consider the $\mathbf{k} = \pmb{0}$ magnon mode in an isotropic ferromagnet represented by Eq.~\eqref{sq:eq:H1}, since it pertains to the transverse components of the total spin $\mathcal{S}_{x,y}$ [Eq.~\eqref{sq:eq:expect1}]. Can we accomplish a reduction of quantum noise in, say $\mathcal{S}_y$? The Heisenberg principle cannot be violated. Hence, such a reduction should accompany an increase in quantum fluctuations of $\mathcal{S}_x$.The so-called squeezed vacuum state $\ket{\eta}$ is defined as~\cite{Gerry2004,WallsNature1983,Walls2008}
\begin{align}\label{sq:eq:sqdef}
\ket{\eta} & = \hat{S}(\eta) \ket{0},  \quad \hat{S}(\eta) \equiv \exp \left[ \frac{1}{2} \left( \eta^* \hat{a}^2 - \eta \hat{a}^{\dagger 2} \right) \right],
\end{align}
with $\eta = r e^{i \theta}$, where $\hat{S}(\eta)$ is called the squeeze operator, $r > 0$ is known as the squeeze parameter, and $\theta$ determines the squeezing direction as clarified below. Employing the unitarity of $\hat{S}(\eta)$ and the identities~\cite{Gerry2004}
\begin{align}
\hat{S}^\dagger (\eta) = \hat{S}(- \eta), \quad \hat{S}^\dagger (\eta) \hat{a} \hat{S} (\eta) = \hat{a} \cosh r - \hat{a}^\dagger e^{i \theta} \sinh r,
\end{align}
we evaluate the quantum fluctuations in $\mathcal{S}_{x,y}$ as
\begin{align}
\Delta \mathcal{S}_x = & \sqrt{\frac{NS}{2}} \sqrt{\cosh^2 r + \sinh^2 r - 2 \sinh r \cosh r \cos \theta} ~ = \sqrt{\frac{NS}{2}} e^{r} \left(\mathrm{taking}~\theta = \pi \right), \label{sq:eq:deltasx} \\
 \Delta \mathcal{S}_y = & \sqrt{\frac{NS}{2}} \sqrt{\cosh^2 r + \sinh^2 r + 2 \sinh r \cosh r \cos \theta} ~ =  \sqrt{\frac{NS}{2}} e^{-r} \left(\mathrm{taking}~\theta = \pi \right). \label{sq:eq:deltasy}
\end{align}
The Heisenberg uncertainty region has thus been squeezed from a circle [Fig.~\ref{sq:fig:mag1}(a)] into an ellipse [Fig.~\ref{sq:fig:sqofmag}(a), upper panel] while maintaining its area. The squeezing of quantum fluctuations takes places along the direction making an angle of $\theta/2$ with the $x$-axis. Further insight into the squeezed vacuum is gained by expressing it using the magnon number states
\begin{align}
\hat{a}\ket{0} = 0, \implies \hat{S}(\eta) \hat{a} \hat{S}^\dagger (\eta) \hat{S}(\eta) \ket{0} = 0, \implies  \left( \hat{a} \cosh r + \hat{a}^\dagger e^{i \theta} \sinh r \right) \ket{\eta} = 0,
\end{align}
which yields the desired expansion
\begin{align}\label{sq:eq:ximag}
\ket{\eta} & = \frac{1}{\sqrt{\cosh r}} \sum_{n=0}^{\infty} (-1)^n ~ e^{i n \theta} ~ \frac{\sqrt{(2 n)!}}{2^n n!} ~ \left( \tanh r \right)^n ~ \ket{2n}.
\end{align}
Thus, the squeezed vacuum state is comprised of even magnon number states [schematically depicted in Fig.~\ref{sq:fig:sqofmag}(a), lower panel] with the corresponding probabilities decreasing with increasing $n$, but independent of $\theta$.

Equation \eqref{sq:eq:ximag} shows that the squeezed vacuum is, strictly speaking, not a vacuum at all. It consists of all even magnon number states, including the true vacuum $\ket{0}$. This also provides a hint regarding generation of such a squeezed vacuum in experiments. We need to create correlated magnon pairs on top of the true vacuum (isotropic ferromagnet ground state), for example via parametric pumping~\cite{DemokritovNature2006,LiPRA2019}. Such states, although commonplace in optical~\cite{SchnabelPR2017} and mechanical~\cite{MeekhofPRL1996,GarrettScience1997,AspelmeyerRMP2014,WollmanScience2015} systems, have not been experimentally observed in ferromagnets thus far. However, two-mode squeezed states (beyond the scope of ongoing discussion) in antiferromagnets have been achieved~\cite{KamraPRB2019,ZhaoPRL2004,ZhaoPRB2006,BossiniPRB2019}. Such states will also be briefly discussed in the context of magnon-magnon entanglement below in Section~\ref{sec_magnonmagnon}.

Often, the quantum nature of phenomena is tested via mathematical inequalities on observables. For example, the violation of Bell's inequality proves quantum entanglement~\cite{Bell1964}. We thus briefly mention one mathematical inequality that is used to characterize ``nonclassical'' states and ascertains squeezed vacuum to be such a state. The Glauber-Sudarshan~\cite{GlauberPR1963,SudarshanPRL1963} P-function $P(\beta)$ for a general state is defined via~\cite{Gerry2004}
\begin{align}
\hat{\rho} = & \int P(\beta) \ket{\beta} \bra{\beta} d^2 \beta,
\end{align}
where $\hat{\rho}$ is the system density matrix, and the integral is over the complex $\beta$ plane (which is also proportional to the phase-space). The P-function acts as a probability distribution in the phase-space and observables can often be written as $P(\beta)$-weighted integral over the $\beta$ plane. Since a classical probability distribution assumes only nonnegative values, a state is nonclassical if its P-function becomes negative in some part of the phase space. Evaluating quantum fluctuations [Eqs.~\eqref{sq:eq:deltasx} and \eqref{sq:eq:deltasy}] in terms of $P(\beta)$, it is seen that squeezing necessitates $P(\beta)$ becoming negative in at least some part of the phase space~\cite{Gerry2004}. This makes the squeezed vacuum a nonclassical state and we can expect various quantum features from it, as discussed further below.

So, coherent states are classical and squeezed states are quantum. Having both in one sounds like an oxymoron. Nevertheless, squeezed coherent state is generated (theoretically) as
\begin{align}
\ket{\beta,\eta} & = \hat{D}(\beta) \hat{S}(\eta) \ket{0}.
\end{align}
We wish to examine dynamics of this squeezed coherent state. A full derivation becomes too lengthy for us to reproduce here~\cite{Gerry2004,Walls2008}. Hence, we directly depict the phase-space picture of this dynamics in Fig.~\ref{sq:fig:sqofmag}(b). Here, we assume that the state $\ket{\beta,\eta}$ is realized using certain drives that are switched off at $t = 0$. Thereafter, the state undergoes its natural evolution due to the single-mode Hamiltonian Eq.~\eqref{sq:eq:H1}. As shown in Fig.~\ref{sq:fig:sqofmag}(b), the uncertainty region center undergoes circular precession. The quantum fluctuations squeezing in $\mathcal{S}_y$ at $t = 0$ appears in $\mathcal{S}_x$ after a quarter of the cycle. Thus, the state harbors phase and time dependent quantum noise~\cite{Gerry2004}. Generally, the quantum nature of this state is due to the Heisenberg uncertainty region squeezing while the circular precession is a classical property. The coherent state can thus be seen as a classical carrier or amplifier of the quantum features allowing the latter to be exploited in macroscopic systems~\cite{SchnabelPR2017}.

{\it Squeezed-magnons in anisotropic ferromagnets.} We now include magnetic anisotropy in describing our ferromagnet, which has a ground state configuration of spins pointing along $-\mathbf{e}_z$ due to the applied magnetic field $H \mathbf{e}_z$. The anisotropy may result from dipolar interaction and/or magnetocrystalline contributions~\cite{KamraPRL2016,KamraPRB2017}. In order to keep our discussion general while focusing on the $\mathbf{k} = \pmb{0}$ mode, we consider~\cite{Skogvoll2021}
\begin{align}
\hat{\mathcal{H}}_{\mathrm{an}} & = \sum_{i} \left[K_x \left( \hat{S}_{ix} \right)^2 + K_y \left( \hat{S}_{iy} \right)^2 + K_z \left( \hat{S}_{iz} \right)^2\right],
\end{align}
which results in the following magnon Hamiltonian
\begin{align}\label{sq:eq:sqham1}
\hat{\mathcal{H}} & = A \hat{a}^{\dagger} \hat{a} + B \left( \hat{a}^2 + \hat{a}^{\dagger 2} \right) = \omega_r \hat{b}^{\dagger} \hat{b},
\end{align}
with $A \equiv \mu_0 g_e \mu_B H + K_x S + K_y S - 2 K_z S$, and $B \equiv S(K_x - K_y)/2$. We assume $B < 0$ for the following discussion. The Hamiltonian above has been diagonalized via a Bogoliubov transform~\cite{HolsteinPR1940}: $\hat{a} = \hat{b} \cosh r  - \hat{b}^{\dagger} e^{i \theta} \sinh r $. We continue to denote the eigenmode energy by $\omega_r$, which now becomes $\omega_r= \sqrt{A^2 - 4 B^2}$. Further, $r$ and $\theta$ are governed by the relations: $\sinh r = 2|B|/\sqrt{(A + \omega_r)^2 - 4 B^2}$ and $\theta = \pi$, due to $B < 0$.

Magnons, represented by $\hat{a}$, are no longer the eigenmodes, which are annihilated by $\hat{b}$ [Eq.~\eqref{sq:eq:sqham1}]. To gain insight into the new ground state and eigenexcitations, we exploit the relations
\begin{align}
\hat{b} = \hat{a} \cosh r  + \hat{a}^{\dagger} e^{i \theta} \sinh r  = \hat{S}(\eta) \hat{a} \hat{S}^\dagger (\eta) , \quad \hat{a} \keta{0} \equiv \hat{a} \ket{0}  = 0,
\end{align}
and typical properties of ladder operators in demonstrating~\cite{KamraPRL2016}
\begin{align}\label{sq:eq:ketb}
\ketb{n} = \hat{S}(\eta) \keta{n} \equiv \hat{S}(\eta) \ket{n},
\end{align}
where $\ketb{n}$ denotes the number state for the new eigenmodes annihilated by $\hat{b}$. This implies that our new vacuum $\ketb{0}$ is the same squeezed magnon vacuum that we have discussed earlier [Eq.~\eqref{sq:eq:sqdef}], with the important difference that it occurs naturally as the ground state of the anisotropic ferromagnet~\cite{KamraAPL2020,KamraPRL2016,ZouPRB2020}. This can be intuitively understood as follows. When $K_y > K_x$ such that $B < 0$, it costs the system more energy to have a nonzero $S_y$ as compared to $S_x$. Thus, the quantum fluctuations in $\mathcal{S}_x$-$\mathcal{S}_y$ phase space due to the Heisenberg uncertainty deform to reduce their component along $y$ direction (and hence, energy), while maintaining the uncertainty region area~\cite{KamraAPL2020,ZouPRB2020,KamraPRB2017}. Thus, the new ground state is squeezed magnon vacuum and adequately represented by Fig.~\ref{sq:fig:sqofmag}(a). On account of Eq.~\eqref{sq:eq:ketb} for $n = 1$, the new eigenmode is termed ``squeezed-magnon''~\cite{KamraPRL2016} and expressed in the magnon number state basis as
\begin{align}\label{sq:eq:sqmagnum}
\ketb{1} = \hat{b}^\dagger \ketb{0} =  \left( \cosh r \hat{a}^\dagger + e^{- i \theta} \sinh r \hat{a}\right) \ket{\eta} = \frac{1}{\sqrt{\cosh^3 r}} \sum_{n=0}^{\infty} (-1)^n ~ e^{i n \theta} ~ \frac{\sqrt{(2 n + 1)!}}{2^n n!} ~ \left( \tanh r \right)^n ~ \ket{2n+1},
\end{align}
whence we see that squeezed-magnon is a composite quasi-particle formed by a superposition of odd magnon number states~\cite{NietoPLA1997,KamraPRB2017,KralJMO1990}. This composite nature underlies various quantum properties that it imbibes as discussed below. It also implies that the average spin carried by squeezed-magnon is larger than 1~\cite{KamraPRL2016,KamraPRB2017}, measurable via magnon spin shot noise~\cite{KamraPRL2016,KamraPRB2016}. The squeezed-magnon spin of $1 + 2 \sinh^2 r$ is evaluated as the expectation value of the magnon number operator $\hat{a}^\dagger \hat{a}$ using Eq.~\eqref{sq:eq:sqmagnum}. Alternately, the same result is obtained more easily by expressing $\mathcal{S}_z$ in terms of squeezed-magnon number ($\langle \hat{b}^\dagger \hat{b} \rangle$)~\cite{KamraPRL2016,KamraPRB2016}, similar to our analysis in Eq.~\eqref{sq:eq:spin1}.

Such an increase in the magnon's magnetic moment was already discussed (indirectly) in the seminal Holstein-Primakoff paper~\cite{HolsteinPR1940} using the thermodynamic definition of magnetic moment $z$-component
\begin{align}
m_z & = - \frac{1}{\mu_0} \frac{\partial \omega}{\partial H},
\end{align}
where $\omega \equiv \omega(H)$ is the magnon energy in an applied field $H \mathbf{e}_z$. It has later been noted that this value is larger than what one expects from a spin-1 magnon~\cite{KaganovPU1997}, alluding to lack of understanding. Hence, this change in the magnetic moment has typically been absorbed into the effective gyromagnetic ratio, treating the magnon spin as fixed. Only recently, with experimental developments in spintronics and theoretical understandings based on squeezing~\cite{KamraPRL2016,KamraPRB2017}, spin and magnetic moment can be clearly separated (e.g., via spin shot noise~\cite{KamraPRL2016}) and the magnetic moment increase has been attributed to the aforementioned squeezed-magnon spin larger than 1 [Eq.~\eqref{sq:eq:sqmagnum}].

\begin{figure}
\begin{center}
\includegraphics[width=0.9\textwidth]{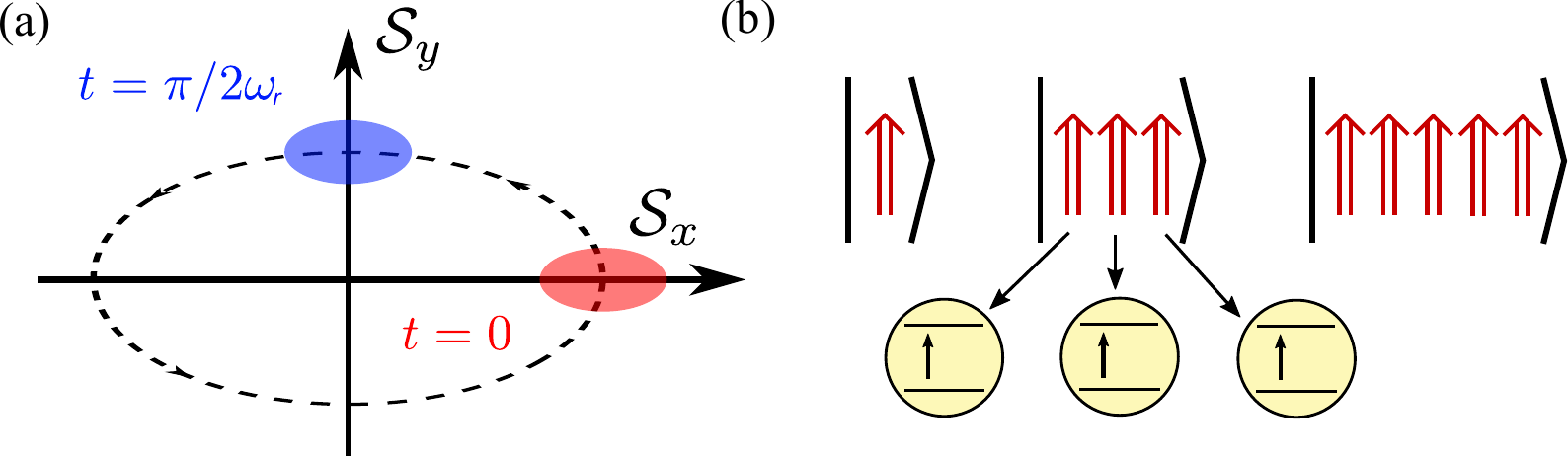}
\caption{(a) Phase space representation of spin dynamics and quantum fluctuations in an anisotropic ferromagnet. This is directly analogous to the isotropic ferromagnet case depicted in Fig.~\ref{sq:fig:mag1}(b). (b) Squeezed magnon is a composite excitation formed by superposition of odd magnon number states [Eq.~\eqref{sq:eq:sqmagnum}]. Thus, it can excite, for example, 3 qubits synchronously thereby generating Greenberger-Horne-Zeilinger and related maximally entangled states: $\left(\ket{ggg} \pm \ket{eee} \right)/\sqrt{2}$. Source: The figure (b) is adapted from Ref.~\cite{Skogvoll2021}.}
\label{sq:fig:sqmag}
\end{center}
\end{figure}

To complete our understanding of the ground state and excitations in an anisotropic ferromagnet, we express the spin components in the eigenbasis [using Eq.~\eqref{HPtransform}]
\begin{align}
\hat{\mathcal{S}}_{x} & = \sqrt{\frac{N S}{2}} \left( \hat{a}^\dagger + \hat{a} \right) = e^r \sqrt{\frac{N S}{2}} \left( \hat{b}^\dagger + \hat{b} \right), \\
\hat{\mathcal{S}}_{y} & = \frac{1}{i} \sqrt{\frac{N S}{2}} \left( \hat{a}^\dagger - \hat{a} \right) = \frac{e^{-r}}{i} \sqrt{\frac{N S}{2}} \left( \hat{b}^\dagger - \hat{b} \right),
\end{align}
whence we conclude that the entire spin operators have been squeezed, and not just the quantum fluctuations. This result allows us to understand various phenomena in anisotropic ferromagnets by simple axes-squeezing of the isotropic case. In particular, classical spin dynamics plus quantum fluctuations in an anisotropic ferromagnet can be depicted in the phase-space as shown in Fig.~\ref{sq:fig:sqmag}(a). As discussed previously, the classical trajectory is traced by the uncertainty region center while the quantum features are encoded in and due to its squeezed shape. A more detailed analysis of this dynamics including a dissipative bosonic bath has been presented in Section~\ref{sec_qm_equation} above.

{\it Equilibrium squeezing.} The emergence of squeezed-magnons and the corresponding vacuum in anisotropic ferromagnets is an equilibrium effect caused by energy minimization~\cite{KamraAPL2020,KamraPRL2016,ZouPRB2020,KamraPRB2017}. Mathematically, it is a consequence of the Bogoliubov transform, making it general to any bosonic system that invokes such transformations. This also gives magnets a niche over other bosonic systems~\cite{KamraAPL2020}, since their excitations are often described via such Bogoliubov transforms. As compared to the nonequilibrium squeezing discussed in isotropic ferromagnets (or typically realized with optical platforms~\cite{WallsNature1983,SchnabelPR2017}), equilibrium squeezing is robust against decay. To distinguish the two, we call it a squeezed state of magnon (represented by $\hat{a}$ above) for the nonequilibrium case and squeezed-magnon ($\hat{b}$) for equilibrium situations and normal excitations~\cite{KamraPRB2019}. This squeezing formalism has allowed us to obtain the wave functions and quantum properties of the ground states and excitations in magnets. See, e.g., Refs.~\cite{MondalPRB2019,BajpaiPRB2021,PetrovPRX2021,MitroPRL2021} and references therein for analysis beyond our linearized dilute magnon gas limit.

{\it Entanglement.} An insight from the obtained wave functions is that the magnetic excitations and ground states are found to harbor entanglement between spatially separated spins~\cite{KamraAPL2020,ZouPRB2020,KamraPRB2019,Yuanmm2020}. Various metrics, such as von Neumann entropy~\cite{KamraPRB2019,HartmannPRB2021} and concurrence~\cite{ZouPRB2020,Yuanmm2020,YuanPRAp2021}, have been employed in quantifying this entanglement content~\cite{AmicoRMP2008,NishiokaRMP2018}, as introduced above in Section \ref{sec_en_measure}. Special focus has been on antiferromagnets~\cite{KamraPRB2019,Yuanmm2020,YuanPRAp2021,MousolouPRB2021,Wuhrerarxiv2021} which manifest very large squeeze parameters, and entanglement, due to exchange interaction causing the squeezing~\cite{KamraAPL2020,KamraPRB2019}. In contrast, relatively weak anisotropy leads to squeezing in ferromagnets. Further, Wuhrer and coworkers~\cite{Wuhrerarxiv2021} have theoretically demonstrated the violation of DGCZ inequalities~\cite{DuanPRL2020} in antiferromagnets, which are continuous variable analogs of the Bell's inequality~\cite{Bell1964}. In this sense, a determination of the wave function for antiferromagnets in the magnon approximation gives insights into their generic entangled nature, which has been well-established in the context of their spin liquid states~\cite{BalentsNature2010,CastelnovoNature2008,SavaryRPP2016}. A more detailed discussion of magnon-magnon entanglement in antiferromagnets is presented in Section~\ref{sec_magnonmagnon} below.

{\it Coupling modification.} Since the magnets typically interact with the outside world via their spin, the composite and high-spin nature of the squeezed-magnon amplifies its coupling via the squeezing effect~\cite{KamraAPL2020,KamraPRB2019}. This has been clarified in the context of magnon-mediated superconductivity enhancement~\cite{ErlandsenPRB2019,JohansenPRL2019} and magnon-magnon ultrastrong coupling~\cite{LiensbergerPRL2019,MakiharaNC2021,KockumNRP2019}, among several others~\cite{KamraPRL2017,Hayashidaarxiv2020,Bambaarxiv2020,ShimPRL2020}. The nonequilibrium equivalent of this amplification was proposed earlier for squeezed states of light~\cite{QinPRL2018,LerouxPRL2018,LuPRL2015} and has been accomplished recently using ion trap systems~\cite{BurdNP2021}. The effect has also been proposed to underlie transient enhancement of phonon-mediated superconductivity~\cite{KnapPRB2016}. However, squeezing can also reduce the magnon coupling with other excitations. Haigh and coworkers~\cite{HaighPRL2021} recently found the squeezing effect to suppress the amplitude for certain optomagnonic transitions in their experiments.

{\it Exploiting nonclassical properties.} The squeezed-magnon vacuum is the magnetic ground state. Thus, extracting entanglement from it is not as straightforward as it is from a squeezed vacuum of light~\cite{OuPRL1992,MilburnPRA1999}. The latter being an excited state can be absorbed by another system thereby transferring its quantum features. To address this issue, several proposals have focused on coupling two qubits with entangled spins of the magnetic ground state~\cite{ZouPRB2020,KamraPRB2019,YuanPRAp2021,FlebusPRB2019}. In this protocol, the qubits need to be decoupled from the magnet after achieving the desired entanglement. On the other hand, a squeezed-magnon is itself highly nonclassical due to its composite nature [Eq.~\eqref{sq:eq:sqmagnum}]. This excitation can thus transfer its nonclassical features more easily to, for example, 3 qubits that jointly absorb it~\cite{Skogvoll2021}, as depicted in Fig.~\ref{sq:fig:sqmag}(b). Generally, an intuitive and not very precise way of understanding the entanglement embodied by squeezed-magnons and their vacuum is as follows. When a wave function is constructed by a superposition of different magnon number states, it involves coherent action of multiple spin flips (each flip corresponding to one magnon) leading to spin entanglement. Thus, this composite nature is fundamental to nearly all the quantum features.

\subsection{Schr{\"o}dinger cat states}\label{sec_cat_magnon}

In 1935, Schr{\"o}dinger proposed a thought experiment according to which a cat should exist in a quantum superposition of being dead and alive, because its state is assumed to be linked to that of an atom, which is itself in a superposition~\cite{Schroedinger1935}. His goal was to argue against the concept of quantum superposition altogether because ``classical'' objects (like a cat) cannot possibly be in two states at once. Here, we consider so-called Schr\"odinger cat states~\cite{Schroedinger1935,DodonovPhysica1974}
\begin{align}\label{sq:eq:catdef}
\ket{\mathrm{cat}} & = \frac{1}{\sqrt{2 + 2 \cos \Phi \exp \left( - 2 |\beta|^2 \right)}} \left( \ket{\beta} + e^{i \Phi} \ket{- \beta}   \right) .
\end{align}
The two coherent states ($\ket{\pm \beta}$) that superpose to form the cat state are themselves classical, as we have discussed before. For large enough $|\beta|$, the two constituent states are macroscopically distinguishable as the ``dead'' and ``alive'' cats. Cat states do not only play a fundamental role in discussions concerning the difference between classical and quantum physics \cite{WinelandRMP2013}, but also serve as an indispensable resource for quantum computing \cite{CochranePRA1999,RalphPRA2003,LundPRL2008}, quantum teleportation \cite{JeongPRA2001} and high-precision measurements \cite{VlastakisScience2013}. Until now, cat states have been realized in optical systems \cite{TakaPRL2008,OurjNature2007,SychevNP2017,HackerNP2019}, but the generation of magnonic cat states has just started to attract interest.
Among various applications of such states~\cite{VlastakisScience2013,OurjoumtsevScience2006,GlancyJOSAB2008} is their ability to emulate Schr{\"o}dinger's thought experiment and characterize quantum decoherence in the system~\cite{BrunePRL1996,MonroeScience1996}, i.e., to show how fast the quantum superposition reduces to a classical statistical mixture. Furthermore, a larger $|\beta|$, corresponding to the constituents ($\ket{\pm \beta}$) being more classical, gives a faster decay of the quantum superposition~\cite{ZurekRMP2003}, consistent with Schr{\"o}dinger's expectation of a completely classical cat loosing the superposition immediately. The Glauber-Sudarshan P-function for the cat states is more singular than a delta function~\cite{BrewsterJMP2018}. This is not allowed for a classical probability distribution and indicates that the cat states are nonclassical.

The even and odd cat states are obtained when $\Phi = 0$ and $\pi$, respectively,
\begin{align}
\ket{\mathrm{cat}}_e & = \frac{1}{\sqrt{2 + 2 \exp \left( - 2 |\beta|^2 \right)}} \left( \ket{\beta} + \ket{- \beta}   \right), \quad \ket{\mathrm{cat}}_o = \frac{1}{\sqrt{2 - 2 \exp \left( - 2 |\beta|^2 \right)}} \left( \ket{\beta} - \ket{- \beta}   \right).
\end{align}
Employing Eq.~\eqref{sq:eq:alphanum}, we see that the even (odd) cat state contains even (odd) number states only
\begin{align}
\ket{\mathrm{cat}}_e & = \frac{2 \exp\left(-\frac{|\beta|^2}{2}\right)}{\sqrt{2 + 2 \exp \left( - 2 |\beta|^2 \right)}} \sum_{m = 0}^{\infty} \frac{\beta^{2m}}{\sqrt{2m!}} \ket{2m}, \quad \ket{\mathrm{cat}}_o  = \frac{2 \exp\left(-\frac{|\beta|^2}{2}\right)}{\sqrt{2 - 2 \exp \left( - 2 |\beta|^2 \right)}} \sum_{m = 0}^{\infty} \frac{\beta^{2m+1}}{\sqrt{(2m+1)!}} \ket{2m+1}.
\end{align}
In this sense, the even and odd cat states are somewhat similar to the squeezed vacuum [Eq.~\eqref{sq:eq:ximag}] and squeezed magnon [Eq.~\eqref{sq:eq:sqmagnum}], which also contain only even and odd number states, respectively. This similarity further provides a convenient method to generate magnon cat states, as proposed recently by Sharma and coworkers~\cite{DaknaPRA1997,SharmaPRB2021}.

\begin{figure}
	\begin{center}
        \includegraphics[width=1.0\textwidth]{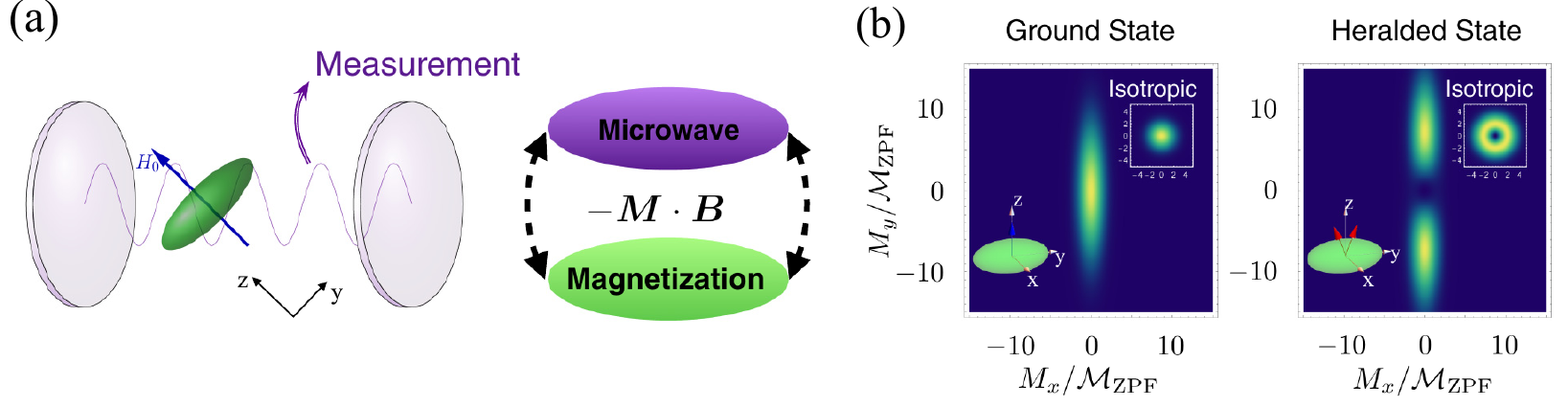}
		\caption{Generation of a magnon cat state by adding a magnon to a squeezed-magnon vacuum. (a) An anisotropic magnet is placed in an optical cavity. Measurement of the one photon state in the cavity projects the magnetic state onto the magnon odd cat state. (b) The left panel depicts the squeezed-magnon vacuum in phase-space. The right panel shows the generated cat state. The insets show the corresponding situation for an isotropic magnet, visibly devoid of the cat. Source: The figures are adapted from Ref.~\cite{SharmaPRB2021}.}
		\label{sq:fig:cat}
	\end{center}
\end{figure}

The basic idea of the method~\cite{DaknaPRA1997,SharmaPRB2021} is that an approximate odd cat state is achieved by adding a single magnon to the squeezed-magnon vacuum [Eq.~\eqref{sq:eq:ximag}], such that the resulting state contains odd-magnon number states only. This magnon addition is accomplished by projecting the combined ground state of an anisotropic magnet-photon cavity ensemble [Fig.~\ref{sq:fig:cat}(a)] to a  one-photon state, achieved via measurement of the photon number. The system Hamiltonian is [Eq.~\eqref{sq:eq:sqham1}]~\cite{SharmaPRB2021}
\begin{align}
\hat{\mathcal{H}}_{t} & = A \hat{a}^{\dagger} \hat{a} + B \left( \hat{a}^2 + \hat{a}^{\dagger 2} \right) + \omega_c \hat{c}^\dagger \hat{c} + g \left( \hat{c} \hat{a}^\dagger + \hat{c}^\dagger \hat{a} \right),
\end{align}
where $\hat{c}$ annihilates the photon and $g$ parameterizes the photon-magnon coupling mediated by the Zeeman interaction. For small $g$, the ground state is approximated by
\begin{align}\label{sq:eq:ketg}
\ket{G} & \approx \left( 1 + \sqrt{P} \frac{\hat{a}^\dagger \hat{c}^\dagger}{\cosh r} \right) \hat{S}(r) \ket{0_a0_c} = \hat{S}(r) \ket{0_a0_c} +  \frac{\sqrt{P}}{\cosh r} \hat{a}^\dagger \hat{S}(r) \ket{0_a1_c},
\end{align}
where $P~(\ll 1)$ is the probability of finding the cavity in its one-photon state, and $\ket{0_a0_c}$ ($\ket{0_a1_c}$) denotes the state with no magnons and (one) photon. The squeeze operator $\hat{S}(r)$ and parameter $r$ are approximately determined by the squeezed-magnon vacuum as discussed earlier [Eq.~\eqref{sq:eq:sqham1}]. $\ket{G}$ is predominantly the squeezed-magnon vacuum plus zero photons, with small contributions from other entangled magnon-photon states. As soon as a one-photon state is detected in the cavity, the squeezed-magnon vacuum is eliminated and $\ket{G}$ is projected onto the state with one photon
\begin{align}\label{sq:eq:ketc}
\ket{C} & = \frac{1}{\cosh r} \hat{a}^\dagger \hat{S}(r) \ket{0_a1_c}.
\end{align}
The ensuing state adds one magnon to the squeezed-magnon vacuum and is approximately a magnon-odd cat state $\sim \left( \ket{\beta_{\mathrm{cat}}} - \ket{- \beta_{\mathrm{cat}}} \right)$ with~\cite{SharmaPRB2021}
\begin{align}
|\beta_{\mathrm{cat}}| \approx \sqrt{\frac{2 B}{\omega_r}},
\end{align}
where $\omega_r$ is the squeezed-magnon energy [Eq.~\eqref{sq:eq:sqham1}]. In obtaining the relatively simple expressions presented here, $B, \omega_c \gg \omega_r \gg g$ has been assumed. A more general analysis appears in Ref.~\cite{SharmaPRB2021}. The Husimi Q-function~\cite{HusimiPPMSJ1940}, defined as $Q(\beta) \equiv \bra{\beta} \hat{\rho} \ket{\beta} / \pi$, for a state described by the density matrix $\hat{\rho}$ is particularly convenient in visualizing cat states in phase space. The Q-function for $\ket{G}$ [Eq.~\eqref{sq:eq:ketg}] and $\ket{C}$ [Eq.~\eqref{sq:eq:ketc}] are shown in Fig.~\ref{sq:fig:cat}(b).

There are several other methods to generate cat states~\cite{GlancyJOSAB2008,YurkePRL1986}. In the context of optomagnonic systems, Sun and colleagues~\cite{SunPRL2021} have suggested a method that does not require equilibrium magnon squeezing. Instead, they suggested a nonequilibrium two-mode squeezing between already generated magnon and photon coherent states. Subsequent remote manipulation and projective measurements on the optical beam enable converting the original magnon coherent state to even or odd cat state. In the near future, the experimental generation and detection of such magnonic cat states will help to determine the potential of low-damping magnetic insulators for quantum information applications.

%
%
\subsection{Quantum many-body states}
\label{sec:quantummanybodystatesofmagnons}
In this section we review some aspects of the many-body physics of magnons, in particular the concepts of magnon BEC and spin superfluidity. For definiteness, we restrict ourselves to solid-state magnets and do not discuss similar physics in liquid helium \cite{BunkovPU2010}. We start out with a brief intermezzo on BEC of interacting particles. This intermezzo is intended to bring the reader up to speed with this topic and may be skipped by the readership that is already familiar with the mean-field description of interacting BEC. Readers who would like to explore more background on these topics may consult, for example, Refs.~\cite{StoofChapter2009,Proudbook2017}.
After this intermezzo we review equilibrium and quasi-equilibrium magnon BEC and spin superfluidity. Throughout, we emphasize the classical versus the quantum aspects of these phenomena. We end this section with some remarks on strongly-correlated many-body states of magnons.

\subsubsection{Intermezzo: Bose-Einstein condensation of interacting particles}
Bose-Einstein condensation is the occupation of the single-particle ground state by a macroscopic fraction of particles. Its textbook version occurs in a system comprised of conserved identical bosons. The simplest model for BEC that includes interactions starts from the Hamiltonian for $N$ identical bosons in a box that interact via short-range interactions. It is given by
\begin{equation}
\label{eq:hamsinglequant}
  \hat{\mathcal{H}} = \sum_{i=1}^N \left( \frac{-\hbar^2 \bm{\nabla}_i^2}{2m}\right) + \frac{1}{2} \sum_{i \neq j} g \delta (\bx_i - \bx_j).
\end{equation}
Here, the bosons have mass $m$, and the short-range interaction is modeled with a delta-function interaction potential as a function of the particle coordinates $\bm{x}_i $ with strength $g$.

At zero temperature, the bosons are expected to occupy the same single-particle ground state $\psi (\bx)$, i.e., the many-body wave function will be given by
\begin{equation}
\label{eq:BECfixedN}
\Psi_N (\bx_1,\bx_2, \cdots, \bx_N) \propto \psi (\bx_1) \psi (\bx_2) \cdots \psi (\bx_N).
\end{equation}
To find the ground state and its energy, one has to minimize the expectation value of the Hamiltonian for the above many-body wave function. This expectation value is given by
\begin{equation}
\label{eq:expecvaluehamBEC}
  \left\langle \Psi_N \right| \hat{\mathcal{H}} \left| \Psi_N \right\rangle = \int d\bx   \psi^* (\bx)\left[ \left ( \frac{-\hbar^2 \bm{\nabla}^2}{2m} \right)  + \frac{g}{2} \left|\psi(\bx) \right|^2\right]\psi(\bx),
\end{equation}
for $N \gg 1$.
Minimizing the above energy yields the Gross-Pitaevskii equation \cite{Griffinbook,RogelEJP2013}
\begin{equation}
\label{eq:GPequation01}
 \left[ \left ( \frac{-\hbar^2 \bm{\nabla}^2}{2m} \right)  -\mu + g \left|\psi(\bx) \right|^2\right]\psi(\bx) = 0,
\end{equation}
where the chemical potential $\mu$ enters as the Lagrange multiplier that enforces the constraint of fixed norm of $\psi (\bx)$. The Gross-Pitaevskii equation yields $\psi (\bx) = \sqrt{n_c} e^{i \bar\theta}$ for $\mu>0$, corresponding to the BEC phase of the system. Here, $n_c=\mu/g$ is the density of particles in the condensate, which, at zero temperature and within the mean-field description developed here, is $n_c = \sqrt{N/V}$, with $V$ the total volume. Furthermore, $\bar\theta$ is the phase of the condensate wave function and is chosen randomly upon each realization of the condensate, corresponding to spontaneous breaking of the $U(1)$ symmetry of Eq.~(\ref{eq:expecvaluehamBEC}). For $\mu<0$ we have that $\psi (\bx)=0$, so that there is no macroscopic fraction of particles in the ground state and therefore no BEC. Hence, the condensate wave function $\psi (\bx)$ is the order parameter for the transition.

The same conclusion is reached by starting from the second-quantized version of Eq.~(\ref{eq:hamsinglequant}), i.e.,
\begin{equation}
\label{eq:hamBEC2ndquant}
  \hat{\mathcal{H}}  = \int d\bx   \hat \psi^* (\bx)\left[ \left ( \frac{-\hbar^2 \bm{\nabla}^2}{2m} \right)  + \frac{g}{2} \left|\hat \psi(\bx) \right|^2\right]\hat \psi(\bx),
\end{equation}
where $\hat \psi (\bx)$ is a bosonic annihilation operator. After insertion of the mean-field approximation $\hat \psi (\bx) = \langle \hat \psi (\bx) \rangle \equiv \psi (\bx) $ into the above, we recover the result in Eq.~(\ref{eq:expecvaluehamBEC}), yielding again the Gross-Pitaevskii equation. In this case, the particle number is fluctuating and the state is a coherent state that was introduced in Section \ref{sec:cohstate}.

The mean-field approximation within the context of second quantization makes it most clear that the BEC for interacting particles at zero temperature is essentially a classical state, albeit one that is described by a complex-valued field. Neglecting the non-commuting character of $\hat \psi (\bx)$, and, hence, neglecting the quantum aspects, is a good approximation in the BEC phase because
\begin{equation}
 \left[\hat \psi (\bx) \hat \psi^\dagger (\bx) - \hat \psi^\dagger (\bx)\hat \psi (\bx)\right]
 \left| \Psi_N \right \rangle = \left( \sqrt{N}\sqrt{N+1} - N  \right) \left| \Psi_N \right \rangle
 	\approx 0,
\end{equation}
for $N \gg 1$.

At finite temperature $T$, quantum mechanics becomes relevant. In this case, the number of particles at the single-particle energy $\epsilon$ is given by the Bose-Einstein distribution function
\begin{equation}
\label{eq:BECdistr}
 n_B (\epsilon-\mu) = \frac{1}
    {e^{   (\epsilon - \mu )/k_B T }-1
 	},
\end{equation}
For energies smaller than the thermal energy, i.e., for  $\epsilon \ll k_B T$, we have that $n_B (\epsilon-\mu) \gg 1$ so that these modes are essentially classical. The Bose-Einstein distribution function approaches the Rayleigh-Jeans distribution function $k_B T/(\epsilon - \mu)$ in this limit. For energies larger than the thermal energy, i.e., $\epsilon \gg k_B T$, we have that $n_B (\epsilon-\mu) \ll 1$, so that quantum-mechanical  effects cannot be neglected. On the contrary, quantum-many body effects are essential for BEC at finite temperature. Namely, the nature of the high-energy part of the Bose-Einstein distribution function is such that for a given negative chemical potential and temperature there is a certain maximum number of particles that can be accommodated by the states with energy $\epsilon >0$. Keeping temperature fixed, this number reaches its maximum when $\mu \rightarrow 0$, at which point BEC occurs. Upon increasing the total number of particles at fixed temperature, the chemical potential thus increases until it becomes zero (measured from the single-particle ground state energy). At this point, the thermal cloud, i.e., the part of the system composed of the particles with energy $\epsilon > 0$, is saturated and all particles that are added increase the number of particles in the ground state, i.e., lead to growth of the BEC. At fixed particle number, decreasing the temperature leads to an increase in chemical potential until the latter becomes zero, leading again to Bose-Einstein condensation. The requirements of low temperature and high density are usually expressed in terms of the phase-space density via the dimensionless number $N \Lambda (T)^3/V$, which should be of order one for BEC to occur. Here, $\Lambda (T) \sim \sqrt{2\pi\hbar^2/m k_B T}$ is the thermal de Broglie wavelength \cite{Huangbook}.

The proper description of interacting BEC requires the inclusion of interactions between the particles in the thermal cloud with the particles in the condensate. Within the Popov approximation, these are described by mean-field shifts of the single-particle energies. This yields the modified Gross-Pitaevskii equation
\begin{equation}
\label{eq:GPequation}
\left[ \left ( \frac{-\hbar^2 \bm{\nabla}^2}{2m} \right)  -\mu + g \left|\psi(\bx \right)|^2+ 2 g n'\right]\psi(\bx) = 0,
\end{equation}
where $n'$ is the density of particles in the thermal cloud that is given by
\begin{equation}
\label{eq:densitythermalcloud}
  n' = \int \frac{d\bk}{(2\pi)^3} n_B \left(\epsilon_\bk + 2 g |\psi|^2 + 2 g n'-\mu\right),
\end{equation}
where the single-particle dispersion is $\epsilon_\bk = \hbar^2 \bk^2/2m$. The critical temperature for BEC is now found as follows. From the modified Gross-Pitaevskii equation it follows that condensation occurs when $\mu = 2 g n'$. At the critical temperature, $\psi =0$, so that the equation for the density of the thermal cloud becomes
\begin{equation}
\frac{\mu}{2 g} = \int \frac{d\bk}{(2\pi)^3} n_B \left(\epsilon_\bk \right).
\end{equation}
Carrying out the integral on the right-hand side of the above, yields the critical temperature
\begin{equation}
\label{eq:Tc}
 T_{\rm BEC} \sim \frac{\hbar^2}{k_B m} \left(\frac{\mu}{g}\right)^{2/3},
\end{equation}
which is valid for $\mu>0$ only. Hence, at zero temperature BEC occurs when $\mu=0$, in agreement with our previous discussion.

In the above, we have expressed the critical temperature as a function of the chemical potential, rather than the usual expression in terms of the total density. As we shall see in the next section, this discussion is convenient for the treatment of equilibrium magnon BEC.

\subsubsection{Equilibrium magnon Bose-Einstein condensation}
The simplest example of equilibrium magnon BEC occurs in systems that are described by the Hamiltonian in Eq.~\eqref{biaxialHam}, for the case that $K_z<0$ and $K_x=0$. These are so-called easy-plane magnets. At zero temperature and at the classical level, this system has a spin-reorientation transition whereby the spin reorients from pointing fully along the external field for large fields, to having a nonzero component in the $x-y$-plane. This reorientation transition may be found by minimizing the Hamiltonian after the mean-field ansatz $\hat {\bf S}_i = \langle \hat {\bf S}_i \rangle$. This ansatz yields in the first instance
\begin{equation}
\label{eq:easyplaneFMH}
\hat{\mathcal{H}} = - J \sum_{<i,j>} \langle \hat {\bf S}_i \rangle \cdot \langle \hat {\bf S}_j \rangle + |K_z| \sum_i \left(\langle \hat S_{i,z}\rangle \right)^2 -\mu_0 g_e \mu_B H \sum_i \langle \hat S_{i,z} \rangle,
\end{equation}
which is minimized by $\langle \hat S_{i} \rangle = S \mathbf{e}_z$ for $H> 2|K_z|S/(\mu_0 g_e \mu_B)$, and by
\begin{equation}
\langle \hat S_{i} \rangle = S \left (\sqrt{1-\frac{\mu_0 g_e \mu_B H}{2|K_z|S}} \cos \varphi, \sqrt{1-\frac{\mu_0 g_e \mu_B H}{2|K_z|S}} \sin \varphi, \frac{\mu_0 g_e \mu_B H}{2|K_z|S} \right)
\end{equation}
for $H< 2 |K_z|S/(\mu_0 g_e \mu_B)$. Here, the angle $\varphi$ is the angle that the magnetization makes in the $x-y$-plane, which is chosen spontaneously during the reorientation transition, in analogy to the phase of the condensate that was discussed in the previous section. Here, the difference is that the absolute phase of a condensate of particles is not directly observable, whereas the magnetization is of course directly observable.

The reorientation transition that occurs at the field $H=2 |K_z|S/(\mu_0 g_e \mu_B)$ can be interpreted as BEC of interacting magnons. To this end, we use the HP transformation for $S_{i,z}$ and use the mean-field ansatz $\langle \hat a_i \rangle = \phi$, where we anticipate that the ground state is homogeneous. Up to a constant, this yields the energy per lattice site
\begin{equation}
\hat{\mathcal{H}}/N = \left( \mu_0 g_e \mu_B H-2|K_z|S\right) |\phi|^2 + |K_z| |\phi|^4,
\end{equation}
which, upon minimization, leads to the Gross-Pitaevskii equation (\ref{eq:GPequation}) with $\mu = 2 |K_z|S - \mu_0 g_e \mu_BH$ and $g=2|K_z|$. The analogy with BEC is understood as follows. For the HP transformation that is used here,  that is, using the $z$-direction as the quantization axis, the creation of a magnon in the single-magnon ground state corresponds semi-classically to the magnetization tilting away from the $z$-axis. A macroscopic number of magnons therefore corresponds to a classical reorientation of the magnetization away from the $z$-direction for this specific setup. A caveat here is in the definition of the magnon itself. The interpretation of the reorientation transition as magnon BEC relies on the HP transformation that uses the $z$-direction as quantization axis. If one, instead, defines the magnon as a linearized excitation around the reoriented ground state for $H < 2 |K_z|S/(\mu_0 g_e \mu_B)$, the interpretation of the reorientation transition in terms of magnon BEC is lost \cite{AndreasPRB2012}. Furthermore, as the tilt of the magnetization away from the $z$-axis increases, the density of magnons becomes too large to be described by the truncated HP transformation. Regardless of one's opinion on whether or not this is BEC of magnons, the analogy with BEC allows one to immediately conclude that: i) there will be gapless excitations for $H < 2 |K_z|S/(\mu_0 g_e \mu_B)$, in analogy to phonons in the BEC, and ii) that at finite temperature the reorientation transition occurs at $T_c \propto J (2S - \mu_0 g_e \mu_BH/|K_z|)^{2/3}/k_B$. Here, we used Eq.~(\ref{eq:Tc}). Finally, we remark that, though the easy-plane ferromagnet is a somewhat academic system as anisotropies will always pin the easy-plane magnetization direction, there are several experiments on materials which realize this type of transition, albeit that the magnetic excitations in these experiments are more complicated than the Holstein-Primakoff magnons that we have introduced in this paper \cite{NikuniPRL2000,RueggNature2003, RaduPRL2005}.

\subsubsection{Quasi-equilibrium magnon Bose-Einstein condensation} Notwithstanding the above-mentioned example of equilibrium condensation, BEC is often a quasi-equilibrium phenomenon and requires thermalization to occur before particles or excitations decay away. Even BEC in cold-atom systems in a trap is of this kind, as e.g., three-particle collisions will cause one atom to escape from the trap that holds the atoms, leaving one molecule behind. The most well-known type of experiments is on quasi-equilibrium magnon BEC in YIG in which pairs of magnons are excited by parametric pumping \cite{DemokritovNature2006}. This saturates the magnon gas and leads to a large occupation of the lowest magnon modes. Because the magnons are pumped at energies that are much smaller than $k_B T$, this situation does not involve the Bose-Einstein distribution function and has been dubbed Rayleigh-Jeans condensation \cite{AndreasPRL2015}. This is different from other types of pumped condensates, such as exciton-polaritons \cite{ByrnesNP2014} and photons \cite{KlaersNature2010}, in which the pumping is at energies on the same order or larger than $k_B T$. Other regimes of magnon BEC have been reached by heating with a current pulse and subsequent fast cooling, and involves all magnons at thermal energies \cite{SchneiderNN2020}. The so-called spin-caloritronic nano-oscillator \cite{SafranskiNC2017}, in which thermal magnons are injected into a magnetic insulator via the spin Seebeck effect, may  be interpreted as quasi-equilibrium magnon BEC \cite{BenderPRL2012}. Very recently, magnon BEC by electric spin injection was reported \cite{Divinskiy2021}. Finally, we remark that in  quasi-equilibrium magnon BEC the condensate is dynamic and leads to time-dependent magnetization dynamics, for example the nano-oscillator mentioned above. An equilibrium magnon BEC, on the other hand, corresponds to a static situation.

\subsubsection{Spin superfluidity} Magnon condensation may underly spin superfluidity\footnote{The connection between BEC and superfluid is a subtle issue. It is possible to have superfluid (BEC) without the presence of BEC (superfluid). More detailed discussion can be found in Ref. \cite{Pethickbook2008}.}. In particular, the magnetization dynamics of the easy-plane ferromagnet, discussed above may --- in the condensed phase --- in the continuum limit be recast in terms of superfluid hydrodynamic equations. That is, inserting the ansatz $\langle \hat {\bf S}\rangle = S (\sqrt{n_c/s} \cos \varphi,\sqrt{n_c/s} \sin \varphi, 1-n_c/s)$, with $s=S/d^3$ the spin density and $d$ the lattice constant, into the LLG equation yields the equations
\begin{eqnarray}
\label{eq:superfluidhydro}
  \frac{\partial \varphi}{\partial t} &=& \mu_0 g_e \mu_B H-2|K_z|S+ 2|K_z| S n_c/s, \nonumber \\
  \frac{\partial n_c}{\partial t} &=& - \bm{\nabla} \cdot {\bf j}_s,
\end{eqnarray}
for $n_c \ll s$. Here we have neglected the Gilbert damping. The spin supercurrent ${\bf j}_s = n_c \bm{\nabla} \varphi$ is proportional to the gradient of the angle $\varphi$ that plays the role of phase of the condensate. The influence of damping and anisotropies on spin superfluidity have been discussed extensively by Sonin \cite{SoninJETP1978, SoninAP2010}. Recent research has focused on spin superfluidity in devices in which heavy metallic leads serve as spin injectors and detectors \cite{TakeiPRL2014} and the first experimental results have been reported \cite{YuanSA2018,BozhkoNC2019, PhysRevLett.123.257201}.

\subsubsection{Strongly-correlated magnon states}
In addition to the states relevant for BEC, magnons may enter more exotic quantum-many body states. The Hamiltonian for the easy-plane magnetic insulator, for example, features a zero-temperature Mott-insulator-to-superfluid transition \cite{FisherPRB1989} for large enough $|K_z|$ \cite{CamiloPRL2018}. That is, for $|K_z| \gg J$, the magnons are localized on each lattice site because of the interactions and the system has a gap $\sim |K_z|$. This state is approximated by
\begin{equation}
\left| \Psi_{\rm MI} \right\rangle \propto \prod_{i=1}^N \left[ \left (\hat a_i^\dagger \right)^{N_f} \right] |0\rangle,
\end{equation}
where $N$ is the number of lattice sites and $N_f$ is the filling of the Mott insulator, i.e., the number of magnons that are localized on-site, which is tuned by the external field. This state should be contrasted with the superfluid, i.e., BEC, phase in which all particles are in the zero momentum state, and which occurs for $J \gg |K_z|$. This state reads for the same number of magnons
\begin{equation}
\left| \Psi_{\rm SF} \right\rangle \propto \left( \sum_{i=1}^N \hat a_i^\dagger \right)^{N_f}|0\rangle.
\end{equation}
Other examples of quantum many-body states are that of pair condensation of magnons, i.e., states with $\langle \hat a \rangle =0 $ while $\langle \hat a \hat a \rangle \neq 0$. This situation corresponds to condensation of pairs or BCS-like state for magnons. Examples of the above states may be found in compounds studied within the field of quantum magnetism \cite{YuNature2012,ZapfRMP2014}. For example, the Mott insulator mentioned above is an example of a gapped quantum paramagnet. It should be noted, however, that in actual quantum magnets the excitations are often more complicated than the HP magnons that we consider here.

This ends our review of the quantum states of magnons. In the next section, we introduce the developments concerning coupling magnons to various quantum platforms, including photons, qubits, phonons, and electrons. Note that the detection of the quantum states of magnons may also require the coupling of magnons with other quantum systems. For example, by counting photons when they are coupled to magnons, one is able to deduce the statistics of magnon distribution \cite{LachScience2020}. By detecting the photoluminescence spectrum of a NV center coupled to the magnetic system, the BEC state of magnons may be measured \cite{FlebusPRL2018}. It would also be meaningful to study the influence of magnon-photon coupling on the phase diagram of the quantum many-body states, which may resemble the situation of a bosonic atomic gas inside a cavity \cite{ChenPRA2016}.

\section{Magnon + X}\label{sec_magnonX}
As information carriers, a great advantage of magnons is that there are various knobs to efficiently control their generation, propagation and detection. For example, magnons are generated by microwaves, spin-orbit torques \cite{MironNature2011,LiuPRL2012,NakayamaPRL2013,ManchonNM2015,LebrunNature2018}, thermal gradients \cite{BeaurPRL1996,KimelNature2005,NishAPL2010,DebPRL2019}, and mechanical strain \cite{ScherPRL2010}, and can be detected by optical means \cite{SebaFP2015}, microwave antennas \cite{VlaPRB2010}, heavy metals through inverse spin Hall effect \cite{CornNP2015} and even spin qubits \cite{LachScience2020}. When we integrate a magnonic system with other quantum systems (denoted as $\mathrm{X}$) such as cavity photons, superconducting qubits, NV centers and mechanical oscillators, it is desirable that the hybrid system benefits from the manipulation tools that have been and are developed in magnon spintronics, which extend the horizon of the coupled quantum systems. At the same time, the introduction of X helps to study quantum states of magnons and push magnonics to the quantum regime. For example, the entanglement between magnons and qubits enables the detection of
single-magnon states \cite{LachScience2020}. The hybrid platforms have already generated multiparticle entanglement among magnons, photons and phonons for continuous variable quantum information and have found inspiring applications to design quantum transducers, quantum memories and high precision measurements. Following this line of mutual benefit between magnons and other quantum systems, we shall review the recent developments concerning hybrid magnon-photon systems in Section \ref{sec_magnon_photon}, magnon-qubit systems in Section \ref{sec_magnon_qubit}, magnon-phonon systems in Section \ref{sec_magnon_phonon}, and magnon-electron systems in Section \ref{sec_magnon_electron}. The field that focuses on the magnon-photon interaction is also dubbed cavity magnonics or spin cavitronics.

\subsection{Hybrid magnon-photon system} \label{sec_magnon_photon}

\subsubsection{Overview}
People's interests in coupling magnets to microwaves can be traced back to the 1960s, where mode splitting from the coupling of the magnetic mode and the microwave mode was observed in ferrimagnet-loaded cavity resonators \cite{AuldJAP1963, Chow1966, Weiner1972}. One popular theory to explain this coupling is coupled-mode theory \cite{AuldJAP1963}. This theory separates the magnetic fields into a superposition of a rotational and irrotational parts, and then expands the hybrid fields in terms of both electromagnetic and magnetostatic modes. This treatment leads to coupled equations for all mode amplitudes, whose solution yields the coupled spectrum. The coupling strength of magnet and microwave is proportional to the overlap of the electromagnetic wave with the magnetostatic mode inside the magnet as, $-\mu_0 \int_V \mathbf{H}\cdot \mathbf{M} d^3r$, where the integral is performed over the whole magnetic body. This coupling phenomenon is purely classical, and is well described by classical electromagnetism.

After a half century of dormancy, this topic was revisited by Soykal et al. in 2010 \cite{SoykalPRL2010}. The interaction of magnon and photon is rewritten in second-quantized form as $\hat{\mathcal{H}}_{\mathrm{int}}=g_{mc}(\hat{a}+\hat{a}^\dagger)(\hat{c}+\hat{c}^\dagger)$, after performing a HP transformation on the magnetization and quantizing cavity's electromagnetic waves. In the rotating wave approximation, the coupling takes the coherent form $\mathcal{H}_{\mathrm{int}}=g_{mc}(\hat{a}^\dagger \hat{c} + \hat{a} \hat{c}^\dagger)$. It was found that the interaction between a YIG sphere and a microwave cavity can reach the strong-coupling regime and that the spin-photon hybrid state has a long spin dephasing time of roughly 14 seconds. In 2013, Huebl et al. \cite{HueblPRL2013} experimentally observed the strong coupling between the uniform magnon mode in YIG and a superconducting coplanar microwave resonator, and found that the energy level splitting is larger than the dissipation of the system. Since then, more experiments demonstrated such a coherent and strong coupling between magnons and photons, using various types of cavities and probing techniques \cite{ZhangPRL2014,CaoPRB2015,GorPRAp2014,TabuchiPRL2014,BaiPRL2015,HouPRL2019,LiPRL2019}. Such strong coupling was also observed in antiferromagnetic systems \cite{BiaekPRAp2021,ShiarXiv2020,GrishuninACS2018,YuanEPL2021}. In 2018, Vahram et al. \cite{VahramPRB2018} theoretically studied the interaction of two electromagnetic waves with one magnet and found that the resulting spectrum displays energy level attraction by tuning the relative phases of the two electromagnetic waves. Such an anomalous spectrum was observed in experiments by Harder et al. in a Fabri-Perot cavity \cite{HarderPRL2018},  Bhoi et al. in an inverted pattern of split-ring resonator \cite{BhoiPRB2019} and Boventer et al. in a reentrant cavity resonator \cite{BoventerPRR2020}. For a consistent understanding of the energy level repulsion and attraction, Yu et al. \cite{YuPRL2019} highlighted the role of travelling-wave modes with strong dissipation in the open cavity. Here the traveling-wave modes couple to both standing-wave photon and magnons and produce an effective complex coupling between magnons and photons. When the coupling is purely real (imaginary), the spectrum displays energy level repulsion (attraction). Yao et al. \cite{YaoPRB2019} also identified the role of the travelling wave mode for the energy level spectrum. The various coupled spectra at room temperature are still mostly classical, and are well described by classical electromagnetism \cite{CaoPRB2015,RameshtiPRB2018,YuPRL2019,VahramPRB2019,WeichaoPRB2020,ShimPRL2020,RairPRAp2021,XiaoPRB2021}. This explains why the numerical solution of the coupled Maxwell and LLG equation, for example, by COMSOL Multiphysics, is sufficient to recover the spectrum observed in the experiments \cite{YuPRL2019}. The use of quantum model that results from quantizing the magnetic excitations and the cavity electromagnetic waves provides an alternative angle to understand the essential physics, but it is not enough to claim quantum aspects of the physics.

Here, we mention some recent reviews on cavity spintronics \cite{LiJAP2020,HarderJAP2021,Rameshti2021cavity}. Li et al. \cite{LiJAP2020} outline the potential of hybrid magnetic systems for transformative applications in devices, circuits and information processing. Harder et al. \cite{HarderJAP2021} focused on the coherent and dissipative coupling characterized by level repulsion and attraction respectively. Rameshti et al. \cite{Rameshti2021cavity} reviewed the coupling of magnons with photons at frequencies ranging from microwaves to optical regimes. In particular, they present a semiclassical approach to model cavity spintronics and further briefly address the coupling of magnons with superconducting qubits. Here, we shall follow the framework sketched in Fig. \ref{qmpicture} to review the recent progress concerning quantum phenomena in the hybrid systems. In particular, we shall focus on the quantum entanglement between magnons and photons. The various energy spectra of the hybrid systems as already reviewed by Rameshti et al. \cite{Rameshti2021cavity} will not be covered in this section.


\subsubsection{Magnon-photon entanglement} \label{sec_mpEn}
\begin{figure}
  \centering
  \includegraphics[width=1.0\textwidth]{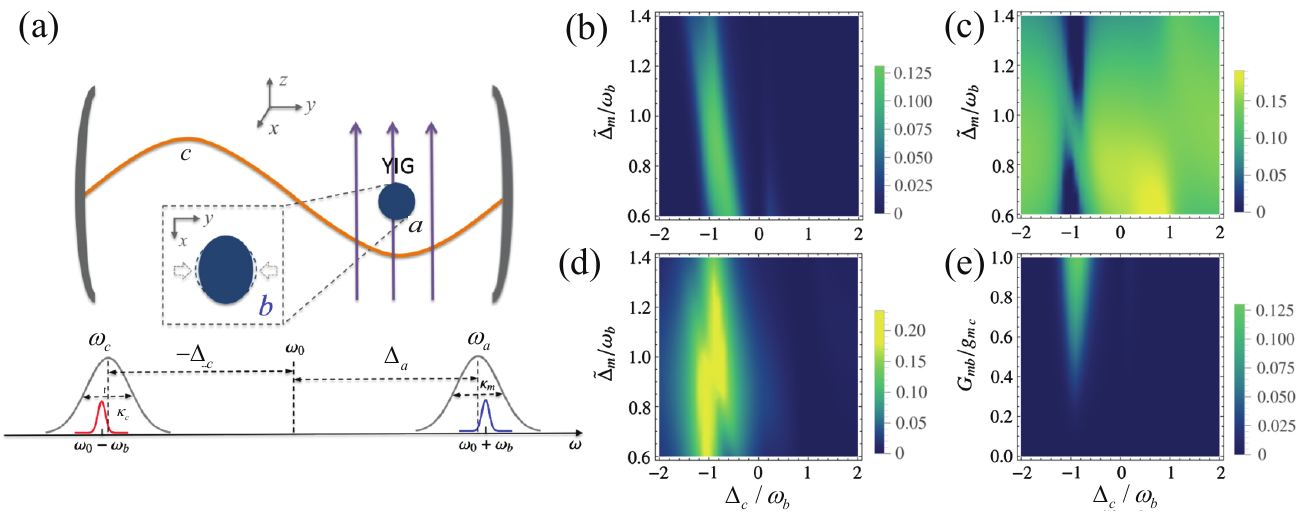}\\
  \caption{(a) Schematic of the hybrid magnet-cavity system. A YIG sphere is placed at the position with maximum cavity magnetic field and an extra static magnetic field is applied to tune the magnon frequency. The bottom panel shows the frequencies and linewidth of magnon mode ($\hat{a}$), cavity mode ($\hat{c}$) and phonon mode ($\hat{b}$). (b) Entanglement of magnon-photon $E_{ac}$, (c) magnon-phonon $E_{ab}$ and (d) photon-phonon $E_{cb}$ degrees of freedom versus detunings $\Delta_c$ and $\Delta_a$. (e) Magnon-photon entanglement versus $\Delta_c$ and $G_{mb}/g_{mc}$. Source: The figures are adapted from Ref. \cite{LiPRL2018}.}
  \label{Limpp2018}
\end{figure}

Li et al. studied steady-state magnon-photon entanglement in a hybrid magnet-cavity system in 2018 \cite{LiPRL2018}. Because the methodology used in this work has wide applicability, we introduce the essential details here. The authors studied the hybrid system shown in Fig. \ref{Limpp2018}(a) that is described by the Hamiltonian
\begin{equation}
\hat{\mathcal{H}}= \omega_c \hat{c}^\dagger \hat{c} + \omega_r \hat{a}^\dagger \hat{a} + \frac{\omega_b}{2} (\hat{q}^2 + \hat{p}^2) + g_{mb} \hat{a}^\dagger \hat{a} \hat{q} + g_{mc}(\hat{a} + \hat{a}^\dagger)(\hat{c} + \hat{c}^\dagger)+i\Omega(\hat{a}^\dagger e^{-i\omega_0t} - \hat{a} e^{i\omega_0 t}),
\end{equation}
where $\hat{c}$ ($\hat{c}^\dagger$) and $\hat{a}$ ($\hat{a}^\dagger$) are respectively the annihilation (creation) operators for the cavity photons and magnons. Furthermore, $\hat{q}$ and $\hat{p}$ are the position and momentum operators for the mechanical oscillations of the magnetic sphere and $g_{mb}$ and $g_{mc}$ are the magnon-phonon and magnon-photon coupling strengths, respectively. Detailed justification of the magnon-phonon coupling term in the Hamiltonian can be found in Ref. \cite{ZhangSA2016} and is briefly introduced when we review the magnon-phonon interaction in Section \ref{sec_magnon_phonon}. Typically, the phonon frequency corresponding to the mechanical deformation of YIG is much smaller than the magnon frequency, i.e., $\omega_b \ll \omega_{c,a}$ such that a strong driving $\Omega$ of the magnon mode is required to induce a sizeable coupling between magnons and phonons. In the rotating frame at the driving frequency $\omega_0$, the quantum Heisenberg-Langevin equations are written as
\begin{subequations}\label{langevin_Li2018}
\begin{align}
&\frac{d\hat{c}}{dt} = -i (\Delta_c -i\kappa_c) \hat{c} -ig_{mc} \hat{a} + \sqrt{2\kappa_c}\hat{c}^{in}, \\
&\frac{d\hat{a}}{dt} = -i (\Delta_a -i\kappa_a) \hat{a} - -ig_{mc} \hat{a} -ig_{mb} \hat{a}\hat{q}+ \sqrt{2\kappa_m}\hat{a}^{in}, \\
&\frac{d\hat{q}}{dt} = \omega_b \hat{p}, \quad \frac{d\hat{p}}{dt} = -\omega_b \hat{q} - \gamma_b \hat{p} - g_{mb} \hat{a}^\dagger \hat{a} + \hat{\zeta},
\end{align}
\end{subequations}
where the detunings $\Delta_c =\omega_c-\omega_0, \Delta_a = \omega_a - \omega_0$, $\kappa_a$, $\kappa_m$ and $\gamma_b$ are the dissipation coefficients of photon, magnon and phonon, respectively. $c^{in}$,\ $a^{in}$ and $\zeta$ denote the respective Gaussian noises associated with these dissipation coefficients. The noise has zero mean and nonzero second-order correlations given by
\begin{subequations}
\begin{align}
&\langle \hat{\nu}^{in}(t) \hat{\nu}^{in \dagger}(t') \rangle= (n_\nu+1)\delta(t-t'),\langle \hat{\nu}^{in \dagger}(t) \hat{\nu}^{in }(t') \rangle = n_\nu \delta(t-t'),\\
&\langle \hat{\zeta}(t) \hat{\zeta}(t') + \hat{\zeta}(t') \hat{\zeta}(t) \rangle=(2n_b +1)\gamma_b \delta(t-t'),
\end{align}
\end{subequations}
where $\hat{\nu} = \hat{c},\hat{b}$, $n_c$,$n_b$, and $n_a$ are respectively the Bose-Einstein distributions for the photons, phonons and magnons, which depend on temperature.

Since the magnons are strongly driven and are strongly coupled to photons, i.e., $|\langle \hat{a} \rangle| \gg 1$, $|\langle \hat{c} \rangle| \gg 1$, the magnon and photon states can be expanded around their mean-field values
$\hat{O}=\langle \hat{O} \rangle + \delta \hat{O}$ ($\hat{O}=\hat{c},\hat{a},\hat{q},\hat{p}$). The linearized form of Eq. (\ref{langevin_Li2018}) that describes the evolution of the quadrature fluctuations is written as, $d\hat{\mathbf{u}}/dt=\mathbf{A}\cdot \hat{\mathbf{u}}+\hat{\mathbf{n}}$, where $\hat{\mathbf{u}}=(\delta \hat{X}, \delta \hat{Y}, \delta \hat{x}, \delta \hat{y}, \delta \hat{q}, \delta \hat{p})^T$, $\hat{\mathbf{n}}=(\sqrt{2\kappa_c}\hat{X}^{in},\sqrt{2\kappa_c}\hat{Y}^{in},\sqrt{2\kappa_m}\hat{x}^{in},\sqrt{2\kappa_m}\hat{y}^{in},0,\hat{\zeta})^T$ is the vector of input noises, and $\mathbf{A}$ is the drift matrix of the system. The quadrature operators are defined as $\delta \hat{X} = (\delta \hat{c} + \delta \hat{c}^\dagger)/\sqrt{2},\delta \hat{Y} = (\delta \hat{c} - \delta \hat{c}^\dagger)/\sqrt{2}i$,$\delta \hat{x} = (\delta \hat{a} + \delta \hat{a}^\dagger)/\sqrt{2},\delta \hat{y} = (\delta \hat{a} - \delta \hat{a}^\dagger)/\sqrt{2}i$. The Gaussian nature of noise guarantees that the steady-state solution to this set of linearized equations is a Gaussian state and thus is completely characterized by a $6\times 6$ covariance matrix $\mathbf{V}$, as introduced in Section \ref{sec_cvs}. The covariance matrix is derived by solving the Lyapunov equation $\mathbf{AV}+\mathbf{VA}=-\mathbf{D}$, where $\mathbf{D}=\text{diag}\{\kappa_c(2n_c+1),\kappa_c(2n_c+1),\kappa_m(2n_a+1),\kappa_m(2n_a+1),0,\gamma_b(2n_b+1)\}$ is the diffusion matrix \cite{VitaliPRL2007}. From the tripartite covariance matrix, one can trace one part and quantify the entanglement between the remaining two parts. In principle, one can also solve the master equation \cite{LindbladCMP1976} of the system numerically to get the steady density matrix, based on which we can also evaluate the covariance matrix by $V_{ij}=1/2 \text{tr}[\rho (u_i u_j + u_j u_i)]$. Using this approach, a proper truncation of the Hilbert space has to be taken to guarantee the convergence of the evolution.

Figures \ref{Limpp2018}(b-d) show the steady entanglement among magnons and photons using the experimentally feasible parameters \cite{ZhangSA2016} while the temperature is taken 10 mK to suppress the decoherence of quantum states. There exists a parameter regime ($\Delta_a \approx \omega_b, \Delta_c \approx -\omega_b$), where all the degrees of freedoms, i.e., magnon-photon, magnon-phonon and photon-phonon, are entangled. Here the nonlinear magnon-phonon interaction ($G_{mb}=i\sqrt{2}g_{mb}\langle m \rangle$) plays a key role to mediate the indirect entanglement between magnons and photons. Without this interaction, the beam-splitter type of interaction between magnons and photons does not generate any finite entanglement, as shown on the bottom of Fig. \ref{Limpp2018}(e) ($G_{mb}=0$).
Besides bipartite entanglement, a parameter regime of finite tripartite entanglement is also identified. In follow-up works, the authors showed that two cavity photons can be prepared in a stationary entangled state if they are respectively resonant with two mechanical side bands of the system \cite{YuPRL2020}. Again, the nonlinearity caused by the magnetostrictive interaction between magnons and phonons is crucial to maintain the robustness of this entangled state. The authors also showed that magnons in a setup similar to Fig. \ref{Limpp2018}(a) can be prepared in a squeezed state via the beam-splitter interaction of magnons and photons, when the cavity is driven by a squeezed vacuum field \cite{LiPRA2019}. Tan \cite{TanPRR2019} identified both bipartite and tripartite EPR steering in such a hybrid system and found that continuous measurement of the cavity fields significantly enhances the bipartite steering.

Yuan et al. \cite{YuanBell2020} found an alternative way to entangle magnons and photons without including nonlinear magnon-phonon interactions in the Hamiltonian explicitly. They considered dissipative coupling between magnons and photons, as already realized in several experiments \cite{HarderPRL2018,BhoiPRB2019,BoventerPRR2020}. This hybrid system is usually described by the non-Hermitian Hamiltonian
 \begin{equation}
\begin{aligned}
\mathcal{\hat{H}}&=  \omega_r \hat{a}^\dagger \hat{a} + \omega_c \hat{c}^\dagger \hat{c}  + g \left ( \hat{c}^ \dagger \hat{a} + e^{i\Phi} \hat{c} \hat{a}^\dagger \right),
\end{aligned}
\label{ham}
\end{equation}
where $\hat{a}~ (\hat{a}^\dagger)$ and $\hat{c}~(\hat{c}^\dagger)$ are respectively, annihilation (creation) operators of magnons and photons.
$\omega_r$ is the magnon frequency, which can be tuned by external magnetic field, $g$ is the effective coupling strength
between magnons and photons, and $\Phi$ is a tunable phase factor due to backaction effects. Such an effective Hamiltonian (\ref{ham}) describes the energy level repulsion and attraction which are observed in experiments very well \cite{HarderPRL2018,BhoiPRB2019,BoventerPRR2020}.

\begin{figure}
  \centering
  \includegraphics[width=1.0\textwidth]{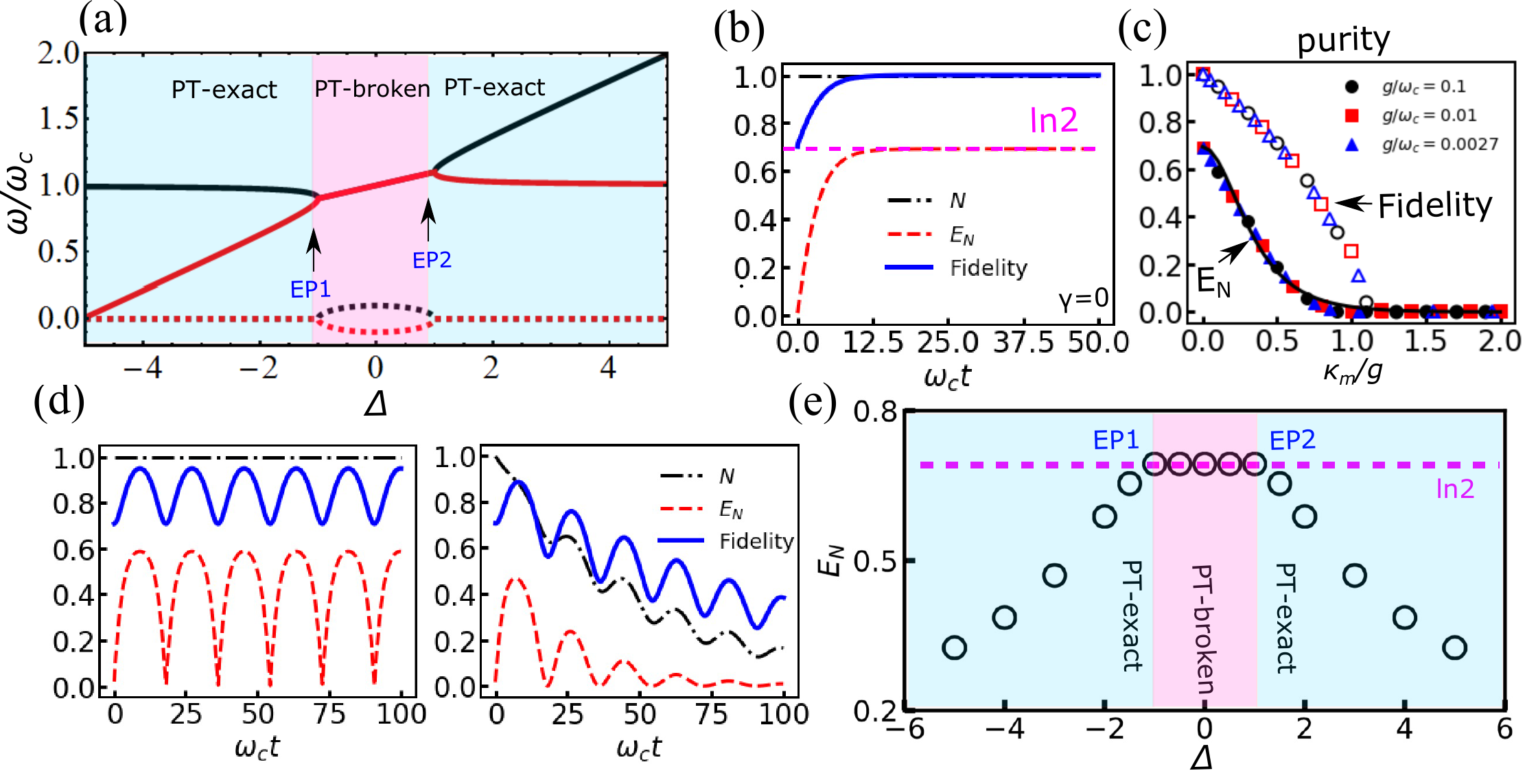}\\
  \caption{(a) Level attraction of a hybrid magnet-cavity described by Eq. (\ref{ham}) for $\Phi=\pi$. The solid and dashed lines represent the real and imaginary parts of the eigenvalues, respectively. (b) Evolution of the entanglement and the fidelity of the temporal state to the Bell state as a function of time. (c) Robustness of the steady Bell state against dissipation with different coupling strengths between magnons and photons ($g$). (d) Evolution of the entanglement for the coherent coupling between magnons and photons without (left) and with (right) dissipations. (e) Magnon-photon entanglement as a function of the detuning between magnons and photons. Source: The figures are adapted from Ref. \cite{YuanBell2020}.}
  \label{bell}
\end{figure}

Figure \ref{bell}(a) shows a typical spectrum corresponding to energy level attraction with $\Phi=\pi$. As the detuning between the magnon frequency and the cavity frequency ($\Delta \equiv  \omega_r-\omega_c$) changes, the system evolves from a $\mathcal{\mathcal{PT}}$ exact phase with purely real eigenvalues to a $\mathcal{PT}$ broken phase with complex eigenvalues. In the transition between these two phases, two exceptional points (EPs) located at $\Delta = \pm 1$ are identified. To study the dynamics of magnons and photons in both $\mathcal{PT}$ exact and $\mathcal{PT}$ broken phases, one decouples the non-Hermitian Hamiltonian into a Hermitian component ($\hat{\mathcal{H}}_H\equiv(\hat{\mathcal{H}} + \hat{\mathcal{H}}^\dagger)/2$) and an anti-Hermitian component ($\hat{\mathcal{H}}_A\equiv(\hat{\mathcal{H}} - \hat{\mathcal{H}}^\dagger)/2$). Here the Hermitian part and anti-Hermitian part characterize the unitary evolution, and gain/loss of the system, respectively. The master equation governing the evolution of the joint density matrix of the system ($\hat{\rho}$) may be written as \cite{KorschJPA1987,MizrahiPS1998,BrodyPRL2012}
\begin{equation}
\frac{\partial \hat{\rho}}{\partial t}=-i[\hat{\mathcal{H}}_H,\hat{\rho}] - i \{\hat{\mathcal{H}}_A, \hat{\rho} \} + 2i \mathrm{tr}(\hat{\rho} \hat{\mathcal{H}}_A)\hat{\rho},
\label{me_nonhermitian}
\end{equation}
where the brackets $[~]$ and $\{~\}$ refer to commutator and anticommutator of two operators, respectively. The third non-linear term on the right-hand side of Eq. (\ref{me_nonhermitian}) is added to preserve $\mathrm{tr}(\hat{\rho})=1$ and thus guarantees the stability of the system. Equation (\ref{me_nonhermitian}) preserves the total probability of the system ($\mathrm{tr}(\rho(t))=1$). A more detailed discussion of this master equation has been presented by Brody et al. \cite{BrodyPRL2012}. Physically, the non-linear effect in a magnonic system may come from the magnon-magnon interactions caused by exchange and dipolar interactions, which will suppress the divergence of the magnon density \cite{RezendeJPCM2010,KlossPRB2010,WangPRL2021,Hulaarxiv2021}. A similar approach was adopted in a single quantum system with $\mathcal{\mathcal{PT}}$ symmetry \cite{WuScience2019}.

By solving the quantum master equation \eqref{me_nonhermitian}, it is found that the hybrid system evolves to a steady state with entanglement $\ln2$, which is the maximum amount of entanglement of two qubits. To understand the structure of this steady state, the authors analytically derived the evolution of the density matrix and found that the eigenstate $|\phi_{s} \rangle = ( | 10\rangle + e^{i \varphi_s} | 01\rangle)/\sqrt{2}$ of the Hamiltonian eventually dominates the steady state, with $\varphi_s= \arccos (\Delta)$. Note that the state $|\phi_{s} \rangle$ is a Bell state and that it has
maximum entanglement $E_N(|\phi_s \rangle \langle  \phi_s |)=\ln2$ as shown in Fig. \ref{bell}(b). One can check the fidelity of the density matrix with respect to the Bell state ($F\equiv \text{tr}\sqrt{\langle \phi_s | \hat{\rho} | \phi_1 \rangle}$) as a function of time, showing that it indeed approaches one in the steady case. This clearly proves that the steady Bell state is generated in the hybrid system. Furthermore, such a state is robust against dissipation ($\kappa_m$) as long as $\kappa_m < g$, as shown in Fig. \ref{bell}(c). Using the experimental parameters that correspond to a millimeter-sized YIG sphere inside a Fabry-Perot cavity \cite{HarderPRL2018}, the fidelity of the generated Bell state is $97.85 \%$. For comparison, in the $\mathcal{PT}$ exact phase, the system keeps oscillating and no steady Bell states are generated, as illustrated in Fig. \ref{bell}(d). This result can also be viewed as Rabi oscillations and it applies to the original proposal put forth in Ref. \cite{SoykalPRL2010}, where the real beam-splitter type interaction cannot generate a steady entanglement between magnons and photons.
A complete phase diagram of the system is found in Fig. \ref{bell}(e). Xie et al. proposed a Bell test using the hybrid magnet-cavity system in a recent preprint \cite{XiearXiv2021}.

\subsubsection{Magnon-magnon entanglement}\label{sec_magnonmagnon}
Besides magnon-photon entanglement, magnon-magnon entanglement is another interesting topic because of its fundamental importance to understand macroscopic quantum phenomena and the classical-quantum transition in magnetic systems, and their potential applications in continuous variable quantum information.
Depending on whether the two magnons are excited in the same or two separate magnets, the study of magnon-magnon entanglement is progressing towards two directions.
The first direction deals with the entanglement between magnons excited in two distant magnetic systems through their simultaneous interaction with a third subsystem, for example, a cavity. Here we will not cover the magnon-magnon coupling in hybrid nanostructures contain two magnetic layers, which can be well understood in the classical regime \cite{StefanPRL2018,ChenPRL2018,QinSR2018,AnPRB2020}. The second direction concerns the entanglement of magnons in ferri/anti-ferromagnets, where the two types of magnons are excited on the two sublattices of a magnet. The two directions are similar in that parametric interactions (or squeezing interactions) between the magnons are essential to generate a finite steady-state entanglement. As briefly discussed in Section~\ref{sec_squeezedmagnon} above, the ground state and magnon modes in an anisotropic ferromagnet were demonstrated to be related to spin-flips in the magnet via single- and two-mode squeeze operators in 2016 \cite{KamraPRL2016,KamraPRB2016}. Similar considerations in ferri- and antiferromagnets have been presented in 2017~\cite{KamraPRB2017}. However, while commenting on the relation between squeezing and entanglement \cite{KamraPRB2016}, the authors
did not study the entanglement, a task having received increasing attention in recent years as discussed further below.

\begin{figure}
  \centering
  \includegraphics[width=1.0\textwidth]{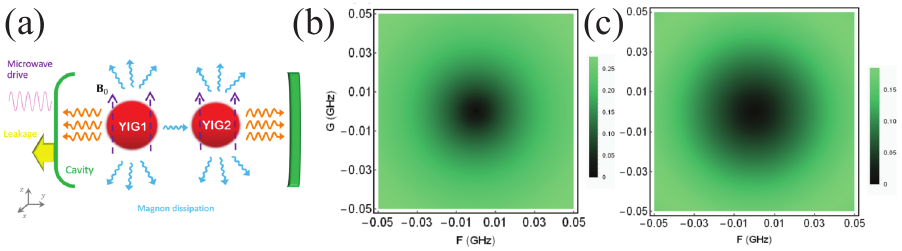}\\
  \caption{(a) Schematic of two distant magnetic YIG spheres inside a cavity entangled through the cavity field. (b-c) Two-dimensional plots that quantify magnon-magnon entanglement(b) and magnon-photon entanglement (c). Here $G=2v \mathrm{Re} \langle \hat{a}_s \rangle^2$, $F=2v \mathrm{Im} \langle \hat{a}_s \rangle^2$, $v_1=v_2\equiv v$, $\langle \hat{a}_s \rangle$ is the steady-state magnon amplitude. Source: The figures are adapted from Ref. \cite{ZhangPRR2020}.}
  \label{ZhangmmEg2019}
\end{figure}

Zhang et al. \cite{ZhangPRR2020} studied the entanglement of two distant magnets in a hybrid magnet-cavity system as shown in Fig. \ref{ZhangmmEg2019}. Here the cavity field mediates the coupling of two well-separated magnets and the effective Hamiltonian of the hybrid system is
\begin{equation}
\hat{\mathcal{H}} = \omega_c \hat{c}^\dagger \hat{c} + \sum_{j=1}^2 \left [ \omega_j \hat{a}_j^\dagger \hat{a}_j +g_j (\hat{c}\hat{a}_j^\dagger + \hat{c}^\dagger \hat{a}_j) + v_j \hat{a}_j^\dagger \hat{a}_j \hat{a}_j^\dagger \hat{a}_j \right ) + i\Omega (\hat{c}^\dagger e^{-i\omega_dt} - \hat{c} e^{i\omega_d}t),
\end{equation}
where $\hat{a}_1$ and $\hat{a}_2$ are the magnon operators on the first and second magnets, respectively, $\hat{c}$ is the cavity photon mode, $v_j$ is the nonlinearity coming from the magnetocrystalline anisotropy of the system, similar to non-linearity that is featured in the discussion in Section \ref{sec_singlemagnon}, and $\Omega$ is the driving strength.
In the strong driving limit, one can expand the magnon and photon operators around their mean-field values and derive the effective Hamiltonian for the magnonic and photonic fluctuations, as introduced in Section \ref{sec_mpEn}. Then the covariance matrix ($V_{6\times 6}$) is numerically obtained by solving the linearized Langevin equation with Gaussian noise. The entanglement between magnons and photons as quantified by log-negativity is calculated and shown in Fig. \ref{ZhangmmEg2019}(b). Firstly, in the absence of Kerr nonlinearities ($v_1=v_2=v\equiv 0,G=F=0$), there is no entanglement among magnons and photons, because all the remaining interactions between magnons and photons are of the beam-splitter type, which cannot generate any steady-state entanglement. With Kerr nonlinearities ($GF\neq 0$), the entanglement of magnons appears, because these interactions generate an effective parametric interaction between magnons and thus create a squeezed state of the two distant magnons with finite entanglement. In addition, it is identified that a finite entanglement emerges between magnons and photons as shown in Fig. \ref{ZhangmmEg2019}(c), which is understood in a similar manner.

Li et al. \cite{LiNJP2019} considered a setup similar to the one depicted in Fig. \ref{ZhangmmEg2019}(a), but focused on the nonlinearity coming from the magnetostrictive interaction in the form $\hat{a}_1^\dagger \hat{a}_1 \hat{q}$, where $\hat{q}$ is the mechanical displacement of the first magnet. The magnetostrictive interaction in the second magnet is not activated upon properly choosing its magnetization direction. They showed that such a nonlinearity generates a finite entanglement between the two distant magnets, which is robust against thermal fluctuation below 0.2 K. Furthermore, the magnon-magnon entanglement can be transferred to the magnon-photon system through beam-splitter type of interactions between magnons and photons. Simon et al.  \cite{SimonNP2007} considered an optical resonator loaded with two laser-cooled atomic ensembles and showed that the optical mode of the resonator can help to generate magnon-number entanglement between the two ensembles. In particular, they first generated a single-magnon state in ensemble A with certain probability and then transferred it to ensemble B with the assistance of photons. At the same time, they suppressed the population of the photonic state using the technique of adiabatic dark-state transfer, and thus a partial transfer of a single magnon between these two atomic ensembles is realized. This results in a superposition in which the two atomic ensembles share one magnon. The photons, as described by the authors, plays the role of a bus to entangle the two atomic ensembles.

Besides amplifying the nonlinearity in the magnetic system under strong driving to induce the entanglement of magnons, Nair et al. \cite{NairAPL2020} presented a scheme to entangle two macroscopic magnets in the absence of nonlinearities. They proposed to place two YIG spheres in a single cavity and to drive the cavity using a quantum field with strong squeezing. They found that the squeezing property of the driving can be transferred to the cavity field \cite{YuanSqueeze2021} and thus mediates entanglement of two magnetic modes by producing an effective parametric interaction between them. In summary, the effective parametric interaction between two magnons is essential to generate steady-state entanglement of magnons, while the methods to generate such a parametric interaction differ between these proposals. Following a different chain of arguments, Kamra et al. previously demonstrated that the dipolar interaction generates equilibrium squeezing of magnons in a ferromagnetic film \cite{KamraPRL2016,KamraPRB2016}. Similarly, if  we look at two or more sublattice ferrimagnets and antiferromagnets, the magnons excited on each sublattice are naturally exchange-coupled by a parametric interaction because of the antiparallel alignment between the two sublattice magnetizations. It is thus expected that they will form squeezed states with strong entanglement. This idea is illustrated in detail as below.

Yuan et al. \cite{Yuanmm2020} and Kamra et al. \cite{KamraPRB2019} independently discussed the squeezing and entanglement properties of the antiferromagnetic ground state. To illustrate these ideas, let us consider an anisotropic two-sublattice ferrimagnet, described by the Hamiltonian
\begin{equation}\label{Ham2subAFM}
\mathcal{H}_{\mathrm{FiM}}=J \sum_{l,\delta}  \hat{\mathbf{S}}_l \cdot \hat{\mathbf{S}}_{l+\delta}
-\sum_l \left [ K_a\left (\hat{\mathbf{S}}_{l}^z \right )^2+ K_b\left (\hat{\mathbf{S}}_{l+\delta}^z \right )^2 \right ]- \mu_0 g_e \mu_B\sum_l \mathbf{H} \cdot \hat{\mathbf{S}}_{l},\\
\end{equation}
where $J>0$ and $K_{a,b}>0$ denotes the antiferromagnetic coupling between the two sublattices and anisotropy coefficients, respectively, and $\mathbf{H}=H\mathbf{e}_z$ is an external magnetic field acting on the magnetization, which is assumed to be along the $z$ axis. Following the standard quantization procedures using the HP transformation around the classical ground state $\mathbf{S}_l= S_a\mathbf{e}_z$ and $\mathbf{S}_{l+\delta}=-S_b\mathbf{e}_z$, the effective Hamiltonian of the magnon excitations is written as
\begin{equation}\label{afm_wtphoton}
\hat{\mathcal{H}}=  \omega_a \hat{a}_1^\dagger \hat{a}_1 + \omega_b \hat{a}_2^\dagger \hat{a}_2 +g_{ab}(\hat{a}_1^\dagger
\hat{a}_2^\dagger + \hat{a}_1 \hat{a}_2),
\end{equation}
where $\hat{a}_1$ and $\hat{a}_2$ are the magnon operators on the two sublattices of an antiferromagnet, $\omega_a=H_{\text{ex,b}} + H_\text{an,a} + \mu_0 g_e \mu_B H, \omega_b=H_{\text{ex,a}} +H_\text{an,b} - \mu_0 g_e \mu_B H$, $H_{\text{ex,$\mu$}}=2ZJS_\mu, ~H_\text{an,$\mu$}=2K_\mu S_\mu$~ ($\mu=a,b$), $S_\mu$ is the magnitude of spin vector, $Z$ is the coordination number, and $g_{\text{ab}}=\sqrt{H_{\text{ex,a}}H_{\text{ex,b}}}$ is the coupling strength of magnon modes on the two sublattices. The parametric interaction between the two types of magnons is understood as follows. The antiferromagnetic interaction between the two sublattices tends to align the spins antiparallel to each other, for example, in the $+\mathbf{e}_z$ and $-\mathbf{e}_z$ directions. The motion of $\mathbf{S}_1$ ($\mathbf{S}_2$) towards the $+\mathbf{e}_z$ ($-\mathbf{e}_z$) direction will each annihilate one magnon, i.e., $\hat{a}_1 \hat{a}_2$. The inverse process will each generate one magnon on the two sublattices, i.e., $\hat{a}_1^\dagger \hat{a}_2^\dagger$.

\begin{figure}
  \centering
  \includegraphics[width=1.0\textwidth]{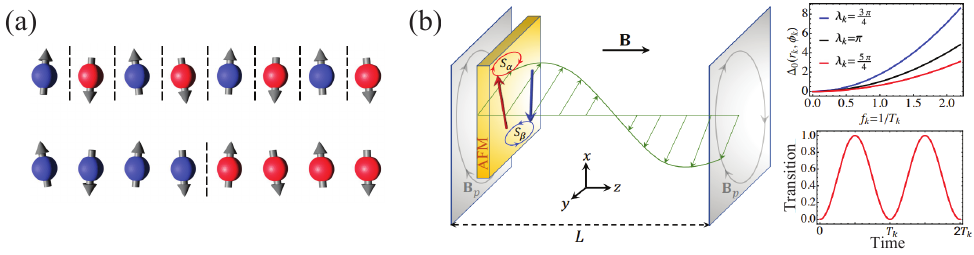}\\
  \caption{(a) An antiferromagnetic system is divided into two sublattices by alternating sites (red and blue) to characterize the intersublattice entanglement (top). The system is partitioned through the middle for the central cut entanglement (bottom). (b) A schematic of magnet-cavity system to measure the entanglement of the magnon modes. The EPR function $\Delta_0$ quantifies the entanglement of magnons in the ground state and it can be inversely deduced by measuring the transition frequency between acoustic magnons and photons ($f_k$) in the cavity. $\lambda_k$ is the scaled wavenumber. Source: The figures are adapted from Refs. \cite{HartmannPRB2021,MousolouPRB2021}.}
  \label{AkashmmEg2021}
\end{figure}

By solving the Heisenberg-Langevin equation for this Hamiltonian (\ref{afm_wtphoton}) in the antiferromagnetic limit ($S_a=S_b\equiv S,K_a=K_b \equiv K$), the covariance matrix of the system is found to be
\begin{equation}
\mathbf{V}=\left ( \begin{array}{cc}
            \mathbf{A} & \mathbf{C} \\
            \mathbf{C}^T & \mathbf{B}
          \end{array}
\right )=\left ( \begin{array}{cccc}
            \Lambda_a & 0& \Lambda_c & \Lambda_d \\
            0& \Lambda_a & \Lambda_d & -\Lambda_c \\
            \Lambda_c& \Lambda_d & \Lambda_b & 0 \\
            \Lambda_d& -\Lambda_c & 0 & \Lambda_b \\
          \end{array}
\right ),
\end{equation}
with
\begin{align}
\Lambda_a &=\Lambda_b= \frac{1}{2} \left ( 1+ \frac{g_{ab}^2}{(H_{\mathrm{ex}} + H_{\mathrm{an}})^2 - g_{ab}^2 + \gamma_m^2} \right ),\nonumber\\
\Lambda_c &= -\frac{1}{2} \frac{g_{ab}(H_{\mathrm{ex}} + H_{\mathrm{an}})}{(H_{\mathrm{ex}} + H_{\mathrm{an}})^2 - g_{ab}^2 + \gamma_m^2},\nonumber\Lambda_d  = -\frac{1}{2} \frac{g_{ab} \gamma_m}{(H_{\mathrm{ex}} + H_{\mathrm{an}})^2 - g_{ab}^2 + \gamma_m^2},
\end{align}
where $\gamma_m$ is the dissipation coefficient of each sublattice. Then the entanglement of the sublattice magnons is quantified as
\begin{equation}\label{gsEnAFM}
E_N = -\log(2\eta^-)=\ln \left (1+ \frac{g_{ab}}{\sqrt{(H_{\mathrm{ex}}+ H_{\mathrm{an}})^2 + \gamma_m^2}} \right).
\end{equation}
This result suggests that the magnons excited on the two sublattices are always entangled, and the amount of entanglement approaches the value $\ln2$ since $g_{ab}\sim H_{\mathrm{ex}}\gg H_{\mathrm{an}},\gamma_m$ in a typical crystalline antiferromagnet. In a van der Waals magnet, the anisotropy field can be comparable to and even larger than the exchange field \cite{ZhangNM2020}, hence the amount of entanglement between sublattice magnons will be reduced. Kamra et al. \cite{KamraPRB2019} quantified the entanglement of spatially uniform magnon modes by the entanglement entropy a system described by Eq. (\ref{Ham2subAFM}) and found it  increases monotonically with the squeezing parameter. Hartmann et al. \cite{HartmannPRB2021} calculated the ground-state entanglement entropy by including the contribution from magnons with finite wavevectors by partitioning the antiferromagnet into two sublattices as shown in Fig. \ref{AkashmmEg2021}(a) and found it scales with the system size in the thermodynamic limit as
\begin{equation}
S_{EE} =  N^\upsilon \frac{(1+2^{\upsilon-1}K/J)^{\upsilon/2}}{2^{-\upsilon/2-1}\pi^{\upsilon/2} \Gamma (\upsilon/2)} I\left (\upsilon,\frac{K}{J}\right),
\end{equation}
where $\upsilon$ is the dimension of the system and $I$ is a special integral function. With this special partition [Fig. \ref{AkashmmEg2021}(a), upper panel], the area law becomes a volume law while the density of entanglement entropy becomes an intensive property. The entanglement density is evaluated to be a universal dimensionality-dependent constant when $K \ll J$. This result was further corroborated by the density matrix renormalization group (DMRG) calculations for a one dimensional spin chain. Wuhrer et al. \cite{Wuhrerarxiv2021} derived the full statistics of the sublattice magnon number with wave number $\mathbf{k}$ in the ground state. They showed that magnons with opposite wave vectors are always created in pairs and their entanglement, as quantified by the DGCZ inequality \cite{DuanPRL2020}, decreases at the corners of Brillouion zone. Mousolou et al. \cite{MousolouPRB2021} identified the relation between bipartite magnon entanglement and the EPR function and proposed a measurement setup based on the interaction of light and antiferromagnets, as shown in Fig. \ref{AkashmmEg2021}(b). Here the strength of magnon entanglement is quantified by the EPR function $\Delta_0$ and it depends sensitively on the magnon-photon transition frequency $f_k$ in the cavity, which can be measured directly. It has also been shown that the Dzyaloshinskii-Moriya (DM) interaction enhances the entanglement for all $\mathbf{k}$ modes \cite{MouPRB2020}.

\begin{figure}
  \centering
  \includegraphics[width=1.0\textwidth]{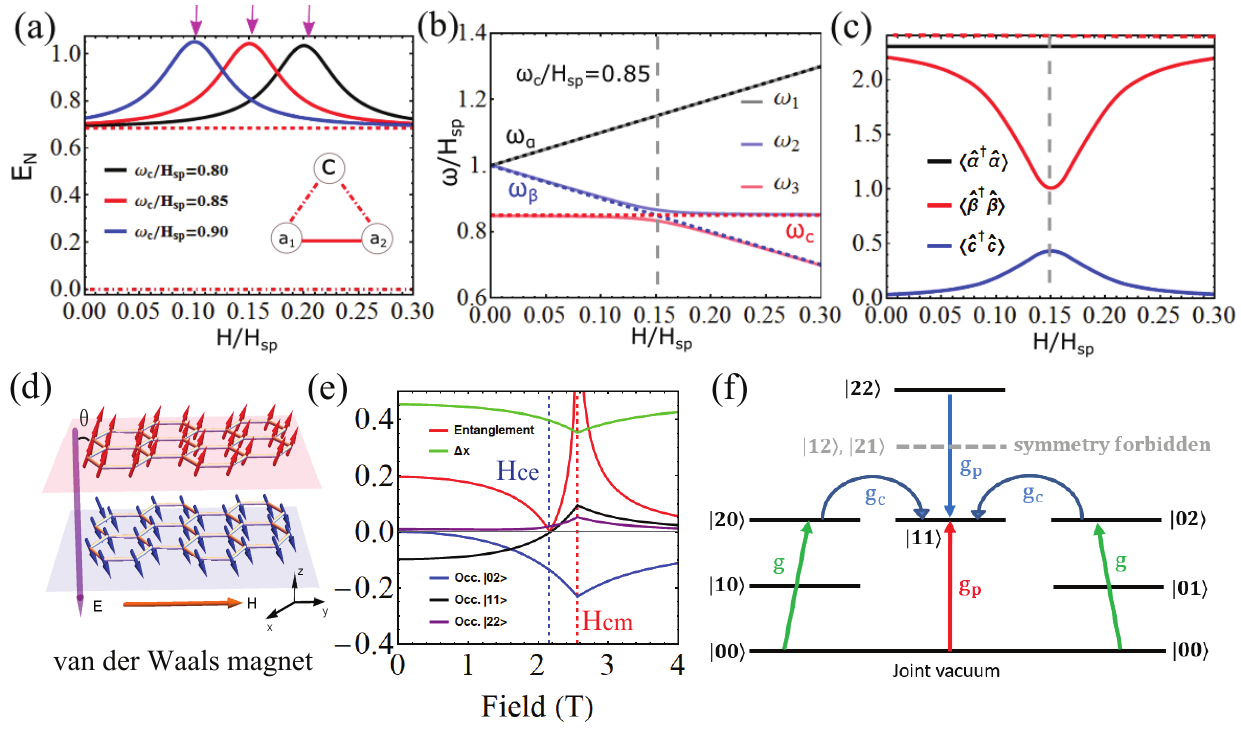}\\
  \caption{(a) Entanglement of magnons in a two-sublattice antiferromagnet as a function of external field. (b) Energy spectrum of the two-sublattice antiferromagnet coupled to cavity photons. The acoustic mode $\omega_\beta$ is hybridized with the photon mode $\omega_c$ with an anticrossing structure near resonance, while the optical mode $\omega_\alpha$ is left unchanged. (c) Magnon and photon population as a function of external field. (d) Schematic of a bilayer van der Waals magnet. An in-plane field is applied to tilt the magnetization. (e) Entanglement (red line) and squeezing (green line) of magnons in the ground state of magnons as a function of external field. (f) Physical picture of the non-monotonic field dependence of the magnon-magnon entanglement.  Source: The figures are adapted from Refs. \cite{Yuanmm2020,YuanPRAp2021}.}
  \label{YuanmmEg2019}
\end{figure}

In reality, one may wish to manipulate the entanglement. Zou et al. \cite{ZouPRB2020} demonstrated that the entanglement of spins can be controlled by external fields. This is not easily done in a crystalline antiferromagnet as the entanglement between magnons does not depend on the external field as shown by Eq. (\ref{gsEnAFM}). To overcome this challenge, Yuan et al. \cite{Yuanmm2020} considered placing the antiferromagnet in a cavity and found that a circular-polarized cavity photon could be coupled to the magnon state as
 \begin{equation}
\hat{\mathcal{H}} =  \omega_a \hat{a}_1^\dagger \hat{a}_1 + \omega_b \hat{a}_2^\dagger \hat{a}_2 +g_{ab}(\hat{a}_1^\dagger
\hat{a}_2^\dagger + \hat{a}_1 \hat{a}_2)  + \omega_c \hat{c}^\dagger \hat{c} + g_{ac} \left ( \hat{a}_1^ \dagger c^\dagger + \hat{a}_1 c \right)
+ g_{bc} \left (  \hat{a}_2^\dagger \hat{c} + \hat{a}_2 \hat{c}^\dagger \right).
\label{FimHm}
\end{equation}
Here only when the magnetization $\mathbf{m}_2$ is pointing along the $-\mathbf{e}_z$ direction does the system absorb the photon's angular momentum $\hbar$ to excite one magnon, resulting in a beam-splitter type interaction  $\hat{a}_2^\dagger \hat{c} + \hat{a}_2 \hat{c}^\dagger$. The interaction of photons with $\mathbf{m}_1$ does not conserve the total angular momentum, and is of the parametric-type $\hat{a}_1^ \dagger \hat{c}^\dagger + \hat{a}_1 \hat{c}$.  Following similar procedures as above, the magnon-magnon entanglement is quantified by log-negativity and its dependence on external field is shown in Fig. \ref{YuanmmEg2019}(a). Clearly, the entanglement of magnons is enhanced by photons and it is maximal at a particular field, at which the photons and acoustic magnons (lower magnon branch) reach resonance as shown in Fig. \ref{YuanmmEg2019}(b). To understand the enhancement of this magnon-magnon entanglement, the authors rewrote the effective Hamiltonian in the dressed basis $\hat{\alpha} = \cosh \theta \hat{a}_1- \sinh \theta \hat{a}_2^\dagger $, $\hat{\beta} = -\sinh \theta \hat{a}_1^\dagger + \cosh \theta \hat{a}_2$, where $\tanh 2\theta = -2g_{ab}/(\omega_a + \omega_b)$
\begin{equation}
\hat{\mathcal{H}} = ~\omega_\alpha \hat{\alpha}^\dagger \hat{\alpha} + \omega_\beta \hat{\beta}^\dagger \hat{\beta} + \omega_c \hat{c}^\dagger \hat{c}  + g_{\alpha c}(\hat{\alpha}^\dagger \hat{c}^\dagger+ \hat{\alpha} \hat{c} )+ g_{\beta c}( \hat{\beta} \hat{c}^\dagger+ \hat{\beta}^\dagger \hat{c} ),
\label{abeta}
\end{equation}
where $\hat{\alpha}$ and $\hat{\beta}$ are respectively the optical and acoustic modes of the magnons.
The beam-splitter type of coupling between acoustic magnons and photons ($g_{\beta c}$) will cool the magnons towards their ground state. Note that the ground state is a two-mode squeezed state  $|\theta \rangle = \hat{S}(\theta) | 0_{a_1}, 0_{a_2} \rangle$ \cite{KamraPRB2019} with entanglement $E_N = 2|\theta| \approx \mathrm{arctanh} (1+H_{\mathrm{an}}/H_{\mathrm{ex}})^{-1}\approx 2.41$, which is larger than the steady-state entanglement $\ln2$ in the absence of light. Therefore, this cooling mechanism will enhance the entanglement of two magnons. On the other hand, the optical mode $\alpha$ is a dark mode which is non-resonant with the photons, hence its role in the cooling process is negligible. The change of magnon and photon population near the resonant field is shown in
Fig. \ref{YuanmmEg2019}(c), which confirms this argument. By extending this result to a two-sublattice ferrimagnet, Zheng et al. \cite{ZhengSC2020} found that EPR steering of the two types of magnons is asymmetric, even when they have identical dissipation. The essential physics is understood to be due to the unbalanced population of acoustic and optical magnons under the cooling effect of cavity photons. Alternatively, Sharma et al. studied the cooling of a magnetic sphere by a monochromatic laser source and found that the magnon distribution in the steady state is controlled by the light intensity through inelastic magnon-photon scattering \cite{SharmaPRL2018}.

An alternative way to tune the entanglement of magnons is to utilize the van der Waals magnets \cite{YuanPRAp2021}. Van der Waals magnets are a class of magnetic materials with only one to a few magnetic monolayers, and the neighboring layers are coupled through van der Waals interactions \cite{GongNature2017,HuangNature2017,BurchNature2018,GibertiniNN2019}. Depending on the stacking order of layers and the external electric field, the interlayer coupling can be either ferromagnetic or antiferromagnetic \cite{WangCongPRL2021}. Compared with bulk crystalline magnets, van der Waals magnets have the advantage of tunable exchange and anisotropy fields by gate voltage, where the exchange and anisotropy can be comparable in strength such that the magnetic ground state can be altered by either electric field or magnetic field. For a canted magnetic state induced by an in-plane magnetic field, as shown in Fig. \ref{YuanmmEg2019}(d), the effective magnon excitation is described by the following Hamiltonian \cite{YuanPRAp2021}
\begin{equation}
\hat{\mathcal{H}}=\sum_{i=1}^2 \left [ \omega_0 \hat{a}_i^\dagger \hat{a}_i + \frac{1}{2}g(\hat{a}_i^\dagger \hat{a}_i^\dagger + \hat{a}_i \hat{a}_i) \right ] +g_c(\hat{a}_1^\dagger \hat{a}_2 + \hat{a}_1 \hat{a}_2^\dagger)+g_p(\hat{a}_1^\dagger \hat{a}_2^\dagger + \hat{a}_1 \hat{a}_2),
\end{equation}
where $\omega_0=(J + K)S \cos 2\theta + KS \cos^2\theta + \mu_0 g_e \mu_B H \sin \theta, g = KS\sin^2\theta, g_p = JS \cos^2\theta,
g_c=JS\sin^2\theta$, with $J$ and $K$ being the interlayer exchange and anisotropy coefficients, respectively. As the external magnetic field increases above a critical value $H_{\mathrm{cm}}=2(J+K)S/(\mu_0 g_e \mu_B)$, the system reorients magnetically from a canted state ($\theta<\pi/2$) to a collinear state ($\theta=\pi/2$).

Now, both the two-mode squeezing ($g_p$) and single mode squeezing ($g$) interactions will contribute to the quantum correlations between the two magnons, and generate an anomalous field-dependence of the magnon-magnon entanglement. Even when the squeezing of the magnons decreases gradually (green solid line) when the field approaches the magnetic reorientation field, the magnon-magnon entanglement does not change monotonically (red solid line), as shown in Fig. \ref{YuanmmEg2019}(e). The essential physics is understood from the schematics of the energy levels as shown in Fig. \ref{YuanmmEg2019}(f). The key influencer of the magnon-magnon entanglement is the occupation of the state $|11\rangle$, because it determines the difficulty to separate the joint wave function into a cross product of the wave function of each sublattice magnon. Here the two-mode squeezing $g_p$ contributes positively to the occupation of $|11\rangle$, while the combined effect of coherent processes $g_c$ and single mode squeezing reduces the occupation of this energy level. Then these two channels will interfere to result in a local minimum of magnon-magnon entanglement before the transition field. This competition between the two channels is still there when the dissipation of magnons is taken into account. A sudden death of entanglement was  observed upon tuning the exchange and anisotropy field by a gate voltage. On the application side, this tunable entanglement channel --- by both electric and magnetic means --- may serve as a bridge to entangle distant qubits, placed on the top and bottom sides of the bilayer system.

\subsubsection{Bistability}\label{sec_bistable}
Several works reviewed above have shown that the hybrid magnet-cavity system provides a promising platform to study the entanglement among magnons, photons and phonons. Nonlinear effects usually play a crucial role to generate non-vanishing steady-state entanglement. On the other hand, nonlinearities often cause dynamic systems to have two, three or even more stable solutions under exactly the same conditions.  This is called bistability, tristability or, more generally, multi-stability. Which state the system evolves to depends on the initial conditions and past dynamics. Note that bistability itself is not a quantum phenomenon exclusively and exists in various physical systems, including optical systems \cite{Gibbsbook1985}, acoustic systems \cite{BrunoPRL2006,OsamaPNAS2017}, electronic systems \cite{MaAPL2002,ChuAM2005}, magnetic systems \cite{SessoliNature1993}, etc. Manipulation of bistable states may find promising applications in data storage and memory devices \cite{KamalAPL2010,KubyPRL2014,AlexoudiLight2020,ShenPRL2021}. Here we review the developments of this direction in hybrid magnet-cavity systems.

\begin{figure}
  \centering
  \includegraphics[width=1.0\textwidth]{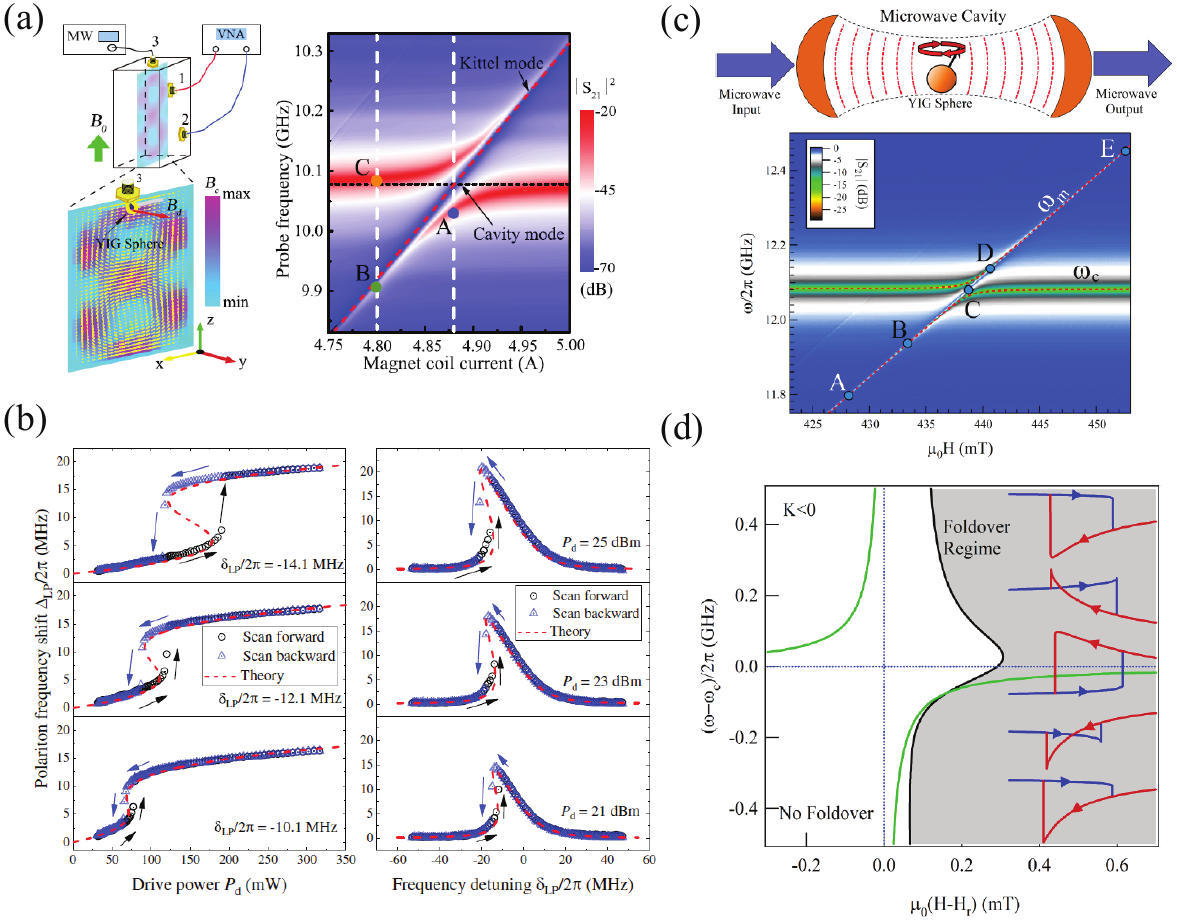}\\
  \caption{(a) Schematic diagram of a 3D cavity with a YIG sphere embedded and the coupling spectrum. Ports 1 and 2 of the 3D cavity are used for transmission spectroscopy and port 3 is for driving the magnetic sphere. (b) The detuning $\Delta_{LP}$ as a function of driving power $P_d$ (left) and as a function of frequency detuning $\delta_{LP}$. The red dashed lines are the theoretical prediction based on Eq. \eqref{eqbistable}. Here the [100] axis of YIG is aligned parallel to the static field and the cavity mode $\mathrm{\mathrm{TE}}_{102}$ is in resonance with magnons. (c) Schematic diagram of a follow-up experiment with driving on the cavity modes. (d) Phase diagram of the hysteresis loops, where the grey color indicates the different types of the bistability (foldover). $H_r$ is the resonance field and $\omega_c$ is the cavity frequency. Source: The figures are adapted from Refs. \cite{WangPRL2018,HydePRB2018}.}
  \label{bistability}
\end{figure}

Wang et al. \cite{WangPRL2018,ZhangSC2019} studied the influence of Kerr nonlinearities on the stability of magnon-polaritons in cavity-magnet systems as shown in Fig. \ref{bistability}(a). To demonstrate nonlinear effects, a microwave with frequency $\omega_d$ (Port 3) is applied to the magnetic sphere to drive the magnon mode, while an external static field was tuned to let magnons get into resonance with photons (Position A in the lower branch of the anticrossing spectrum).
Figure \ref{bistability}(b) first shows the frequency shift of the lower energy branch $\Delta_{\mathrm{LP}}$ as a function of the driving power $P_d$ at different driving frequency detuning ($\delta_{\mathrm{LP}}=\omega_{\mathrm{LP}}-\omega_d$). Clearly, a hysteretic behavior of the frequency shift is identified in both forward and backward scanning of the driving frequency. This suggests bistability of the magnon polariton. Moreover, when the frequency detuning is scanned at a given driving power, a hysteresis loop is also observed, where the loop area decreases when reducing the driving power $P_d$.

To understand the essential physics, the authors considered the Hamiltonian of the hybrid system as \cite{WangPRB2016}
\begin{equation}
\hat{\mathcal{H}} = \omega_c \hat{c}^\dagger \hat{c} + \omega_r \hat{a}^\dagger \hat{a} + v (\hat{a}^\dagger \hat{a})^2 + g(\hat{c}^\dagger \hat{a} + \hat{c}\hat{a}^\dagger)
+ \Omega_d(\hat{c}^\dagger e^{-i\omega_d t} + \hat{c} e^{i\omega_d t}),
\end{equation}
where $\hat{a}$ and $\hat{c}$ are the magnon and photon creation operators, respectively, $v$ is the Kerr nonlinearity caused by magnetocrystalline anisotropy of the system, so that $v$ is inversely proportional to the volume of the magnet, which is consistent with the model we discussed in the context of magnon antibunching (Section \ref{sec_singlemagnon}). $\Omega_d$ is the driving strength. Following a quantum Langevin approach, the authors derived a cubic equation for the frequency shift of the magnon-polariton $\Delta_{\mathrm{LP}}$ in the steady state as
\begin{equation}\label{eqbistable}
\left [ (\Delta_{\mathrm{LP}} + \delta_{\mathrm{LP}})^2 + \left ( \frac{\gamma_{\mathrm{LP}}}{2} \right )^2 \right ] \Delta_{\mathrm{LP}} - \varrho P_d = 0,
\end{equation}
where $\gamma_{\mathrm{LP}}$ is the damping rate of the lower energy branch of the polariton and $\varrho$ is a fitting parameter that characterizes the coupling strength between the driving microwave and the polariton. Equation (\ref{eqbistable}) has two stable solutions, which correspond to the bistability of the system. The prediction of the frequency shift $\Delta_{\mathrm{LP}}$ fits well with the experimental observations, as shown by the red dashed lines in Fig. \ref{bistability}(b). Furthermore, the authors demonstrated the appearance of bistability for off-resonant conditions [Points B and C in the energy spectrum Fig. \ref{bistability}(a)]. A follow-up work by Hyde et al. \cite{HydePRB2018} studied the bistability of the magnon-polariton by driving the cavity mode instead of the magnon mode with a high-power microwave, as shown in Fig. \ref{bistability}(c). They identified several types of bistability including (counter-)clockwise, (reversed) butterflylike, and clockwise hysteresis loops, as shown in Fig. \ref{bistability}(d), which depend on the driving frequency as well as the magnon frequency. Tuning the bistable behavior by external field and driving power may have potential applications in designing low-energy switching devices. In a follow-up work \cite{ShenPRL2021}, the authors experimentally observed the tristability in a three-mode hybrid magnet-cavity system, which is composed of two YIG spheres strongly coupled to a microwave cavity. The three separable stable states can be used to build a logic gate with zero, moderate and high driving powers.

Recently, Bi et al. \cite{BiPRB2021} further considered the nonlinearity of photons and found that their interplay with the magnon density saturation under strong driving induces tristability of the magnon polariton. Nair et al. \cite{NairPRB2020} proposed to place two YIG spheres into a single cavity and to drive one of them to study the spin excitation of the other undriven sphere. They found that the spin excitation in the driven sphere can migrate to the undriven one. The indirect coupling channel is engendered by the cavity photons and shows multistable behaviors. They \cite{NairPRB2021} also found that the threshold power to induce the bistability can be reduced by a factor of 5, when both the magnon and photon modes are connected to a common thermal bath and dissipation effects dominate the coherent coupling. Yang et al. \cite{YangPRR2021} evaluated the magnon-photon entanglement in the nonlinear regime of this hybrid system and found that the entanglement can jump from one state to the other near the switching points of the bistable regime.

\subsection{Hybrid magnet-qubit system}\label{sec_magnon_qubit}
A qubit is a two-state quantum-mechanical system and it is the basic information carrier in quantum information science and technology. Various types of qubits have been implemented to realize quantum information tasks, including superconducting qubits \cite{MartinisPRL2002,MakRMP2001}, photonic qubits \cite{KokRMP2007,SergAPR2019}, atoms \cite{KaneNature1998}, trapped ions \cite{CiracPRL1995}, NV centers \cite{ChildressNV2006}, SV centers \cite{LachPRL2014,PingPRL2014}, and anyons \cite{NayakRMP2008}, among others. Magnons have been successfully integrated with the superconducting qubit and NV center either by direct dipolar coupling or mediated by cavity photons. On the one hand, qubit systems are able to help to detect and image magnetic textures through magnetorelaxometry, where the spins are coupled with the qubit by dipolar interactions. Moreover, the qubits are capable of detecting the quantum magnonic state in both collinear and noncollinear magnetic structures, since it is very sensitive to magnetic noise. On the other hand, magnons are able to control the qubit state via dipolar interaction and further mediate the entanglement of two or more distant qubits. This takes advantage of the long lifetime of magnons in magnetic insulators and the high tunability of magnonic systems by electric and magnetic means. In this section, we will discuss these two sides and review the recent developments.

\subsubsection{Magnon-superconducting qubit inside one cavity}

\begin{figure}
  \centering
  \includegraphics[width=1.0\textwidth]{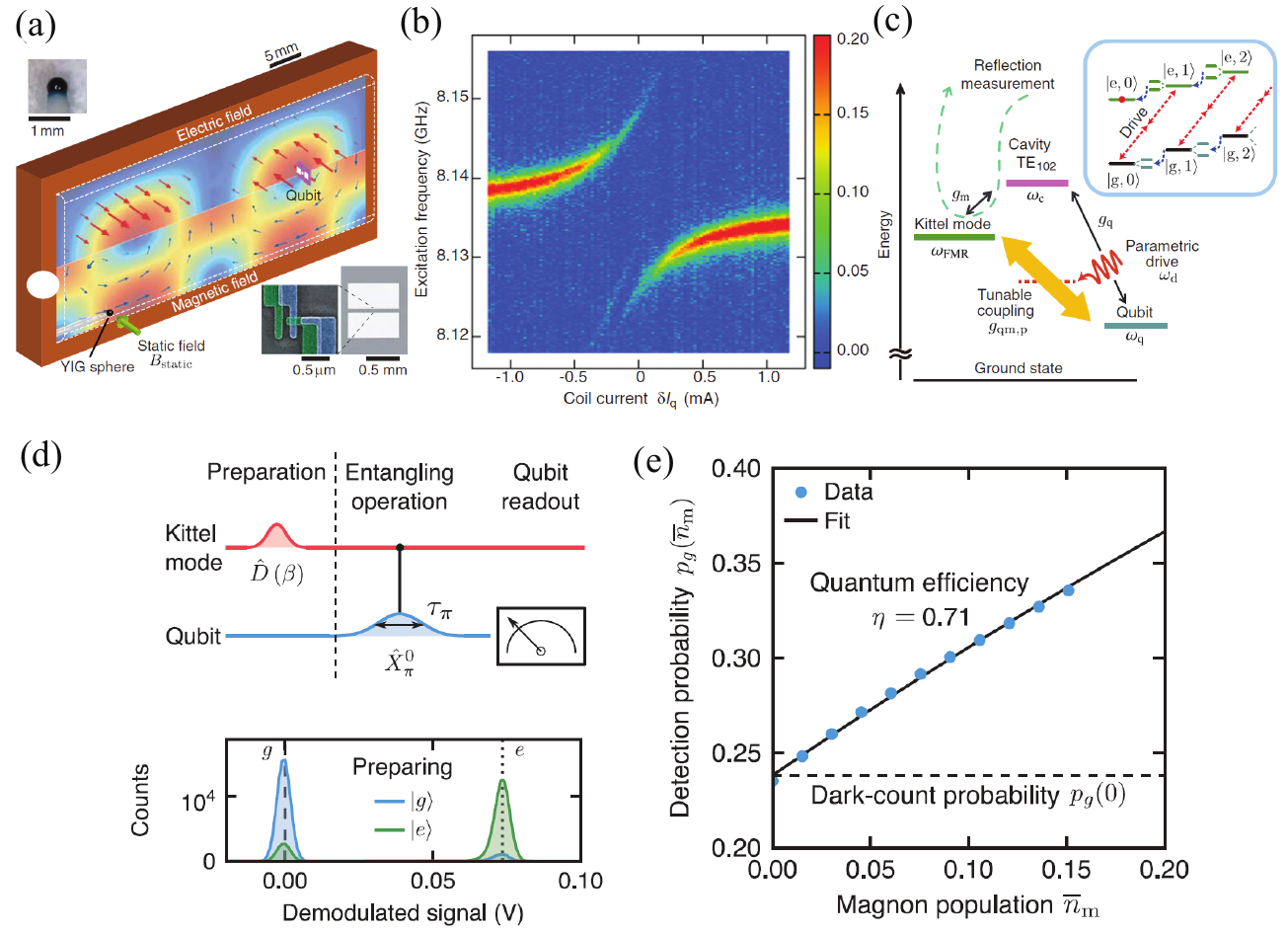}\\
  \caption{(a) Schematic of a hybrid magnet-qubit-cavity system. (b) Rabi splitting due to the indirect coupling between magnons and qubits mediated by cavity photons. (c) Energy level distribution of the hybrid system to explain the off-resonant manipulation of the coupling strength between magnons and qubits. (d) Procedure for single-magnon detection using superconducting qubits. (e) Detection probability of magnons as a function of magnon population. Source: The figures are adapted from Refs. \cite{LachScience2020,TabuchiScience2015}.}
  \label{Tabuch_mqubit}
\end{figure}

Tabuchi et al. first experimentally demonstrated a hybrid magnet-qubit-cavity system that consists of a ferromagnetic insulator (YIG sphere) and a superconducting qubit, as shown in Fig. \ref{Tabuch_mqubit}(a) \cite{TabuchiScience2015}. The state of the qubit is well described by a two-level system while the magnon excitations in the magnet form a continuous variable system. Here magnons and qubits are magnetically and electrically coupled to the cavity photons ($\mathrm{TE}_{102}$ mode), and hence there is an indirect coupling between magnons and photons by exchange of a virtual photon. This leads to an interaction of the form $\hat{\mathcal{H}}_{\mathrm{int}}=g_{mq} \hat{\sigma}^+ \hat{a} + g_{mq}^* \hat{\sigma}_- \hat{a}^\dagger$, where $\hat{\sigma}^{\pm}$ describes the qubit, $g_{mq}=g_m g_q/\Delta$ is the effective magnon-qubit coupling strength, with $g_q$ and $g_m$ respectively the coupling strength of cavity photons to qubit and magnons, and $\Delta$ is the detuning between the qubit (magnon) and cavity mode. To demonstrate the magnon-qubit coupling, the authors read the qubit state by another cavity mode ($\mathrm{TE}_{103}$ mode) and found that the transmission spectrum takes the typical form of energy level repulsion as shown in Fig. \ref{Tabuch_mqubit}(b), which is a manifestation of the coherent coupling between magnon and qubit. The effective coupling strength $g_{mq}/2\pi =10~ \mathrm{MHz}$ is consistent with the theoretical estimate of 11.8 MHz, and it is much larger than the dissipation of either magnons or qubits, such that the coupling falls into the strong coupling regime. The authors further showed that parametric driving applied to the qubit effectively tunes the coupling strength between the qubit and magnons. Now, the hybrid system oscillates between the ground state $|g,0\rangle$ and  excited state $|e,1\rangle$ as shown in Fig. \ref{Tabuch_mqubit}(c), where $|g\rangle, |e\rangle$ are the qubit ground and excited states and $|n\rangle$ are the magnonic Fock states. The effective gap between these two states is manipulated by the driving power as well as by the detuning between the driving and the average of the magnon and qubit frequency.

Based on the strong coupling of magnon and qubit, Lachance-Quirion et al. tuned the system into a strongly dispersive regime, where the detuning between the qubit and magnon is much larger than their coupling, i.e., $|\omega_q-\omega_m| \gg g_{mq}$, which produces an effective nonlinear interaction between magnons and photons in the form of $\hat{\mathcal{H}}_{\mathrm{int}} = \chi_{qm} \hat{a}^\dagger \hat{a} \hat{\sigma}_z$ \cite{LachScience2020,LachSA2017,WolskiPRL2020}. This implies that the excitation probability of the qubit sensitively depends on the number of excited magnons. The higher the number of magnons, the larger the frequency shift of the qubit frequency. Therefore, the number of magnons can be deduced by measuring the shift of the qubit frequency. On the other hand, by measuring the probability of the qubit to be in the excited states, one determines the excitation probability of the magnons.
Figure \ref{Tabuch_mqubit}(d) shows the procedure for single-magnon detection. First, a coherent state of magnons was prepared. Second, a conditional
excitation $X_\pi^0$ pulse is applied on the qubit system to entangle the qubit and magnons. Finally, the qubit state is read out by using a high-power technique \cite{ReedPRL2010}. Figure \ref{Tabuch_mqubit}(e) shows the detection probability of magnons as a function of magnon number, where the detection probability is given by $p_g(n_m)=\bar\eta(1-e^{-n_m}) + p_g(0)$, with $\bar\eta$ and $p_g(0)$ being respectively the quantum efficiency and the dark-count probability, which can be obtained by fitting the experimental data. When the magnet is in the vacuum state, the probability of the detector click is $p_g(0)$. When the magnonic system is in the first excited state $|1\rangle$, the detector clicks with a probability $\bar\eta + p_g(0)$. Then one can readily detect single-magnon state by counting and analyzing the detector clicks. To guarantee the efficiency of detection, the measuring time has to be lower than the decoherence time of qubit, which is around 200 ns. This proof-of-concept experiment establishes the qubit state as single-magnon detector, which has promising applications to detect weak magnon excitation with high precision. The highest fidelity of single-magnon detection realized in this experiment corresponds to a precession of the magnetization with an angle of $10^{-17}$ degree in a millimeter-sized YIG sphere.

Following these experimental works, Liu et al. \cite{LiuPRA2021} studied the steady-state behavior of the hybrid magnet-qubit-cavity system and found that the dispersive coupling between magnons and qubits induces bistability of the system. Wu et al. \cite{WuPRA2021} considered a model with the superconducting qubit replaced by a three-level fluxonium qubit and found that both single and multi-magnon blockade is achieved and interchanged by properly choosing the parameters. Luo et al. \cite{LuoOL2021} showed that nonlocal entanglement between two remote magnet-qubit-cavity systems can be realized.
Janss{\o}nn et al. theoretically demonstrated the propagation of superconducting signatures to a distant ferromagnet mediated by their mutual coupling to a cavity electromagnetic mode \cite{JansonPRB2020}.

\subsubsection{Manipulating one or more qubits by magnons}
\begin{figure}
  \centering
  \includegraphics[width=1.0\textwidth]{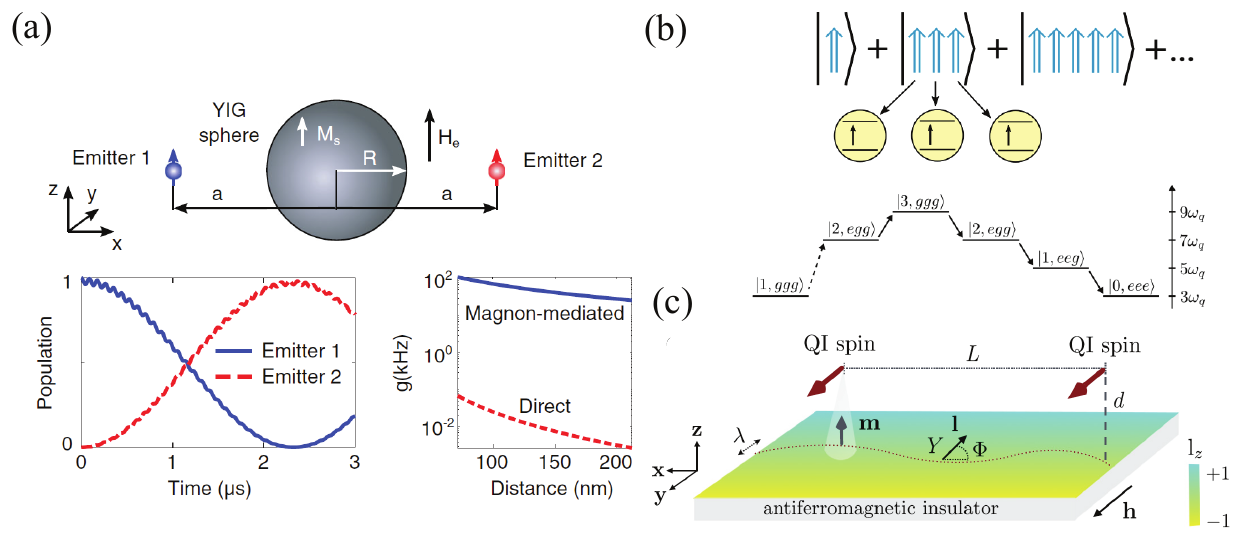}\\
  \caption{(a) (Top) Schematics depiction of two qubits coupled by a magnon mode in a nanometer-sized magnetic sphere. (Bottom left) The population of the two qubits as a function of time. (Bottom right) Effective coupling strength of two qubits mediated by magnons as a function of the distance between the two qubits and its comparison with the direct coupling in the absence of magnets. (b) Schematic of the squeezed magnon with only odd number occupation and the transition $|1,ggg\rangle$ to $|0,eee\rangle$ mediated by a set of virtual states. (c) Schematic of the coupling of two qubits via the magnon mode in an antiferromagnetic insulator. Here QI is the abbreviation of quantum impurity. Source: The figures are adapted from Refs. \cite{FlebusPRB2019,NeumanPRL2020,Skogvoll2021}.}
  \label{magnon2qubit}
\end{figure}

Trifunovic et al. \cite{TrifPRX2013} first proposed to couple two spin qubits separated by a relatively larger distance (several $\mathrm{\mu m}$) through magnonic excitations in a ferromagnet with a large gap. Here the two qubits are coupled to the ferromagnet through dipolar interactions, hence the qubit has to be placed close to the magnets, i.e., on the order of several tens of nanometers, to guarantee the coupling strength is sufficiently strong. The on-and-off switching of the qubit-qubit interaction is achieved by tuning the qubits on and off resonance with the ferromagnets. The decoherence time of the qubit $T_2$ depends sensitively on $k_BT/\Delta_F$, where $\Delta_F$ is the magnon gap. Hence it is desirable to choose materials with a small magnon gap and to perform the operations at low temperature. Neuman et al. \cite{NeumanPRL2020} considered a nanomagnet to mediate the long-range coupling of two qubits as shown in Fig. \ref{magnon2qubit}(a). Here each qubit is coupled to the magnon mode inside the nanomagnet and their detunings relative to the magnon frequency are brought out of resonance to enhance their mutual coupling via the transfer of a virtual magnon. They found population exchange between the two qubits, which clearly demonstrates the indirect coupling of the two qubits. The effective coupling strength is three to four orders of magnitude larger than the direct coupling strength.

The generation of quantum entangled states of two or more qubits via magnons was recently theoretically proposed. Skogvoll et al.~\cite{Skogvoll2021} theoretically showed that three qubits simultaneously coupled to a squeezed magnon mode can jointly absorb the latter and robustly generate a 3-qubit Greenberger-Horne-Zeilinger (GHZ) state \cite{GHZarxiv2007}. Here the squeezed magnon state is a superposition of odd magnon-number excitations only [Eq.~\eqref{sq:eq:sqmagnum}]. Taking the case of three qubits as an example, this feature enables the system to transition from 1 squeezed magnon and 3 ground-state qubits $|1,ggg\rangle$ to 0 squeezed magnons and 3 excited qubits $|0,eee\rangle$ via a set of virtual states, as shown in Fig.~\ref{magnon2qubit}(b). The generated GHZ and related states may be useful for implementing Shor's algorithm \cite{ShorPRA1995} for quantum error correction. Ramos et al. designed a chiral quantum network, where the interactions between spins are mediated by magnon excitations in a 1D XX spin chain \cite{RamosPRA2016}. Fukami et al. \cite{Fukamiarxiv2021} theoretically showed that two qubits that are separated by $1-2~\mu$m can become entangled due to exchange of magnons in a ferromagnetic YIG bar. The estimated cooperativity exceeds $10^4$ for on-resonance conditions and the ratio of the effective qubit-qubit interaction to the decoherence rate of the qubit is around $10^3$ away from resonance, which may be useful to implement entangling gate protocols. Zou et al. \cite{Zouarxiv2021} showed that the two qubits can be entangled via dissipative coupling to a magnetic medium, in the absence of any coherent coupling. A proper measurement and post-selection can produce the maximally entangled Bell states of the two qubits. Xiong et al. \cite{Xiongarxiv2021} showed how distant quantum information tasks including two-qubit iSWAP gates and charger-battery devices can be potentially realized in such a hybrid system.

Besides using magnonic excitation of a collinear magnetic structure as illustrated above,  Flebus et al. \cite{FlebusPRB2019} theoretically showed that an antiferromagnetic domain wall is able to mediate a tunable coherent coupling between two qubits separated by around a micrometer as shown in Fig. \ref{magnon2qubit}(c). Here the planar rotation of the staggered order parameter, i.e., the spin superfluid mode, is excited and simultaneously coupled two qubits placed above the domain wall. We note that the experimental demonstration of qubit-qubit coupling through magnon modes is still lacking. Nevertheless, current studies have already enabled the coupling between a single qubit and a magnon \cite{AndrichNPJ2017,KikuchiAPE2017,WolfNC2016,Candido2020} as briefly described below.

\begin{figure}
  \centering
  \includegraphics[width=1.0\textwidth]{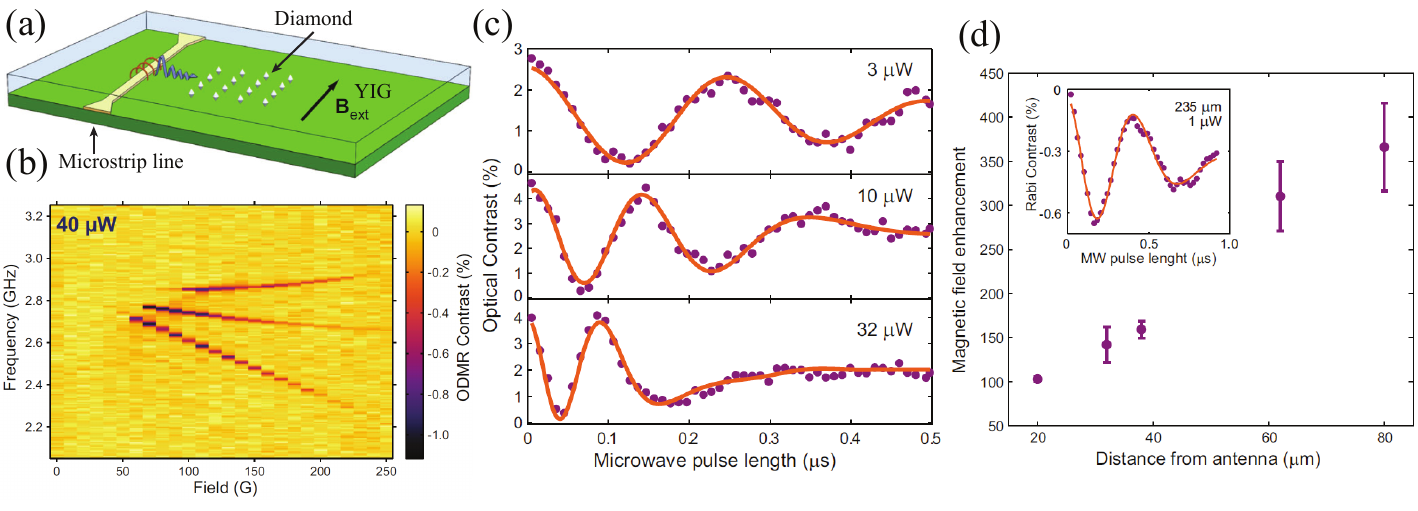}\\
  \caption{(a) Schematic of the hybrid magnet-NV center system. Surface spin-waves are excited from the microstrip lines and propagate toward the nanodiamonds. Each nanodiamond contains hundreds of NV centers. (b) Optically detected magnetic resonance spectra as a function of the external magnetic field and driving frequency at microwave power $40~\mathrm{\mu W}$. (c) Spin-wave mediated Rabi oscillations of NV centers at a fixed field 120 G. (d) Microwave magnetic field enhancement as a function of the nanodiamond's distance from the antenna. Source: The figures are adapted from Ref. \cite{AndrichNPJ2017}.}
  \label{Andrich_nv}
\end{figure}

Andrich et al. \cite{AndrichNPJ2017} developed a hybrid YIG-nanodiamond (ND) system as shown in Fig. \ref{Andrich_nv}(a). Here each ND contains hundreds of NV centers, which can be viewed as a qubit bus at room temperature. They showed that the surface SW mode, which is excited using a micro-strip line in the YIG film, strongly interacts with the NV center ensembles in NDs. Specifically, the authors detected the photoluminescence spectrum change of the NV center by a nano-particle, located $40~\mu$m away from the antenna. The results at driving power of $40~\mu$W are shown in Fig. \ref{Andrich_nv}(b). At such low power, the antenna's microwave field is negligible at the ND's location. Then, the observed discrete modes of the NV centers are likely to come from the coherent coupling between the propagation of SWs and NV centers. To strengthen this conclusion, the authors measured the Rabi oscillation signal of the NV centers as shown in Fig. \ref{Andrich_nv}(c). Here, the larger the driving power, the larger the SW amplitude, and thus the faster the Rabi oscillation. Quantitatively, the frequency oscillation scales with the square root of the driving power. Finally, the authors studied the enhancement of coherent microwave fields by comparing the Rabi frequencies in the low-field regime (antenna microwave dominates, no significant coupling between SW and NV) and high-field regime (SW dominates the coupling with NV centers), and found that the magnetic field can be enhanced by two orders of magnitude as shown in Fig. \ref{Andrich_nv}(d). Similar SW mediated excitation of NV centers was also reported by An et al. \cite{KikuchiAPE2017}.

Wolf et al. \cite{WolfNC2016} experimentally showed that the magnetic fields produced by an oscillating magnetic vortex core can coherently manipulate the Rabi oscillation of NV centers placed above the magnetic film at a distance of a few nanometers. Here, the vortex core is moved and repositioned on a time scale of 100 ns, which is the same scale as the coherence time of the NV centers and thus allows a coherent operation of the qubits. Using similar principles, Yuan et al. \cite{YuanNV2019} theoretically showed that an electric field can drive the breathing motion of magnetic skyrmions by periodically changing the anisotropy of the system. The breathing skyrmions generate oscillating dipolar fields outside the magnetic film that can manipulate an NV center located a few nanometers above the magnetic film.

\subsubsection{Detecting magnetism by single-spin qubits}\label{sec_magnon_NV}
Besides manipulating the state of spin qubits by magnons excited in a ferromagnetic system, spin qubits in quantum impurities, including NV centers and SV centers, can serve as a sensor to image magnetic textures, such as magnetic vortices \cite{RondinNC2013}, domain walls \cite{TetNC2015,TetScience2014}, skyrmions \cite{DovNC2018}, and noncollinear antiferromagnets \cite{GrossNature2017}. Furthermore, spin qubits can detect magnonic excitations above these magnetic textures \cite{SarNC2015} and the spin chemical potential of magnons \cite{DuScience2017}. Up to now, most of these efforts involve NV centers and a systematic review was presented in the early 2018 by Casola et al. \cite{CasolaReview2018}, while SC centers draw little attention \cite{BejIEEE2021}. In the last few years, this field has been continuing to develop, featuring the detection of magnetic states in 2D van der Waals magnets and the study of SWs and spin current transport via spin qubit relaxometry. We will focus on the most recent developments of this field.

\begin{figure}
  \centering
  \includegraphics[width=1.0\textwidth]{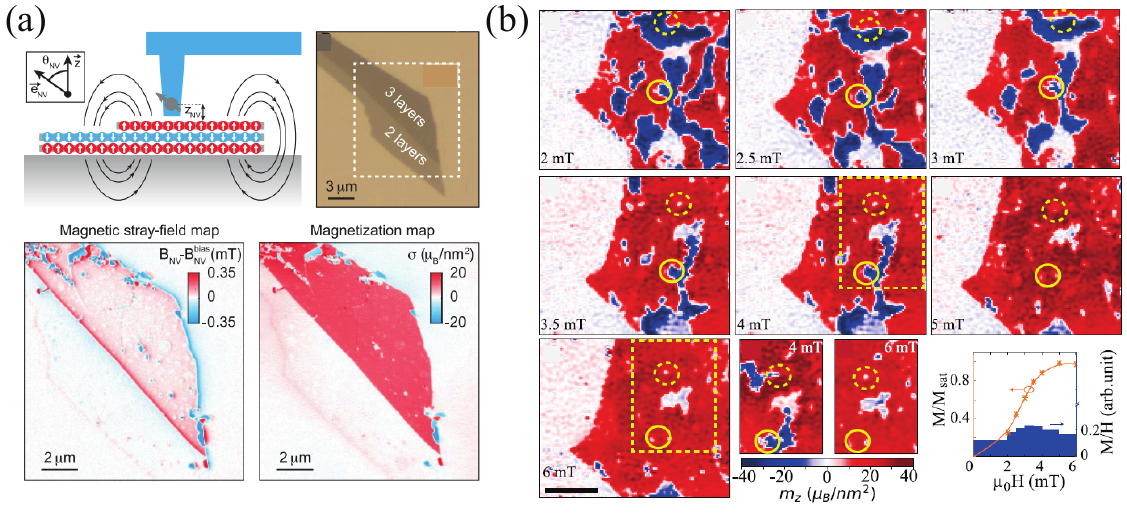}\\
  \caption{(a) (Top) Schematic of the single-spin magnetometry technique for imaging magnetism in van der Waals magnets $\mathrm{CrI}_3$ with even and odd number of stacking layers. Strong and weak stray fields in the magnetic stray-field map emerge from trilayer and bilayer flakes, respectively. (Bottom) The corresponding magnetization distribution is recovered by the reverse propagation method.  (b) Magnetic domain evolution of a $\mathrm{CrBr}_3$ bilayer upon increasing the magnetic field. The solid and dashed yellow circles represent two pinning sites. Source: The figures are adapted from Refs. \cite{ThieScience2019,SunNC2021}.}
  \label{NV_2D_magnet}
\end{figure}

Ever since the discovery of ferromagnetic order in 2D materials $\mathrm{CrI_3}$ and $\mathrm{Cr_2Ge_2Te_6}$ \cite{GongNature2017,HuangNature2017}, 2D magnetism has gained significant research interest and the characterization of the magnetization in these 2D materials is an important issue. Compared to the traditional techniques to characterize 2D magnetism, including Kerr microscopy and magnetic circular dichroism, a sensor based on a quantum impurity has a high spatial resolution of a few tens of nanometers, which enables an accurate measurement of the magnetic-moment distribution \cite{FernScience2019}. Thiel et al. \cite{ThieScience2019} imaged the magnetic moments of the van der Waals magnet $\mathrm{CrI}_3$ by sensing the magnetic stray fields generated by the 2D magnet, as shown in Fig. \ref{NV_2D_magnet}(a). Using reverse-propagation protocols, the magnetic-moment distribution in the van der Waals magnet is recovered from its stray field distribution. The authors found that the total magnetization of an odd number of layers is $16~ \mu_B /\mathrm{nm^2}$, while the magnetization of an even number of layers is nearly zero. These findings corroborated the antiferromagnetic nature of the interlayer coupling in layered $\mathrm{CrI}_3$.  Sun et al. \cite{SunNC2021}  improved the magnetic field sensitivity by using a pulsed measurement scheme and successfully imaged the domain reversal in few layer $\mathrm{CrI}_3$ samples, as shown in Fig. \ref{NV_2D_magnet}(b). A wide field NV-microscope with lower resolution ($\sim 500$ nm) was used to detect magnetization reversal by external magnetic field \cite{BroadAM2020}. Such techniques can track the domain structure of individual flakes, while it sacrifices the spatial resolution.

\begin{figure}
  \centering
  \includegraphics[width=1.0\textwidth]{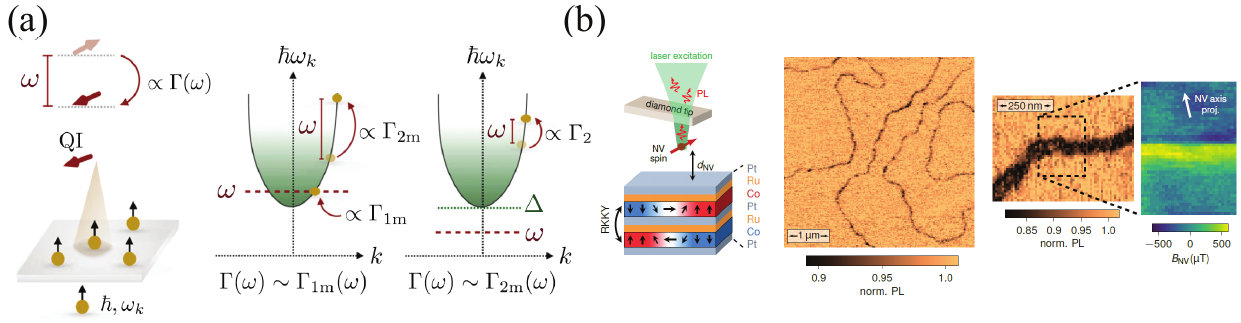}\\
  \caption{(a) Quantum-impurity relaxation via one and two-magnon scattering processes.  When the quantum impurity splitting $\omega$ is larger than the spin-wave gap $\Delta_F$, one magnon process dominates the relaxation rate of the quantum impurity. When $\omega < \Delta$, two magnon process dominates the relaxation. (b) Imaging of a magnetic domain wall in a synthetic antiferromagnet ($\mathrm{Co/Ru/Pt/Co}$) by a scanning-NV magnetometer. The overview of the domain and domain wall structures is obtained by measuring the photoluminescence of NV center. Source: The figures are adapted from Refs. \cite{FlebusPRL2018,AvinPRB2020}.}
  \label{spin_relaxometry}
\end{figure}

In addition to the sensitivity of spin qubits to static magnetic fields, the relaxation process of the qubit is influenced considerably by magnetic noise of the environment. By measuring the relaxation rate of the qubit, one is able to gain information on the dynamic excitation of magnetic textures which induce magnetic noise nearby. The magnetic noise generated by the magnetic system with a magnon gap lower than the resonance frequency of the qubit was addressed in Refs. \cite{TetNC2015,DuScience2017,CasolaReview2018}. To address the applicability of the technique to a wider class of ferromagnets with large magnon gap caused by strong anisotropy or external field, Flebus et al. \cite{FlebusPRL2018,FlebusJAP2021} theoretically generalized this method to the detection of magnetic noise using a qubit with subgap frequencies. In particular, as shown in Fig. \ref{spin_relaxometry}(a), the two-magnon process contributes to the qubit relaxation rate, which is different from the one-magnon process in the normal case. Note that the two-magnon process contributes to the longitudinal noise of the magnetic field, which can be seen from the HP transformation $\hat{S}_z=S-\hat{m}^\dagger \hat{m}$. By evaluating the longitudinal susceptibility of the system, the authors derived the longitudinal noise's influence on the qubit's relaxation rate ($\Gamma$) as
\begin{equation}
\Gamma \sim \frac{16\mu_B^4}{A^3 \beta^2}\ln \frac{A}{d_q^2(\Delta_F - \mu)},
\end{equation}
where $A$ is the exchange coefficient, $\beta=1/(k_BT)$, $d_q$ is the distance of the qubit from the magnetic sample, $\Delta_F$ is the magnon gap, and $\mu$ is magnon chemical potential. In the equilibrium state, $\mu=0$, and the relaxation time $\Gamma^{-1} \sim 10$ ms for parameters of YIG, while near BEC, $\Gamma^{-1} \sim 100$ $\mu$s, which is much shorter than the intrinsic relaxation time of the NV center. Therefore, magnon BEC may be probed via two-magnon noise through spin qubit relaxometry. Similar ideas apply to BEC in antiferromagnets, and were extended to propose the detection of magnon modes in antiferromagnetic domain walls \cite{FlebusPRB2018}. One advantage of this method is that it can image magnetic structures with zero magnetization, such as a compensated antiferromagnet.

Finco et al. \cite{FincoNC2021} experimentally demonstrated this proposal by successfully imaging various spin textures in synthetic antiferromagnets, including domain walls, spin spirals, and skyrmions as shown in Fig. \ref{spin_relaxometry}(b). Rustagi et al. \cite{AvinPRB2020} further addressed the influence of chiral magnetic noise on the qubit relaxation. In particular, the left-moving and right-moving magnons produce unequal magnetic noise and they are respectively coupled to the $m_s=0 \rightarrow -1$ transition ($\Gamma_-$) and $m_s =0 \rightarrow +1$ transitions ($\Gamma_+$) in an NV center. When the chiral symmetry of SWs is broken, for example by surface SWs, the magnitudes of the magnetic noise for left-moving and right-moving SWs are different, hence the relaxation rate $\Gamma_-$ and $\Gamma_+$ are unequal. The authors proposed a general formalism to evaluate this difference and verified the prediction experimentally. Furthermore, it has been experimentally demonstrated that the NV center spins can effectively probe coherent SW transport driven by microwaves, spin torques, and thermal gradients \cite{SolyomNL2018,BertScience2020,Dwi2021,Bertarxiv2021}.

\subsection{Hybrid magnon-phonon system} \label{sec_magnon_phonon}
The coupling between magnon and the quanta of mechanical oscillations is another interesting direction. The traditional approaches mainly address the interplay of magnons and phonons and their influence on magnon transport properties, for example, excitation of hybrid magnon-phonon polarons \cite{KamraPRB2015,KikkawaPRL2016,LiPRL2020}, magnon Seebeck effect driven by a thermal gradient \cite{BauerSSE2012}, phonon-mediated spin transport \cite{AndreasPRL2020}, and surface acoustic waves mediated coherent magnon excitation \cite{MaekawaAIP1976,WeilerPRL2011, DreherPRB2012,WeilerPRL2012,XuSA2020}. A review of recent developments is performed by Bozhko et al. \cite{BozLTP2020}. For quantum magnonics, the entanglement of magnons with phonons is manipulated. In particular, here, phonons are not limited to the deformation of the magnetic system itself, and the coupling of magnetic modes to the oscillations of a distinct mechanical system is discussed as well.

\begin{figure}
  \centering
  \includegraphics[width=1.0\textwidth]{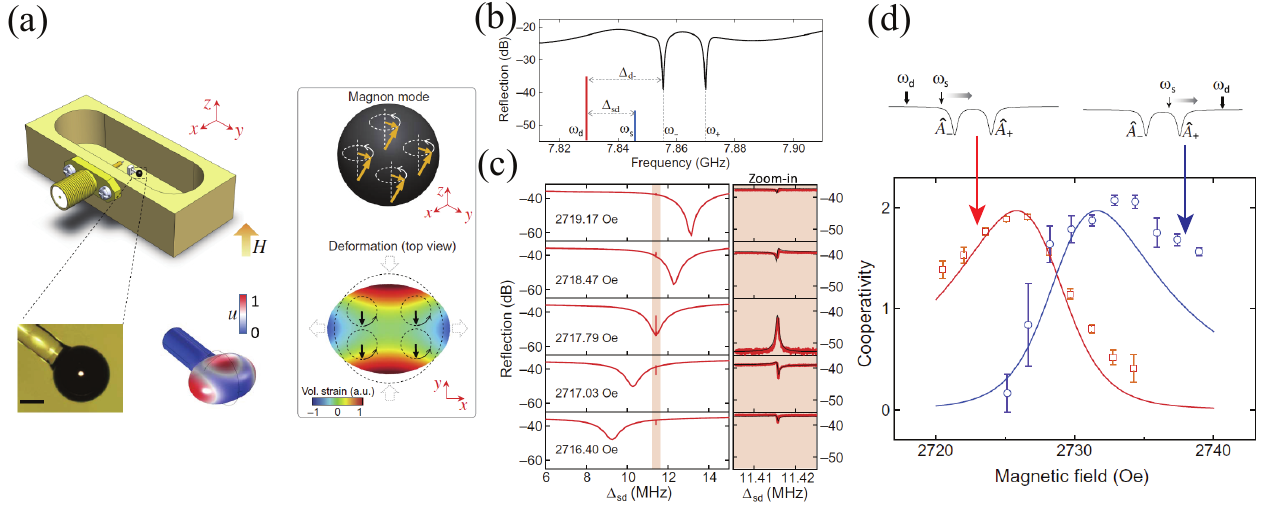}\\
  \caption{(a) Schematic device containing a three-dimensional copper cavity and a YIG sphere. (b) The reflection spectrum when the magnon is on resonance with the cavity photon.  (c) Measured reflection spectra with a red-detuned driving between pump and probe photons near the lower hybrid mode $A_-$. A Fano-like lineshape is identified, accompanied by a transparency peak when $\Delta_{sd}=\omega_b$, where $\omega_b$ is the phonon frequency. (d) The magnonmechanical cooperativity as a function of bias magnetic field. The red and blue lines refer to the red and blue detuning between the pump and probe frequency, respectively. Source: The figures are adapted from Ref. \cite{ZhangSA2016}.}
  \label{magnon_mechanics}
\end{figure}

{\it Coupling of magnons with deformation of magnets.} Zhang et al. \cite{ZhangSA2016} studied a hybrid cavity-magnet system by placing a YIG sphere into a copper cavity, as shown in Fig. \ref{magnon_mechanics}(a). The magnetization oscillation of the YIG sphere induces mechanical deformation via magnetostrictive interactions of the form \cite{AndreasPRB2014}
\begin{equation}\label{mphonon_ZhangSA}
\hat{\mathcal{H}}_{\mathrm{int}} = b_1 \int d^3 r \left[ \hat{m}_x^2 \epsilon_{xx} +  \hat{m}_y^2 \epsilon_{yy} - (\hat{m}_x^2 + \hat{m}_y^2) \epsilon_{zz} \right ]
=\frac{b_1M_s}{V} \hat{a}^\dagger \hat{a}  \int d^3 r ( \epsilon_{xx} +  \epsilon_{yy} - 2\epsilon_{zz} ),
\end{equation}
where $b_1$ is magnetoelastic constant, $\epsilon_{ij} \equiv (\partial_j u_i + \partial_i u_j)/2$ are the components of the strain tensor, $\mathbf{u}$ is the atomic displacement, and $V$ is the volume of the magnet.
In this model, the magnon-phonon interaction is dispersive, as the magnon frequency is shifted by the phonons. After quantizing the phonon operators, the effective coupling between the magnon and phonon modes is written as $\hat{\mathcal{H}}_{\mathrm{int}}=g_{mb} \hat{a}^\dagger \hat{a} (\hat{b}+\hat{b}^\dagger)$, where $\hat{b}~(\hat{b}^\dagger)$ is the annihilation (creation) operator for phonons. This justifies the magnon-phonon interaction that was used to study  magnon-phonon entanglement in Section \ref{sec_mpEn}.

When the magnon frequency is tuned to be close to the photon frequency, two hybrid magnon-photon modes at energy $\omega_{\pm}$ with eigenvectors $\hat{A}_+ = \cos \theta \hat{c} + \sin \theta \hat{a}$ and  $\hat{A}_- = -\sin \theta \hat{c}+ \cos \theta \hat{a}$ are found as shown in Fig. \ref{magnon_mechanics}(b). Here, the magnon component of the hybrid mode is determined by the magnon-photon coupling $g_{mc}$ and the magnon-photon detuning $\Delta_{mc}$ as, $\tan 2\theta = 2g_{mc}/\Delta_{mc}$. The phonon mode does not play a role here because its MHz frequency is far below the magnon frequency, which is in GHz regime. To bring the magnon and phonon frequencies closer, a parametric driving $\omega_d$ is applied to drive the cavity field.  In the dressed basis, the interaction component of the Hamiltonian is rewritten as $\hat{\mathcal{H}}_{\mathrm{int}} = g_{mb}(\hat{b}+\hat{b}^\dagger)\left (\sin^2 \theta \hat{A}^\dagger_+ \hat{A}_+ +\cos^2\theta \hat{A}^\dagger_- \hat{A}_- \right )$. Now both the beam-splitter-type interaction ($\hat{A}_{\pm}^\dagger \hat{b} + \hat{A}_{\pm} \hat{b}^\dagger$) and parametric interaction ($\hat{A}_{\pm}^\dagger \hat{b}^\dagger + \hat{A}_{\pm} \hat{b}$) between the phonon and the hybrid magnon-photon modes are realized and the coupling strength can be manipulated by the detuning between magnon and photon ($\Delta_{mc}$ or angle $\theta$). For red-detuned driving, the driving field only strongly interacts with the lower branch $\omega_-$ and a typical Fano lineshape is observed with a transparent mode at the resonance, as shown in Fig. \ref{magnon_mechanics}(b). In the blue-detuned case, the driving is strongly coupled with the higher energy branch ($\omega_+$), yielding a dip in the reflection spectra. By tuning the external magnetic field or magnon frequency, one is able to achieve the transition between these two situations, as shown in Fig. \ref{magnon_mechanics}(c).  This is a distinct advantage of this hybrid system.

Based on this accessible platform, various theoretical proposals have come up to study the interplay between magnons, phonons and photons. Sarma et al. \cite{SarmaNJP2021} proposed a stimulated Raman adiabatic passage-like coherent transfer between the cavity mode and phonon mode, by properly tuning the driving and frequency detuning between cavity photon and magnon. The phonon mode with lower damping may be used to store the quantum state for a long time. Moreover, as introduced in Section \ref{sec_mpEn}, nonlinear magnon-phonon interactions enable the existence of steady-state entanglement among magnons, photons, and phonons at low temperature.  Huai et al. \cite{HuaiPRA2019} implemented gain to the magnonic system and constructed a $\mathcal{PT}$ symmetric-like system with a passive cavity and active magnon modes. Lu et al. \cite{LuPRA2021} studied the role of EPs in the hybrid system and showed flexible on-off optical transmission and slow-to-fast light switching near the EPs.

\begin{figure}
  \centering
  \includegraphics[width=1.0\textwidth]{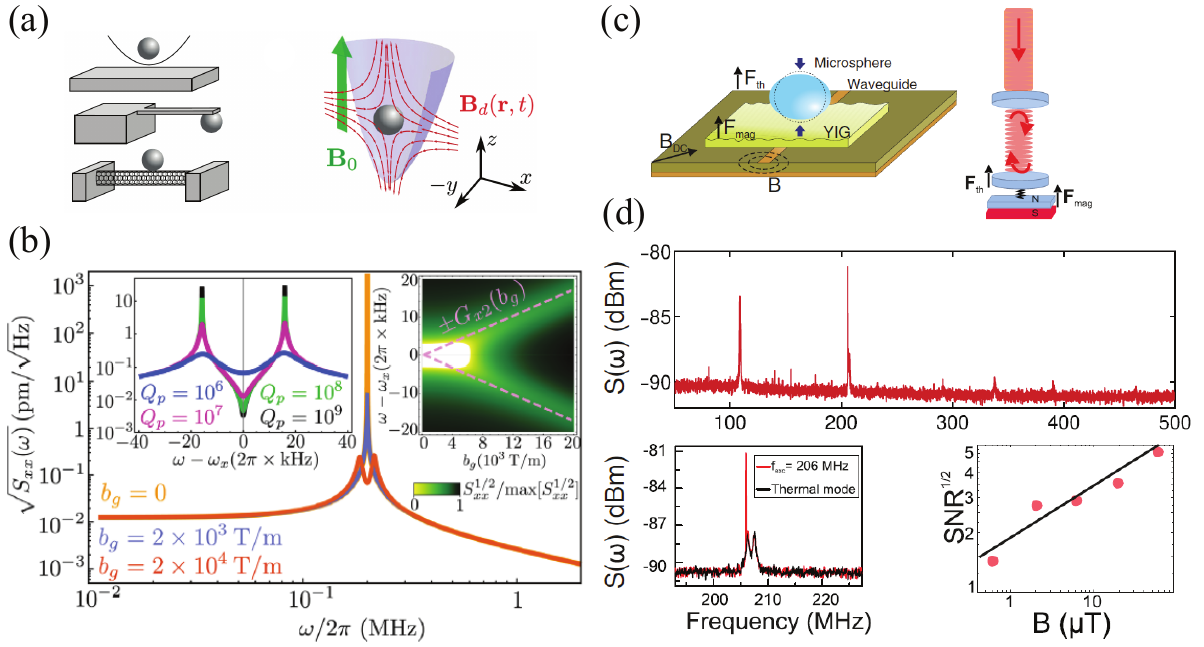}\\
  \caption{(a) Schematic illustrations of a micromagnet trapped by a mechanical potential. (b) Power spectral density of the system at different acoustic-mechanical couplings ($b_g$). (c) Schematic representation of the magnetometer, containing a microsphere, a ferromagnetic YIG film and a microstrip waveguide. (d) Spectral response of a microsphere excited by applying an oscillating field of $2.3~ \mathrm{\mu T}$ at 206 MHz (red line). The black curve is the response when the sphere is lifted from the YIG film.  SNR is short for signal-to-noise ratio of the measurement. Source: The figures are adapted from Refs. \cite{CarlosPRL2020,ColomPRL2020}.}
  \label{magnon_mechanics2}
\end{figure}

{\it Coupling between magnons and external mechanical oscillators.} Alternatively, Gonzalez-Ballestero et al. \cite{CarlosPRL2020,CarlosPRB2020} considered a spherical micromagnet (for example, YIG) trapped in a harmonic potential, as shown in Fig. \ref{magnon_mechanics2}(a). The magnetic sphere experiences an external static field $\mathbf{B}_0=B_0\mathbf{e}_z$ and an oscillating gradient field $\mathbf{B}_d=b_g(-x\mathbf{e}_x+z\mathbf{e}_z)\cos (\omega_d t)$. The static field tunes the magnon frequency while the inhomogeneous oscillatory field induces a coupling between the magnon mode and the center-of-mass motion of the magnet. The hybrid system is described by the Hamiltonian
\begin{equation}
\hat{\mathcal{H}}= \omega_x \hat{b}^\dagger \hat{b} + \omega_r \hat{a}^\dagger \hat{a} + \omega_p \hat{c}^\dagger \hat{c} + g (\hat{a}^\dagger \hat{c} + \hat{c}^\dagger \hat{a}) +
G_x \cos (\omega_d t) (\hat{a} + \hat{a}^\dagger) (\hat{b} + \hat{b}^\dagger),
\end{equation}
where $\hat{b}$, $\hat{a}$, $\hat{c}$ are respectively the bosonic operators for the center-of-mass motion along the $x-$axis, the magnon mode and the acoustic mode with frequency close to the magnon mode. The coupling strength $G_x$ is proportional to the amplitude of the oscillating field $b_g$. To study how the hybrid magnon-phonon mode couples with the center-of-mass motion, this Hamiltonian is partially diagonalized in the subspace of magnon and phonon through a Bogoliubov transformation and the result reads
\begin{equation}
\hat{\mathcal{H}}= \omega_x \hat{b}^\dagger \hat{b} + \sum_{\alpha=1}^2 \Delta_\alpha \hat{d}_\alpha^\dagger \hat{d}_\alpha + (\hat{b}^\dagger +\hat{b})\sum_{\alpha=1}^2 (G_{x\alpha} \hat{d}_\alpha^\dagger + h.c.),
\end{equation}
where $\hat{d}_\alpha$ ($\alpha=1,2$) is a superposition of the magnon ($\hat{a}$) and acoustic ($\hat{c}$) mode, $\Delta_\alpha$ is the detuning between $i-$th eigenmode and the driving frequency, and the reduced couplings are $G_{x1}=G_x/(2\lambda)$ and $G_{x2}=-\chi G_{x1}$, with $\chi = -2g/(\Delta - \sqrt{\Delta^2 + 4g^2})$ and $\lambda=\sqrt{1+\chi^2}$. To maximize the interaction between the acoustic mode and the center-of-mass motion, the detuning of the magnon frequency with respect to the driving frequency is adjusted ($\Delta_2=\omega_x$) to guarantee that the $\hat{d}_2$ mode is $99.99\%$ acoustic. In this way, an effective acoustomechanical system is realized. The authors further showed that this strong and tunable acoustomechanical coupling helps to probe the acoustic magnon excitation by measuring the mechanical displacement of the magnet. As shown in Fig. \ref{magnon_mechanics2}(b), without acoustic magnon coupling ($b_g=0$), the power spectrum of the magnet has a single peak at the intrinsic frequency $\omega_x$. A strong enough coupling (larger $b_g$ in the figure) splits the peaks into two subpeaks. Such a tunable quantum platform can be realized using micromagnets levitated or deposited on a clamped nanomechanical oscillator, and it may open new possibilities for using acoustic phonons as a quantum memory and to study out-of-equilibrium quantum mesoscopic physics in a levitated particle.

Later, Colombano et al. \cite{ColomPRL2020} experimentally realized a hybrid magnet-mechanical system, by depositing a glass microsphere optical cavity on top of a YIG thin film, as shown in Fig. \ref{magnon_mechanics2}(c).
When an oscillating field acting on the magnet excites the magnon mode in YIG, it will also excite a deformation of the YIG sphere through magnetostrictive interactions. Such a deformation will immediately couple to the mechanical breathing mode of the microsphere. By detecting the mechanical motion of the sphere, the oscillating magnetic field can be indirectly recovered. Figure \ref{magnon_mechanics2} (d) shows the typical response of the system under an oscillating field of $2.3\ \mu$T at $206$ MHz. Clearly, a sharp peak response (red line) is observed, which disappears once the microsphere is levitated from the YIG film (black line), which demonstrates the mechanical origin of the signal. The authors estimated the field sensitivity of this technique at a few $\mathrm{nT\ Hz}^{-1/2}$ at 206 MHz, which is close to the highest sensitivity of known magnetometers. The working frequency of this proposal can be extended to the GHz regime, by tuning the magnon frequency with an external field. This finding provides exciting opportunities to design a high-performance magnetometer.

Furthermore, Li et al. \cite{LiPRXQ2021} considered a hybrid cavity-magnet-mechanical oscillator system and showed that both classical and quantum states of magnons can be transferred to and stored in a distant long-lived mechanical resonator, through the assistance of optical pulses. The nonlocal entangled state of magnons and mechanical modes can also be prepared in this setup. Tan et al. \cite{TanPRR2021} considered such a hybrid system and proposed a scheme to establish EPR steering channel between the magnon and phonon modes. Proskurin et al. \cite{ProskPRB2018} studied the coupling of magnetic domain walls pinned by a potential well with the cavity photons through magneto-optical interactions and found that the oscillating motion of the domain wall can be effectively coupled with the cavity photons in a nonlinear way as $\hat{\mathcal{H}}_{\mathrm{int}}= ig(\hat{b}-\hat{b}^\dagger)\hat{a}^\dagger \hat{a} $, where $\hat{b}$ ($\hat{b}^\dagger$) is the quantization of the oscillating motion of the domain wall.
This type of nonlinear interaction resembles the magnon-phonon coupling in Eq. (\ref{mphonon_ZhangSA}) and thus most of the physics presented there can be translated to such a system. This extension may generate nonclassical states of domain walls, which will significantly enrich current research efforts concerning the classical dynamics of domain walls.

\subsection{Hybrid magnon-electron systems}\label{sec_magnon_electron}
Finally, we discuss the coupling of magnons with electrons. We restrict ourselves to magnetic insulators because the magnetic quality of these materials, in particular, that of YIG, is much larger than that of metallic magnets. Magnons in magnetic insulators interact with electrons in adjacent normal metals via an interfacial exchange coupling $\propto \hat {\bf S} \cdot \hat {\bf s}$, where $\hat {\bf S}$ corresponds to the spin density in the insulator and $\hat {\bf s}$ to that of the metal. After rewriting the latter in terms of electronic creation and annihilation operators $\hat \psi_\sigma$ and $\hat \psi_
\sigma^\dagger$, with $\sigma \in \{\uparrow, \downarrow \}$ labelling the electronic spin state, and performing a HP transformation on  $\hat {\bf S}$, we find that the interface coupling becomes $\propto \hat a \psi^\dagger_\uparrow \psi_\downarrow + {\rm h.c.}$. Here, we restrict ourselves to the linearized version of the HP transformation and focus on terms that exchange spin between metal and insulators. The exchange of spin is clear from the fact the interface coupling describes that a spin flip in the normal metal is accompanied by the creation or annihilation of a magnon in the magnetic insulator, thereby leading to the exchange of one quantum of spin angular momentum between metal and insulator.

The physics that results from this coupling has been extensively studied in heterostructures with YIG as the magnetic insulator and Pt as the normal metal. Here, Pt is used because of its ability to convert spin current into charge current (and vice versa) via the (inverse) spin Hall effect that results from its large spin-orbit coupling. In setups consisting of a single Pt electrode on top of YIG, the so-called spin Seebeck effect, a magnon spin current that is driven by a thermal gradient and that is injected from YIG into Pt, was discovered \cite{uchida2010}. In nonlocal setups with two Pt leads, electric actuation and detection of magnon spin current were achieved \cite{ab9817b8a4964a7ebb40e9945e0961de}, and it was shown that the spin diffusion length of magnons is on the order of 10 $\mu$m in YIG at room temperature. As discussed in Section~\ref{sec:quantummanybodystatesofmagnons}, such nonlocal setups were used to probe signatures of spin superfluidity.

Both the spin Seebeck and nonlocal transport experiments probe thermal magnons for which quantum properties --- other then quantum statistics --- are not relevant. Kamra and Belzig \cite{KamraPRL2016} and Bender et al. \cite{BenderPRL2019} proposed to probe quantum properties, squeezing and temporal correlations, respectively, of magnons via spin-current noise in setups of normal metals on top of magnetic insulators. These theoretical proposals are waiting to be followed up experimentally. In a parallel development, spin current injection from insulating materials into normal metals has been used to probe various properties --- such as the magnetic phase diagram --- of quantum materials, in addition to the spin current itself. For a recent review, see Ref.~\cite{Han2020}.

To conclude, we have reviewed recent developments concerning the coupling between magnons with other quantum platforms. These hybrid platforms provide an avenue to study non-Hermitian physics, and suggest applications in quantum memories, high precision measurements, and quantum logic gates, as we shall introduce in the following two sections.

\section{Non-Hermitian physics and $\mathcal{PT}$ symmetry}\label{sec_PT}

In recent years, non-Hermitian physics and $\mathcal{PT}$ symmetry have attracted considerable attention. It was first discovered by Bender and Boettcher in 1998 \cite{BenderPRL1998} that non-Hermitian Hamiltonians allow for an entirely real spectrum as long as the combined parity ($\mathcal{P}$) and time ($\mathcal{T}$)-reversal symmetries are respected. $\mathcal{PT}$ symmetry has been studied in optics \cite{FengNM2013, PengNP2014, WenJPB2018}, microwaves \cite{BittnerPRL2012}, acoustics \cite{ZhuPRX2014}, mechanics \cite{BenderAJP2013}, electric circuits \cite{SchindlerPRA2011} and, recently, in magnonics \cite{YangPRL2018, YuPRB2020} and magnon-polaritonics \cite{ZhangNC2015,WangOE2018}. Physical systems respecting $\mathcal{PT}$ symmetry demonstrate exotic features. For example, a $\mathcal{PT}$ symmetric ferromagnetic bilayer structure displays either a real spectrum or complex conjugate eigenfrequency pairs \cite{YangPRL2018}. In the real-spectrum regime (termed the $\mathcal{PT}$ exact regime), the eigenstates respect $\mathcal{PT}$ symmetry and are equally distributed over both loss and gain ferromagnetic layers, and the whole system displays ferromagnetic behaviour. In the latter case (termed the $\mathcal{PT}$ broken regime), the eigenstates deviate from equilibrium and the intensity is more concentrated either in the loss or gain layer associated with a phase transition from ferromagnet to antiferromagnet. The phase transition point has degenerate eigenvalues and eigenstates and is called an exceptional point (EP). Operating systems in different $\mathcal{PT}$ phases or around EPs leads to new dynamics and unusual phenomena, with most achievements being reported in optics, such as unidirectional invisibility \cite{FengNM2013}, loss-induced laser suppression \cite{PengScience2014}, laser mode selection \cite{FengScience2014}, and EP enhanced sensing \cite{WiersigPRL2014, ChenNature2017, HodaeiNature2017}. Notably, the property that non-Hermitian Hamiltonian has real eigenvalues is not limited to $\mathcal{PT}$ symmetry, but belongs to a larger class of ``pseudo-Hermitian'' systems. EPs generally stand for degeneracy points of any non-Hermitian system, like anti-$\mathcal{PT}$ symmetric systems \cite{YangPRL2020,ZhaoPRAp2020,NairPRL2021}. New physics and mechanisms in this emerging field are expected to be explored, for both fundamental interest and practical innovations.

\subsection{Preliminary knowledge of $\mathcal{PT}$ symmetry}
%
$\mathcal{PT}$ symmetry in quantum mechanics requires that the potential satisfies $V(\hat{r})=V^*(-\hat{r})$ \cite{BenderPRL1998}, where $\hat{r}$ is the position operator. The parity operator $\hat{\mathcal{P}}$ reverses the sign of $\hat{r}\to -\hat{r}$ and the time reversal operator $\hat{\mathcal{T}}$ changes  $i\to-i$. The operator $\hat{\mathcal{P}}$ is unitary and $\hat{\mathcal{T}}$ is anti-unitary, and they commute with each other. As an example, we consider a two-level system describing two coupled sites with gain and loss
\begin{equation}
\hat{\mathcal{H}}=  \left(
    \begin{array}{cc}
      \omega_1-i\kappa_1 & g \\
      g & \omega_2+i\kappa_2 \\
    \end{array}
  \right),~~~~~~~
  \hat{\mathcal{H}}|\phi_{1,2}\rangle=E_{1,2}|\phi_{1,2}\rangle,
  \label{PT_Ham}
\end{equation}
where $\omega_{1,2}$ is the single site energy, $\kappa_{1,2}$ is the loss/gain parameter of the two sites, and $g$ is the coupling strength. When $\kappa_{1,2}=0$, the Hamiltonian is Hermitian and the eigenvalues are real. For general $\kappa_{1,2}$, the Hamiltonian becomes non-Hermitian, and the eigenvalues are complex. To satisfy the $\mathcal{PT}$ symmetry, we first define the parity operator as
\begin{equation}
\hat{\mathcal{P}}=  \left(
    \begin{array}{cc}
      0 & 1 \\
      1 & 0 \\
    \end{array}
  \right)=\hat{\mathcal{P}}^\dagger,
\end{equation}
and the time-reversal operator corresponds to complex conjugation. Then, the Hamiltonian respects $\mathcal{PT}$ symmetry,
i.e., satisfies $[\hat{\mathcal{P}}\hat{\mathcal{T}},\hat{\mathcal{H}}]$, if $\omega_1=\omega_2=\omega_0$ and $\kappa_1=\kappa_2=\kappa$. We therefore obtain the eigenvalues and eigenstates
\begin{equation}
  E_{1,2}=\omega_0\pm\sqrt{g^2-\kappa^2},~~~~~~~
  \phi_{1,2}=\left(\frac{-i\kappa\pm\sqrt{g^2-\kappa^2}}{g}, 1
             \right).
\end{equation}

In the $\mathcal{PT}$ exact regime, i.e., when the non-Hermiticity parameter $\kappa<g$, the eigenvalues are completely real. Symmetry-breaking happens at $\kappa=g$, beyond which the eigenvalues become complex ($\mathcal{PT}$ broken). The phase transition point occurs, as mentioned before, at an EP.
The eigenstates are no longer orthogonal, $\langle \phi_1|\phi_2\rangle \neq 0$, but follow a biorthogonal relation $\langle \phi_1^*|\phi_2\rangle =0$ \cite{BrodyJPA2013}. In the $\mathcal{PT}$ exact regime, the probability density of the eigenstates is equally distributed on the two sites because $|(-i\kappa\pm\sqrt{g^2-\kappa^2}/g)|^2=1$. However, at the EP, $\phi_1=\phi_2=[-i,1]^T$. In the $\mathcal{PT}$ broken regime, two probability density of the eigenstates is unevenly distributed on site 1 and site 2.

{\it Exceptional point.} The non-Hermitian degeneracy point or the EP is not limited to $\mathcal{PT}$ symmetry. The order $\mathcal{N}$  of the EP is determined by the number of degenerate eigenstates. The lifting of the non-Hermitian degeneracy is a sub-linear effect: the eigenvalue splitting for an external perturbation $\epsilon$ at the EP is proportional to $|\epsilon|^{1/\mathcal{N}}$, which thus provides a way to  significantly enhance the sensitivity. This feature was recently confirmed in $\mathcal{PT}$ symmetric optical laser system \cite{WiersigPRL2014, ChenNature2017, HodaeiNature2017}. The very presence of the EP would strongly affect the wave dynamics, by causing non-reciprocal energy transfer \cite{XuNature2016}, eigenmode switching when encircling the EP \cite{DopplerNature2016, DeyOC2021}.

{\it Anyonic-parity-time symmetry.} Another type of interesting symmetry in non-Hermitian systems is the counterpart of $\mathcal{PT}$ symmetry, called anti-parity-time ($\mathcal{APT}$) symmetry. A Hamiltonian respecting $\mathcal{APT}$ symmetry anti-commutes with the combined parity and time-reversal operators, i.e., $\{\hat{\mathcal{P}}\hat{\mathcal{T}},\mathcal{H}\}=0$. For simplicity, we consider the following $\mathcal{APT}$ Hamiltonian
\begin{equation}
\hat{\mathcal{H}}_\mathrm{APT}=  \left(
    \begin{array}{cc}
      \omega_0-i\kappa & ig \\
      ig & -\omega_0-i\kappa \\
    \end{array}
  \right).
\end{equation}
An $\mathcal{APT}$-symmetric system is lossy in all sites with imaginary coupling.
The eigenvalues of the $\mathcal{APT}$-symmetric Hamiltonian are $E_{1,2}=-i\kappa\pm\sqrt{\omega_0^2-g^2}$.
In contrast to the $\mathcal{PT}$ case, when $\omega_0>g$, the system is in the regime of $\mathcal{APT}$ symmetry, two eigenvalues are complex with the same imaginary parts. When $\omega_0<g$ and the system is turned into the $\mathcal{APT}$ broken regime, the eigenvalues become purely imaginary.

In the general case, the complex coupling between the two sites can be tuned in phase, and the system respects an anyonic-$\mathcal{PT}$ symmetry \cite{Arwasarxiv2021}. The $\mathcal{PT}$ symmetry and $\mathcal{APT}$-symmetry are special cases when the coupling strengths are purely real and purely imaginary, respectively. The anyonic-$\mathcal{PT}$ Hamiltonian satisfies
\begin{equation}\label{anionic}
  \hat{\mathcal{P}}\hat{\mathcal{T}}\hat{\mathcal{H}}=e^{-2i\phi}\hat{\mathcal{H}}\hat{\mathcal{P}}\hat{\mathcal{T}}.
\end{equation}
When $\phi=0 ~\mathrm{or}~ \pi$, the Hamiltonian is $\mathcal{PT}$ symmetric and satisfies bosonic commutation relations, while $\phi=\pm \pi/2$ corresponds to the $\mathcal{APT}$ symmetry and the Hamiltonian fulfills the fermionic commutation relation. An anyonic-$\mathcal{PT}$ symmetric (laser) system is realized by manipulating the complex coupling from purely dispersive to purely dissipative \cite{Arwasarxiv2021}. The anyonic-$\mathcal{PT}$ Hamiltonian is written as
\begin{equation}
\hat{\mathcal{H}}_\mathrm{Anyonic-\mathcal{PT}}=  \left(
    \begin{array}{cc}
      \delta & ge^{i\phi} \\
      ge^{i\phi} & -\delta\\
    \end{array}
  \right),
\end{equation}
where $\delta=\Delta\omega-i\Delta\kappa$, with the frequency detuning $\Delta\omega$ and the loss difference $\Delta\kappa$ of each laser (relative to their average values).  One can easily verify that the Hamiltonian satisfies the commutation relation (\ref{anionic}) if $\tan\phi=\Delta\omega/\Delta\kappa$. For the $\mathcal{PT}$ symmetric case $\phi=0 ~\mathrm{or}~ \pi$, the EP is located at $\Delta\kappa=\pm g, \Delta\omega=0$, while for $\mathcal{APT}$ symmetric one $\phi=\pm \pi/2$, the EP rotates $\pi$ in $\Delta\omega-\Delta\kappa$ plane, at $\Delta\omega=\pm g, \Delta\kappa=0$. By dynamically changing the coupling phase, the EP itself can move along a circle in parameter space, which corresponds to the relative dynamics of encircling the EP \cite{XuNature2016, DopplerNature2016, DeyOC2021}.

\subsection{$\mathcal{PT}$ symmetry in spintronics}
Despite that $\mathcal{PT}$ symmetry was originally developed in quantum mechanics, this concept has been employed in classical systems, such as photonic crystals, then extended to mechanical membranes and electrical circuits. As we have mentioned in this review, SWs have many properties analogous to electromagnetic waves, such as BEC \cite{NikuniPRL2000, BenderPRL2012}, squeezed states \cite{KamraPRB2019, ZhaoPRL2004}, frequency combs \cite{WangPRL2021}, and quantum entanglement \cite{Yuanantibunch2020}. The success of exploring $\mathcal{PT}$ symmetry and non-Hermiticity in optics stimulates these research activities in magnonic systems.
We shall first review the studies of $\mathcal{PT}$ symmetry in coupled magnets in Section \ref{sec_coupled_magnets_PT}. The physics here is mostly classical and it is included as a relevant and additional feature of the coupled magnets system discussed in Section \ref{sec_magnonmagnon}. Then we shall proceed to discuss the $\mathcal{PT}$ symmetry in hybrid magnon-photon system in Section \ref{sec_cavity_magnonics_PT}. We expect purely quantum effects to manifest themselves as one goes to the low temperature regime or reduces the number of spins in the system, as discussed below.

\subsubsection{$\mathcal{PT}$ symmetry in coupled magnets} \label{sec_coupled_magnets_PT}

\begin{figure} %
  \centering
  \includegraphics[width=0.9\textwidth]{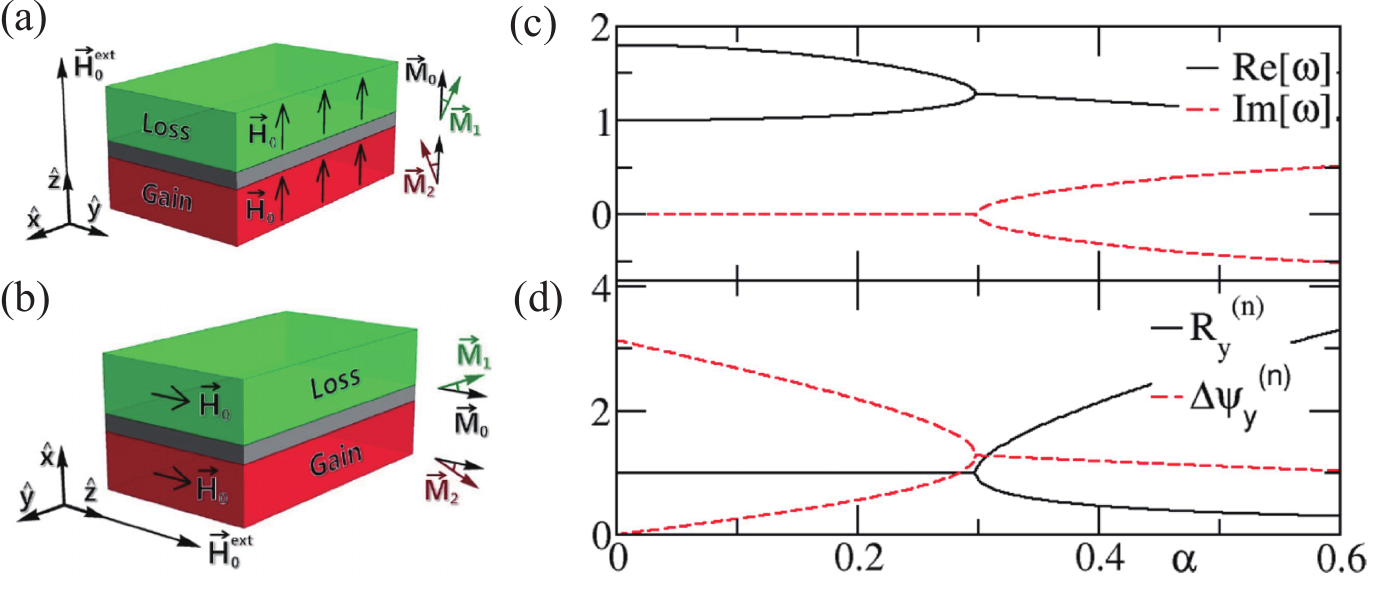}
  \caption{Two exchange-coupled ferromagnetic thin films in an external magnetic field with two geometries: (a) out-of-plane and (b) in-plane. (c) Real and imaginary parts of eigenvalues of the hybrid system as a function of the Gilbert constant. (d) The ratio of $y$-components of the magnetization in the two layers $R_y^{n}$ and their relative phase difference $\Delta\psi_y^{n}$ versus the Gilbert constant. Source: The figures are adapted from  Ref. \cite{LeePRB2015}.}\label{Lee2015}
\end{figure}

{\it $\mathcal{PT}$ symmetric magnetic bilayers.} Two coupled ferromagnetic films, one with loss and the other with a balanced gain, satisfy $\mathcal{PT}$ symmetry. Interesting dynamics for the magnetization emerges in both $\mathcal{PT}$ symmetry exact and broken regimes. The coupled dynamics are described by the LLG equations, as introduced in Section \ref{sec_llg},
\begin{eqnarray}
  \frac{\partial \mathbf{m}_1}{\partial t}
  &=&-\gamma \mathbf{m}_1\times\mathbf{H}_1-\gamma J \mathbf{m}_1\times\mathbf{m}_2
  +\alpha\mathbf{m}_1\times\frac{\partial \mathbf{m}_1}{\partial t}, \\
  \frac{\partial \mathbf{m}_2}{\partial t}
  &=&-\gamma \mathbf{m}_2\times\mathbf{H}_2-\gamma J \mathbf{m}_2\times\mathbf{m}_1-\alpha\mathbf{m}_2\times\frac{\partial \mathbf{m}_2}{\partial t},
\end{eqnarray}
where the second layer has magnetic gain by via the reverse sign of the Gilbert damping constant $\alpha$. Here, $\mathbf{m}_{1,2}$ are the normalized magnetizations, $\mathbf{H}_{1,2}$ are the local effective magnetic fields (considering only the external field and the demagnetizing field $\mathbf{H}_{1,2}=\mathbf{H}_\mathrm{ext}-4\pi \hat{N} M_s \mathbf{m}_{1,2}$ with $\hat{N}$ the demagnetizing tensor), $M_s$ is the saturation magnetization, and $J$ is the exchange coupling constant. The two equations are invariant under  combined parity $\hat{\mathcal{P}}$ and time reversal $\hat{\mathcal{T}}$  operations, which means they are unchanged by interchanging $1\leftrightarrow2$ together with $t\leftrightarrow-t$. In the linear approximation, taking $\mathbf{m}_{1,2}=\mathbf{e}_z+\delta \mathbf{m}_{1,2}$, and considering identical effective magnetic fields $\mathbf{H}_1=\mathbf{H}_2=H\mathbf{e}_z$, as shown in Fig. \ref{Lee2015}(a), one obtains the eigenvalues in the out-of-plane geometry as
\begin{equation}
    \omega_{1,2}=\frac{\omega_H+\omega_J\pm\sqrt{\omega_J^2-\alpha^2\omega_H(\omega_H+2\omega_J)}}{1+\alpha^2},
\end{equation}
where $\omega_H=\gamma H$ and $\omega_J=\gamma J M_s$. The real and imaginary parts of the eigenvalues as a function of the Gilbert constant $\alpha$ are plotted in Fig. \ref{Lee2015}(c). The EP emerges at the critical Gilbert constant and frequency
\begin{equation}
  \alpha_\mathrm{cr}=\frac{\omega_J}{\sqrt{\omega_H(\omega_H+2\omega_J)}},~~~~~
  \omega_\mathrm{cr}=\frac{\omega_H(\omega_H+2\omega_J)}{\omega_H+\omega_J}.
\end{equation}

The ratio of the $y$-components  $R_y^{(n)}=|\delta m^{(n)}_{1y}/\delta m^{(n)}_{2y}|$ ($n=1,2$ denote the two eigenmodes) at $\omega_J=0.4\omega_H$ as a function of the Gilbert constant is plotted in Fig. \ref{Lee2015}(d), and so is the phase difference $\Delta\psi_y^{(n)}$. For $\alpha=0$, the phase differences are $0$ and $\pi$, indicating a symmetric mode $\delta \mathbf{m}_1=\delta \mathbf{m}_2$ and an antisymmetric one $\delta \mathbf{m}_1=-\delta \mathbf{m}_2$, respectively. The two modes coalesce at the EP. A new type of steady-state magnetization dynamics is thus found: the precessional polar angle is bounded and neither attenuates (as in the case of loss) nor amplifies (as in the case of gain).

The notion of $\mathcal{PT}$ symmetry in magnetic structures has been introduced by assuming a negative Gilbert constant. Pursuing a smaller Gilbert damping corresponds to a higher quality of the magnetic solid. How to realize a Gilbert damping arbitrarily close to zero or even across the border (negative damping) is an interesting and challenging problem \cite{Caoarxiv2020}. It could effectively be implemented by injection of spin current via a spin-transfer torque. This latter scenario was used by Flebus et al. \cite{FlebusPRB2020} to propose an array of coupled spin-torque oscillators that displays non-Hermitian topological phases and edge-state lasing.

\begin{figure} %
  \centering
  \includegraphics[width=1.0\textwidth]{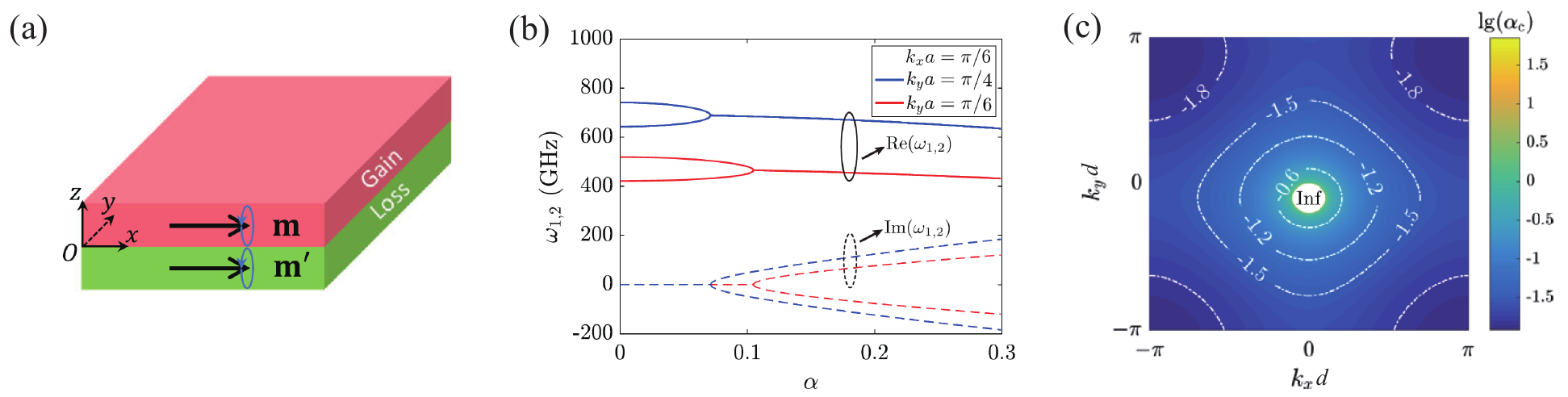}
  \caption{(a) Schematic plot of a bilayer ferromagnet with balanced
gain and loss. The magnetization of all spins is initially along the $x$-direction. (b) Evolution of $\omega_{1,2}$ as a function of $\alpha$ for two representative spin-wave
modes $\mathbf{k}=[\pi/6d,\pi/4d]$ (blue curves) and $[\pi/6d,\pi/6d]$ (red curves).
(c) Contour plot of the mode dependence on $\alpha_c$. $\mathcal{PT}$ symmetry is
never broken in the white region. Source: The figures are adapted from Ref. \cite{YangPRL2018}.}\label{Yang2018}
\end{figure}

{\it Magnonic EPs.} Magnonic EPs are non-Hermitian degeneracies of SWs, which were first considered in ferromagnetic bilayers respecting $\mathcal{PT}$ symmetry \cite{YangPRL2018}, as sketched in Fig. \ref{Yang2018}(a). Different from the macrospin case, the intralayer exchange coupling energy $-\sum_{\langle ij\rangle}J\mathbf{m}_i\cdot\mathbf{m}_j$ is considered with lattice constant $d$, on top of interlayer exchange $-\sum_{i}\lambda J\mathbf{m}_i\cdot\mathbf{m}'_i$, and the interfacial DM interaction
$-\sum_{\langle ij\rangle}\mathbf{D}_{ij}\cdot(\mathbf{m}_i\times\mathbf{m}_j)$. One then derives the dispersion relation of SWs
\begin{equation}
  \omega_{1,2}(\mathbf{k})=\lambda+2\bar\zeta(\mathbf{k})
  \pm\sqrt{\lambda^2-4\alpha^2\bar\zeta(\mathbf{k})\left[\lambda+\bar\zeta(\mathbf{k})\right]},
\end{equation}
where $\bar\zeta(\mathbf{k})=2-\cos k_x d-\cos k_y d+(D/J)\sin k_y d$. For any $\mathbf{k}$, the two eigenfrequencies approach each other as the gain-loss parameter $\alpha$ increases, and coalesce at a critical value $\alpha_\mathrm{c}$, i.e., at the second-order magnonic EP,
\begin{equation}
  \alpha_c(\mathbf{k})=\frac{\lambda}{2\sqrt{\bar\zeta(\mathbf{k})\left[\lambda+\bar\zeta(\mathbf{k})\right]}},~~~~~
  \omega_c(\mathbf{k})=\frac{\gamma J}{\mu_0 M_s a^3}\frac{\lambda+2\bar\zeta(\mathbf{k})}{1+\alpha_c^2(\mathbf{k})},
\end{equation}
and then bifurcate into the complex plane [Fig. \ref{Yang2018}(b)]. In the $\mathcal{PT}$ broken regime, an antiferromagnetic state appears in the gain layer. We point out that for $\mathcal{PT}$ symmetric SW modes, a parameter region $-\lambda\leq \zeta(\mathbf{k})\leq 0$ exists in which the $\mathcal{PT}$ symmetry is never broken in the framework of linear SW theory. This is in contrast to conventional $\mathcal{PT}$ symmetric systems that always suffer from a symmetry breaking at the EP.

\begin{figure} %
  \centering
  \includegraphics[width=1.0\textwidth]{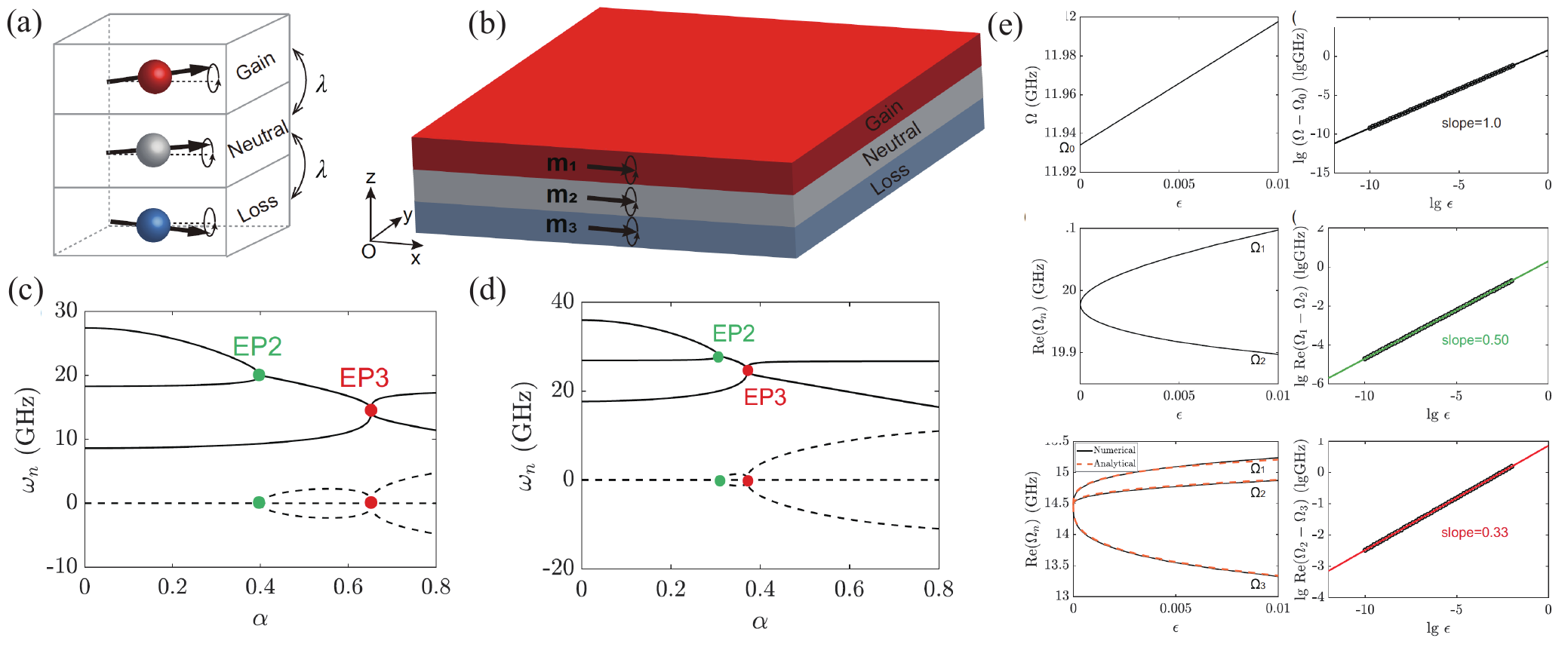}
  \caption{(a) Three exchange-coupled macrospins
consisting of a gain (red), neutral (gray), and (balanced-)loss (blue)
spin. (b) Schematic plot of a ferromagnetic heterostructure with a
gain, neutral, and loss layer, denoted by red, gray, and blue colors,
respectively. The magnetizations of all layers are initially along the $x$
direction.  Eigenvalues of the hybrid system versus Gilbert constant for (c) macrospin version and (d) spin-wave version with respect to wave vector $(k_x,k_y)=(\pi/30d,0)$. (e) The splitting of eigenfrequencies vs the perturbation follows linear, $\epsilon^{1/2}$ and $\epsilon^{1/3}$ responses, respectively, at FMR, EP2, and EP3. Source: The figures are adapted from Ref. \cite{YuPRB2020}.}\label{Yu2020}
\end{figure}

Higher-order magnonic EPs are found as well in the $\mathcal{PT}$ symmetric multilayer ferromagnetic systems. To illustrate the principle, Yu et al. consider a ferromagnetic trilayer structure \cite{YuPRB2020}, consisting of a gain, neutral, and loss spin (ferromagnetic layer), as shown in Figs. \ref{Yu2020}(a) and (b). The Hamiltonian contains Zeeman energy, magnetic anisotropy, and interlayer exchange coupling, and is given by
\begin{equation}
  \hat{\mathcal{H}}=-\mu_0\sum_{n=1}^3 \mathbf{H}\cdot \mathbf{M}_n-\frac{1}{2}\sum_{n=1}^3 K_n(m_n^x)^2 - \lambda\mu_0\mathbf{M}_2\cdot(\mathbf{M}_1+\mathbf{M}_3),
\end{equation}
where $\mathbf{M}_n (\mathbf{m}_n=\mathbf{M}_n/M_n)$ is the normalized spin vector, with $n=1,2,3$ labeling the layer. The external field is along the $x$ axis $\mathbf{H}=H{\bf e}_x$, $K_n>0$ is the uniaxial anisotropy, and $\lambda>0$ is the ferromagnetic exchange coupling strength. The coupled magnetization dynamics is described by the LLG equations
\begin{eqnarray}
  \frac{\partial \mathbf{m}_1}{\partial t} &=& -\gamma\mathbf{m}_1\times \mathbf{H}_\text{eff,1}
  -\alpha \mathbf{m}_1\times  \frac{\partial \mathbf{m}_1}{\partial t},\\
  \frac{\partial \mathbf{m}_2}{\partial t} &=& -\gamma\mathbf{m}_2\times \mathbf{H}_\text{eff,2}, \\
  \frac{\partial \mathbf{m}_3}{\partial t} &=& -\gamma\mathbf{m}_3\times \mathbf{H}_\text{eff,3}
  +\alpha \mathbf{m}_3\times  \frac{\partial \mathbf{m}_3}{\partial t},
\end{eqnarray}
with the local effective magnetic fields $\mathbf{H}_\text{eff,n}=-\mu_0\partial\hat{\mathcal{H}}/\partial \mathbf{M}_n$. In the linear approximation, the above equations yield eigenmodes and frequencies. The evolution of eigenfrequencies as a function of the gain-loss parameter $\alpha$ is plotted in Fig. \ref{Yu2020}(c), where the solid and dashed lines represent the real and imaginary parts of the eigenfrequencies respectively. It shows that EP2 and EP3 separately emerge at different critical $\alpha$. In the vicinity of the non-Hermitian degeneracy at the EPs, a sub-linear response to external perturbations is displayed as $\propto \epsilon^{1/\mathcal{N}}$, with $\mathcal{N}$ the order of the EPs. We consider a perturbation $\epsilon$ being added on the top macrospin by a small external magnetic field. The separation between the eigenfrequencies as a function of the strength of the perturbation is shown in Fig. \ref{Yu2020}(e). We observe a linear, $\epsilon^{1/2}$ and $\epsilon^{1/3}$ responses, respectively, for FMR, EP2 and EP3. To investigate the effect of fluctuations on the sensing performance \cite{LangbeinPRA2018, MortensenOptica2018}, we consider a Gaussian distribution of the perturbation signal with a noise level $\sigma$, i.e.,
\begin{equation}
    P(\epsilon-\epsilon_0)=\frac{1}{\sqrt{2\pi}\sigma}\exp\left[-\frac{1}{2}
    \left(\frac{\epsilon-\epsilon_0}{\sigma}\right)^2\right],
\end{equation}
where the signal $\epsilon_0$ is to be detected. We calculate the ensemble-averaged sensitivity and find the sensor performs well when $\epsilon_0/\sigma>1$. The estimated magnetic sensitivity at EP3 is about $3\times 10^{-14}~\mathrm{T/Hz}^{1/2}$ for a typical FMR frequency $5$ GHz and the FMR linewidth $5$ MHz. This value is three orders of magnitude higher than for a conventional magnetic sensor based on a magnetic tunnel junction \cite{CardosoMT2014}.

The SW solution can be obtained by considering intralayer exchange coupling, as shown in Fig. \ref{Yu2020}(b). The evolution of the eigenfrequencies for wavevector $(k_x,k_y)=(\pi/30d,0)$ with respect to the gain-loss parameter $\alpha$ is shown in Fig. \ref{Yu2020}(d). Magnonic EP2 and EP3 emerge at different critical Gilbert constant. The sublinear response at the higher-order EP leads to an outstanding magnetic sensitivity. Hence, high-order magnonic EPs in all-magnetic structures are promising for designing ultra-high-sensitive magnetometers.

\subsubsection{$\mathcal{PT}$ symmetry in cavity magnonics}\label{sec_cavity_magnonics_PT}

\begin{figure} %
  \centering
  \includegraphics[width=0.9\textwidth]{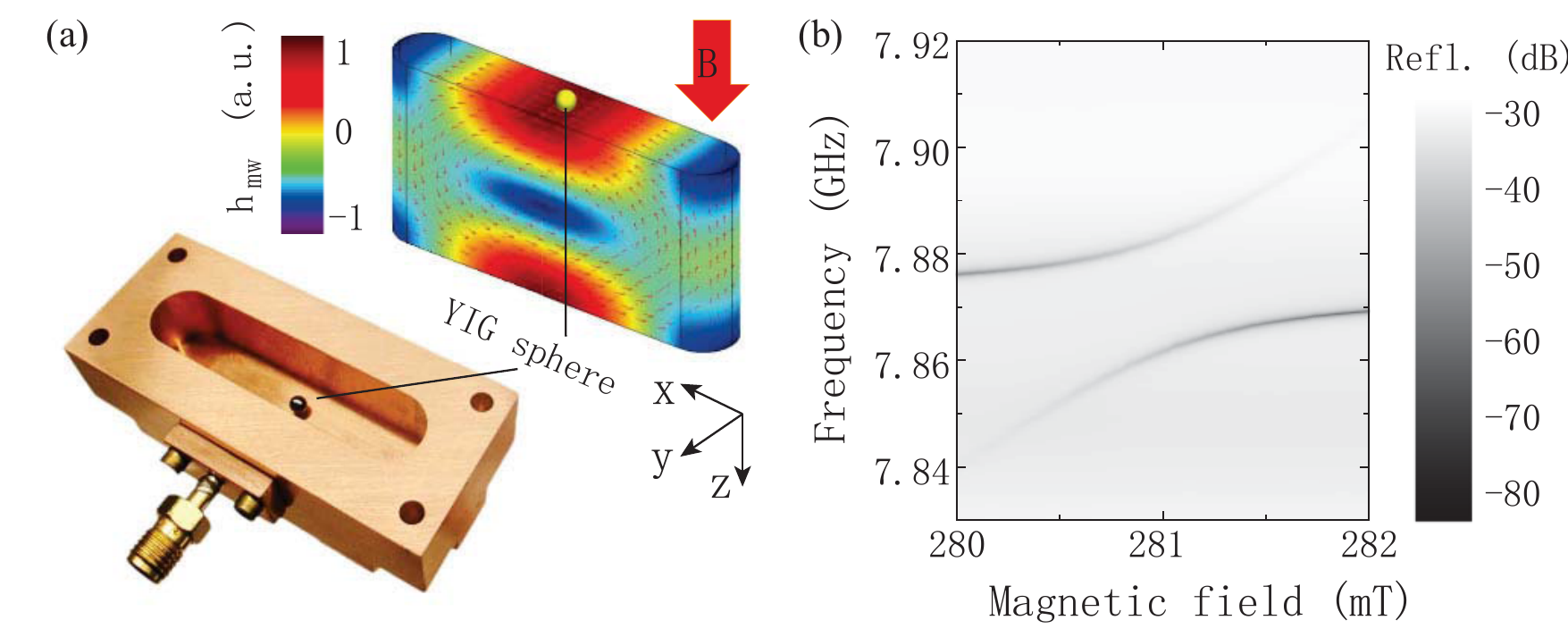}
  \caption{(a) A typical setup for detecting CMPs consists of a YIG sphere placed inside a 3D-microwave cavity. (b) Observation of CMPs in the anti-crossing output spectrum. Source: The figures are adapted from Ref. \cite{ZhangPRL2014}.}\label{Zhang2014}
\end{figure}

Cavity magnonics---the emerging interdiscipline of cavity quantum electrodynamics and magnonics---has been rapidly developing in recent years \cite{HueblPRL2013,ZhangPRL2014,GorPRAp2014,TabuchiPRL2014,BaiPRL2015,BhoiJAP2014,ZhangNPJ2015,FlaigPRB2016}.
An archetypal setup of the field consists of a YIG sphere placed in a microwave cavity, see Fig. \ref{Zhang2014}(a). As discussed in Section \ref{sec_magnon_photon}, a new type of bosonic quasi-particles, called cavity magnon-polaritons (CMPs), have been detected and characterized via the anti-crossing transmission spectrum, as shown in Fig. \ref{Zhang2014}(b), which results from the hybridization of magnons and cavity photons. The polaritonic eigenmodes are usually produced due to the strong light-matter interaction, that lies at the heart of cavity quantum electrodynamics and quantum information science. The entangled spin orientation and photon number state in CMPs enables an efficient quantum information transfer between photon and magnon via Rabi oscillation, which is promising for quantum computing. Recently, non-Hermitian singularities \cite{HarderPRL2018,ZhangNC2017,ZhangPRXQ2021} have been experimentally observed in cavity-magnonic systems. $\mathcal{PT}$ symmetric CMPs and, generally, non-Hermitian CMPs, offer exciting opportunities from the point of view of both scientific curiosity and new technological innovations.

\begin{figure} %
  \centering
  \includegraphics[width=1.0\textwidth]{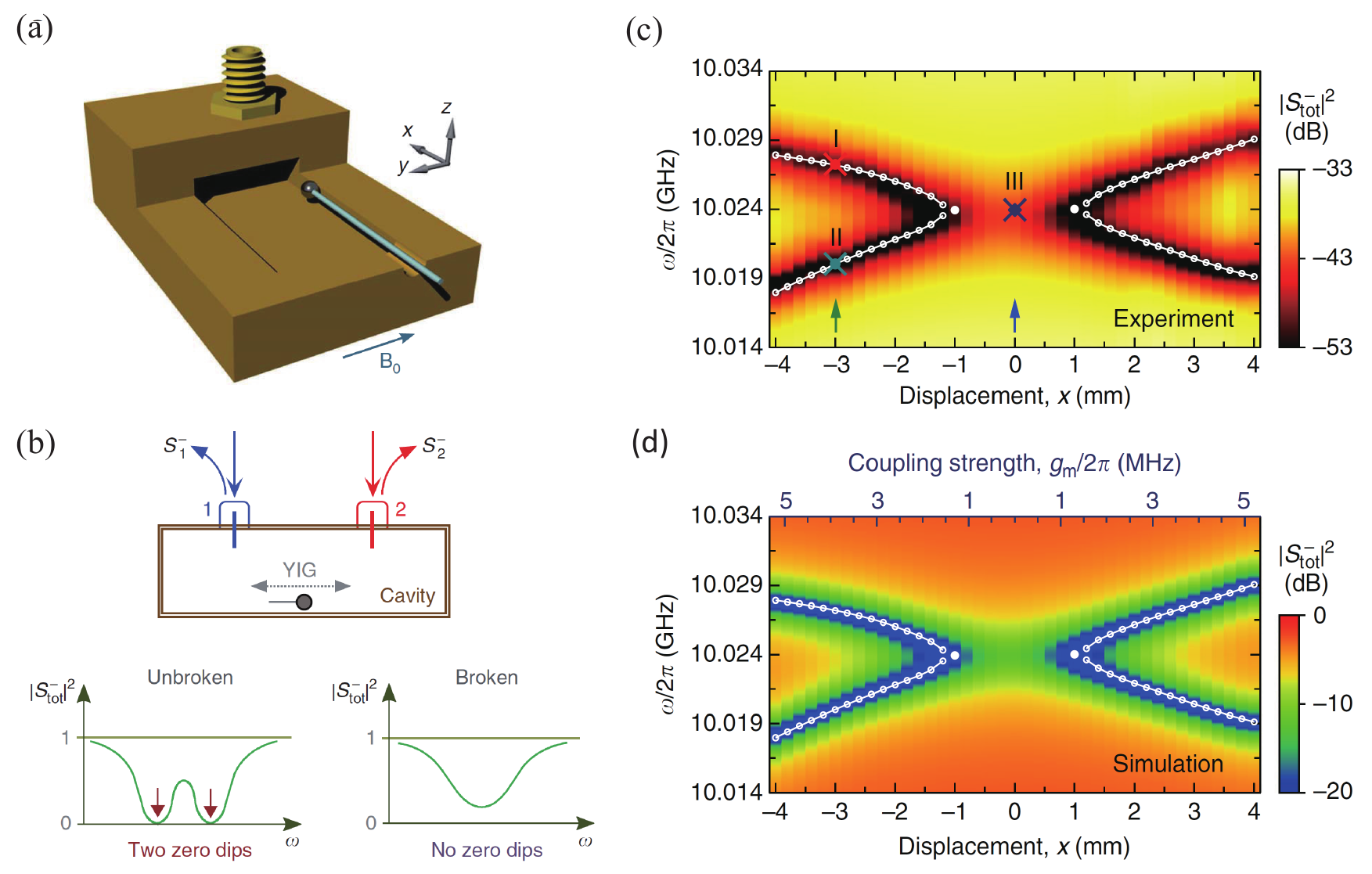}
  \caption{(a) Sketch of the experimental setup, where a YIG sphere is glued on a wooden rod that is inserted into a 3D rectangular cavity. (b) The cavity has 2 ports for both measurement and feeding microwaves into the cavity. The total output spectrum is the sum of outputs from both ports. The output spectrum has zero dips at zero when the system is in the unbroken-symmetry regime, corresponding to the coherent perfect absorption frequencies, which disappears in the broken-symmetry regime. Measured (c) and calculated (d) output spectrum versus displacement of the YIG sphere in the cavity. Source: The figures are adapted from Ref. \cite{ZhangNC2017}.}\label{Zhang2017-1}
\end{figure}

{\it Non-Hermitian CMPs.} Conventional CMPs have a finite lifetime due to the intrinsic losses of photons and magnons. Pumping the cavity to compensate this dissipation may extend the coherent time and lead to a dynamical equilibrium. Such a pumped system is described by a non-Hermitian Hamiltonian, with the special features as discussed in previous sections. These features may significantly modify the properties of the polaritonic modes and the transmission spectrum. The experiment setup is sketched in Fig. \ref{Zhang2017-1}(a), where a YIG sphere is inserted into a 3D rectangular cavity. The essential physics is modelled by an archetypical two-level $\mathcal{PT}$ symmetric Hamiltonian
\begin{equation}
    \hat{\mathcal{H}}_\mathrm{CPA}=(\omega_0+i\gamma_m)\hat{c}^\dagger \hat{c}+(\omega_0-i\gamma_m)\hat{a}^\dagger \hat{a}+g_{mc}(\hat{c}\hat{a}^\dagger+\hat{c}^\dagger \hat{a}),
\end{equation}
where the subscript CPA denotes polaritonic coherent perfect absorption, and $\hat{c}^\dagger (\hat{c})$ and $\hat{a}^\dagger (\hat{a})$ are the creation (annihilation) operators of a microwave photon and magnon, respectively. Both magnon and cavity photon are tuned on resonance with each other at the frequency $\omega_0$, $g_{mc}$ is the coupling strength, and the photon gain is balanced with the magnon loss $\gamma_m$. Figure \ref{Zhang2017-1}(b) shows that the cavity has two ports for both measurement and feeding microwaves into the cavity. The total output spectrum is the sum of outputs from both ports. The Hamiltonian commutes with the combined parity and time-reversal operator, i.e., $[\hat{\mathcal{P}}\hat{\mathcal{T}},\hat{\mathcal{H}}_\mathrm{CPA}]=0$, and hence satisfies $\mathcal{PT}$ symmetry. The corresponding eigenvalues are solved as $\omega_{1,2}=\omega_0\pm\sqrt{g_m^2-\gamma_m^2}$.
When $g_m>\gamma_m$, the system is in the $\mathcal{PT}$ exact regime with two entirely real eigenvalues, which results in two dips at zero in the output spectrum corresponding to the CPA frequencies. For $g_m<\gamma_m$, a pair of complex conjugate eigenvalues appear. The symmetry is spontaneously broken and the CPA then disappears in this regime. The measured output spectrum is shown in Fig. \ref{Zhang2017-1}(c) versus the displacement of the YIG sphere in the cavity (the coupling strength $g_{mc}$ depends linearly on the displacement), which is consistent with the simulation shown in Fig. \ref{Zhang2017-1}(d). In the output spectrum, two real eigenvalues are located at zero output signal and coalesce at the EP when $g_m=\gamma_m$. The central regime located around $x=0$ in the spectrum without any zero dips represents the $\mathcal{PT}$ broken phase.

\begin{figure} %
  \centering
  \includegraphics[width=1.0\textwidth]{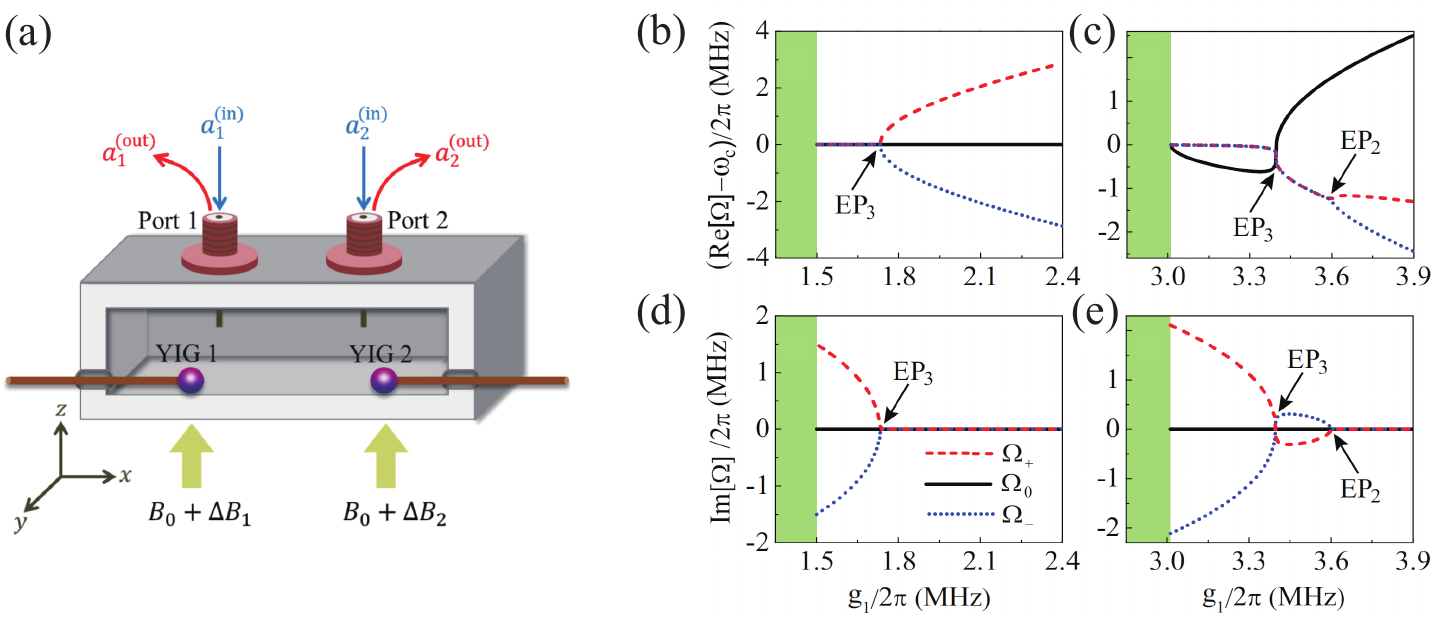}
  \caption{(a)Two YIG spheres with losses coupled to a microwave cavity with gain. Real and imaginary parts of eigenvalues of the hybrid system versus the coupling $g_1$ for the symmetric case (b) and (d), as well as the asymmetric case (c) and (e). Source: The figures are adapted from Ref. \cite{ZhangPRB2019}.}\label{Zhang2017-2}
\end{figure}

{\it Pseudo-Hermiticity and higher-order EPs.} A cavity magnonic system without $\mathcal{PT}$ symmetry but obeying pseudo-Hermiticity supports higher-order (e.g., third-order) EPs \cite{ZhangPRB2019}. As an example, Zhang et al. considered a system consisting of two lossy YIG spheres coupled to a gain microwave cavity, as shown in Fig. \ref{Zhang2017-2}(a). The non-Hermitian Hamiltonian of the hybrid system reads

\begin{equation}
\hat{\mathcal{H}}_\mathrm{eff}=\left(
      \begin{array}{ccc}
        \omega_\mathrm{c}+i\kappa_g & g_1 & g_2 \\
        g_1 & \omega_1-i\gamma_1 & 0 \\
        g_2 & 0 & \omega_2-i\gamma_2 \\
      \end{array}
    \right),
\end{equation}
in which the Kittel frequencies of the YIG spheres are denoted by $\omega_{1,2}$, and their losses with parameters $\gamma_{1,2}>0$.  The gain cavity has resonant frequency $\omega_c$ and a gain coefficient $\kappa_g>0$. Such a ternary system does not respect $\mathcal{PT}$ symmetry but can exhibit a real spectrum as well as second- and third-order EPs [see Figs. \ref{Zhang2017-2}(b)-(e)]. The eigenfrequencies of the system are obtained by solving the determinant $\mathrm{Det}(\hat{\mathcal{H}}_\mathrm{eff}-\Omega I)=0$. According to the spectrum property of the pseudo-Hermitian Hamiltonian, the complex-conjugate equation $\mathrm{Det}(\hat{\mathcal{H}}_\mathrm{eff}^*-\Omega I)=0$ should have the same solutions. By comparing the corresponding coefficients of these two polynomial equations, one can derive the following constraints
\begin{subequations}\label{constrains}
  \begin{eqnarray}
  \kappa_g -\gamma_1-\gamma_2&=& 0,\\
  \Delta_1\gamma_1+\Delta_2\gamma_2 &=& 0,\\
  (\Delta_1\Delta_2-\gamma_1\gamma_2)\kappa_g+g_1^2\gamma_2+g_2^2\gamma_1 &=& 0,
\end{eqnarray}
\end{subequations}
where $\Delta_{1(2)}=\omega_{1(2)}-\omega_c$ is the frequency detuning.
Requiring pseudo-Hermiticity means ensuring that the total gain and loss are balanced in the whole system as indicated by Eq. (\ref{constrains}a). For convenience, one can introduce the ratio of losses and couplings $\eta=\gamma_1/\gamma_2$ and $\delta=g_2/g_1$. It is found that the detunings should satisfy $\Delta_2=-\eta\Delta_1$ according to Eq. (\ref{constrains}b). The last constraint Eq. (\ref{constrains}c) leads to a minimal coupling strength $g_\text{min}$ to achieve an entirely real spectrum of the system. The emergence of the third-order EP thus does not require identical cavity-magnon couplings and magnon losses.

For the symmetric case, two Kittel modes have the same damping rates $\gamma_1=\gamma_2~(\eta=1)$. One then analytically obtains the condition to realize the third-order EP as
\begin{equation}
    g_1=g_2=g_\mathrm{EP3}=\frac{2}{\sqrt{3}}\gamma_2, ~~~~~ \Delta_1=-\Delta_2=\Delta_\mathrm{EP3}=\frac{1}{\sqrt{3}}\gamma_2.
\end{equation}
The corresponding three eigenvalues read $\omega_c, \omega_c\pm\sqrt{3g_1^2-4\gamma_2^2}$ and their real and imaginary parts versus the coupling $g_1$ are plotted in Figs. \ref{Zhang2017-2}(b) and (d). The green regime, i.e., $g_1<g_\mathrm{min}$, was not considered. This result shows that $g_\mathrm{EP3}>g_\mathrm{min}$ is experimentally accessible in the symmetric case.

For the asymmetric case, the magnonic losses $\gamma_1\neq\gamma_2~(\eta\neq 1)$. One can similarly obtain the parameters according to Eq. (\ref{constrains})
\begin{equation}
    g_1=g_\mathrm{EP3}=\left[\frac{1+\eta \delta^2}{(1+\eta)\eta}+\frac{3(1+ \delta^2)}{1+\eta+\eta^2}\right]^{-1/2}2\gamma_2, ~~~ \Delta_1=\Delta_\mathrm{EP3}=\left[\frac{1+\eta \delta^2}{(1+\eta)\eta}g_\mathrm{EP3}^2-\gamma_2^2\right]^{1/2}.
\end{equation}
Numerical results for the spectra for $\gamma_1=2\gamma_2$ are plotted in Figs. {\ref{Zhang2017-2}}(c) and (e), respectively, for the real and imaginary parts versus the coupling strength $g_1$. One observes the second- and third-order EPs separately at $g_\mathrm{EP2}=3.600$ MHz and $g_\mathrm{EP3}=3.394$ MHz.

We note that such exotic spectra can be experimentally detected via a transmission measurement as long as the cavity quality is high enough \cite{GardinerPRA1985,Walls2008,Meystre2007}. This provides a route to study  pseudo-Hermiticity and higher-order EPs in a hybrid spincavitronic system.

\begin{figure} %
  \centering
  \includegraphics[width=1.0\textwidth]{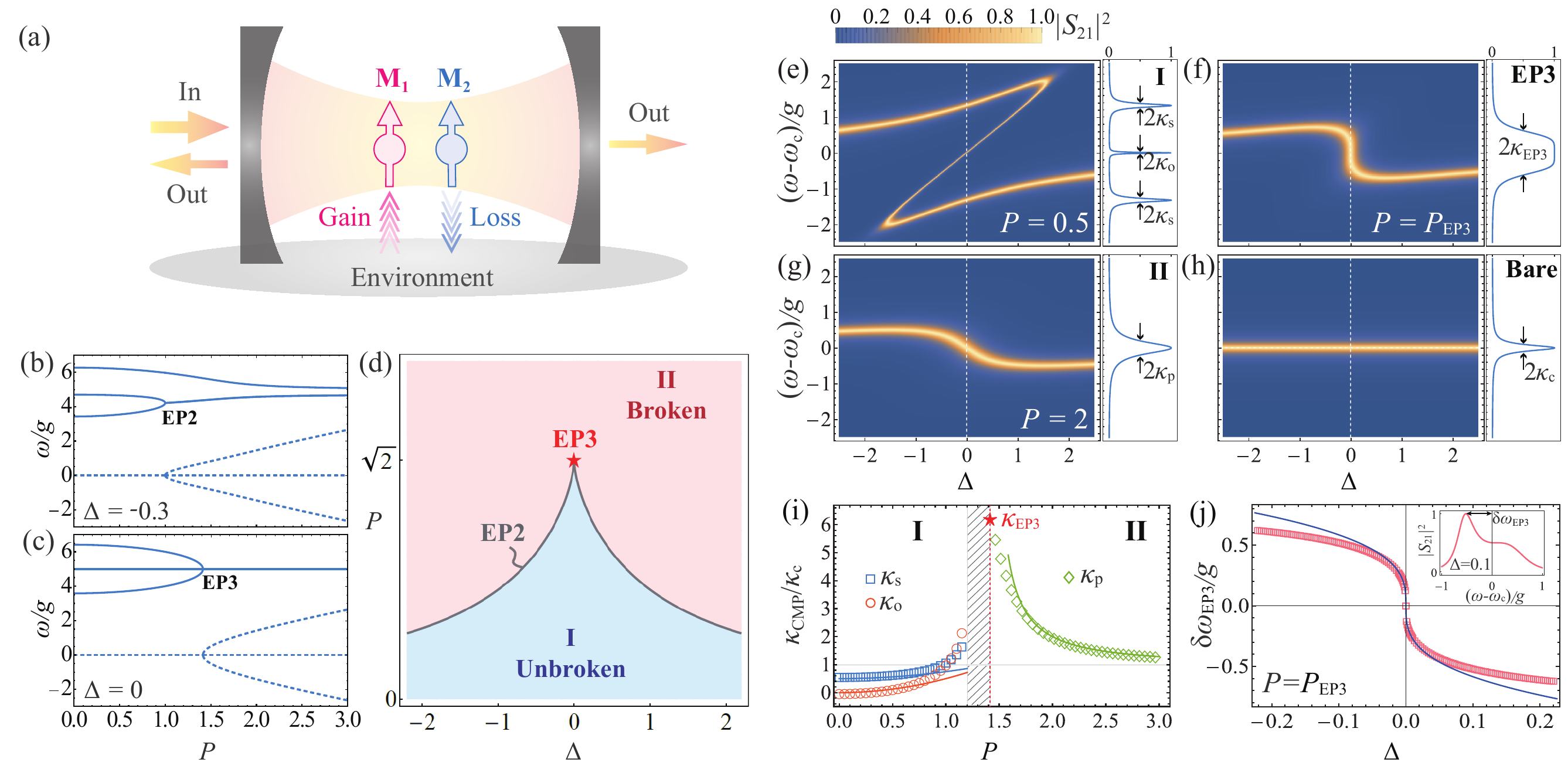}
  \caption{(a) Schematic illustration of photon scattering by two magnons with balanced gain and loss in a microwave cavity.
  (b-c) Evolution of eigenfrequencies as a function of the gain-loss parameter $P$, where the solid and dashed curves respectively represent the real and imaginary part of the eigenfrequencies.
  The detuning parameters are chosen to be (b) $\Delta=-0.3$ and (c) $\Delta=0$.
  The cavity frequency is set as $\omega_\mathrm{c}/g=5$.
  (d) $\mathcal {P}\mathcal {T}$ symmetric phase transition diagram.
  Transmission spectrum for different gain-loss parameters: (e) $P=0.5$, (f) $P=\sqrt{2}$, and (g) $P=2$. (h) Transmission spectrum of a bare cavity. The right panels in (e)-(h) show the zero-detuning spectrum. (i) Linewidth of CMP modes $\kappa_\mathrm{CMP}\in\{\kappa_\mathrm{s},\kappa_\mathrm{o},\kappa_\mathrm{EP3},\kappa_\mathrm{p}\}$ as a function of the gain-loss parameter $P$ at the zero detuning point. Symbols are numerical results and solid curves are the asymptotic formulas (\ref{Eq-width}). The dashed area is not accessible because of the strong overlap between modes. (j) Sensitivity at $P=P_\mathrm{EP3}$. Symbols denote numerical results and the blue curve represents the analytical formula (\ref{Sensitivity}). (Inset) Transmission spectrum as a function of the mode frequency at detuning $\Delta=0.1$. The calculations use $\kappa_{\mathrm{c}}/g=0.1$. Source: The figures are adapted from Ref. \cite{CaoPRB2019}.}\label{PTCMP-1}
\end{figure}

{\it $\mathcal{PT}$ symmetric CMP and exceptional magnetic sensitivity.} The ternary system with $\mathcal{PT}$ symmetry is implemented by a lossless cavity and balanced gain and loss magnonic modes. A setup consisting of two magnetic bodies with balanced gain and loss placed inside a microwave cavity [Fig. \ref{PTCMP-1}(a)], has been proposed to realize an ultrahigh sensitivity magnetometer,  based on the higher-order EP \cite{CaoPRB2019}. The system is described by the following non-Hermitian bosonic Hamiltonian
\begin{equation}\label{Eq-Ham}
  \hat{\mathcal{H}}=\omega_\mathrm{c}{\hat{c}^\dagger}\hat{c}
  +(\omega_\mathrm{s}+i\bar\beta){\hat{s}^\dagger_1}\hat{s}_1
  +(\omega_\mathrm{s}-i\bar\beta){\hat{s}^\dagger_2}\hat{s}_2
  + g \left[{\hat{c}^\dagger}(\hat{s}_1+\hat{s}_2)+h.c.\right],
\end{equation}
where $\hat{c}^\dagger(\hat{c})$ and $\hat{s}^\dagger_{1,2}(\hat{s}_{1,2})$ are the photon and magnon creation
(annihilation) operators, respectively, $\omega_\mathrm{c}$ is the cavity's resonance frequency, $\omega_\mathrm{s}$ denotes the Zeeman splitting, $\bar\beta>0$ describes the energy dissipation/amplification rate, and $g$ represents the magnon-photon coupling strength. Under the combined operation of parity $\hat{\mathcal{P}}$ ($\hat{s}_1\leftrightarrow \hat{s}_2$) and time reversal $\hat{\mathcal{T}}$ ($i\rightarrow-i$, $\hat{s}_{1(2)}\rightarrow-\hat{s}_{1(2)}$, and $\hat{c}\rightarrow-\hat{c}$), it is straightforward to find that Eq. (\ref{Eq-Ham}) is invariant and thus respects $\mathcal {P}\mathcal {T}$ symmetry. The direct exchange coupling between magnons is assumed to be absent in the present model.

One considers single particle processes, so that the three states $\left\{\hat{c}^\dagger|0\rangle, ~\hat{s}^\dagger_1|0\rangle, ~\hat{s}^\dagger_2|0\rangle\right\}$ constitute a complete basis, where $|0\rangle$ represents the vacuum state. The Hamiltonian can therefore be expressed in the following matrix form
\begin{equation}
\hat{\mathcal{H}}=\left(
      \begin{array}{ccc}
        \omega_\mathrm{s}+i\bar\beta & 0 & g \\
        0 & \omega_\mathrm{s}-i\bar\beta & g \\
        g & g & \omega_\mathrm{c} \\
      \end{array}
    \right).
\end{equation}
By solving $\hat{\mathcal{H}}|\phi\rangle=\omega|\phi\rangle$, we obtain the following cubic equation for the eigenvalues
\begin{equation}\label{Eq-Cubic}
\left(\Omega^2+P^2\right)\left(\Omega+\Delta\right)-2\Omega=0,
\end{equation}
with $\Omega=(\omega-\omega_\mathrm{s})/g$, $\Delta=(\omega_\mathrm{s}-\omega_\mathrm{c})/g$ the frequency detuning, and $P=\bar\beta/g$ being the balanced gain-loss parameter.

The evolution of the three eigenvalues is demonstrated in Figs. \ref{PTCMP-1}(b) and (c), where the solid and dashed lines correspond to real and imaginary parts, respectively. The third-order EP3 appears in the case of resonance between photon and magnon, i.e., $\Delta=0$. The phase diagram [plotted in Fig. \ref{PTCMP-1}(d)] is determined by the sign of the discriminant
\begin{equation}
\Lambda=P^2\Delta^4+(2P^4+10P^2-1)\Delta^2+(P^2-2)^3
\end{equation}
of Eq. (\ref{Eq-Cubic}). $\Lambda<0$ gives the $\mathcal {P}\mathcal {T}$ exact phase, in which all three eigenvalues are real and the eigenvectors satisfy the so-called biorthogonal relation $\langle\phi_i^*|\phi_j\rangle=\delta_{ij}$ with $i,j=1,2,3$. For $\Lambda>0$, only one real eigenvalue survives and the other two become complex conjugates to each other, which corresponds to the $\mathcal {P}\mathcal {T}$ broken phase. EP2 occurs along the critical curve $\Lambda=0$ but with $\Delta\neq0$ [see the grey curve in Fig. \ref{PTCMP-1}(d)]. EP3 emerges when both $P=P_\text{EP3}=\sqrt{2}$ and $\Delta=0$ are simultaneously satisfied [see the red star in Fig. \ref{PTCMP-1}(d)].

For a conventional CMP system, strong coupling is usually identified from the gap of the transmission spectrum at the resonance point. From input-output theory \cite{Walls2008}, the scattering coefficient is analytically obtained as
\begin{equation}\label{Eq-S21}
    S_{21}=\frac{\kappa_\mathrm{c}}{i(\omega-\omega_\mathrm{c})-\kappa_\mathrm{c}+\Sigma(\omega)},
\end{equation}
where the total self-energy $\Sigma(\omega)=\Sigma^\mathrm{gain}(\omega)+\Sigma^\mathrm{loss}(\omega)$ from the magnon-photon coupling includes two parts: $\Sigma^\mathrm{gain/loss}(\omega)=g^2/[i(\omega-\omega_\mathrm{s})\pm\beta]$ for gain ($+$) and loss ($-$), respectively. One notes that $\Sigma(\omega)$ now is purely imaginary, leading to a fully transparent transmission at resonance.

Figures \ref{PTCMP-1}(e)-(g) show the transmission spectrum $|S_{21}|^{2}$ as a function of the mode frequency $\omega$ and the cavity detuning $\Delta$, for different gain-loss parameters $P$. As a reference, one plots the bare-cavity spectrum in Fig. {\ref{PTCMP-1}}(h). For $P<P_\mathrm{EP3}$, one finds that the transmission spectrum displays a novel ``Z''-shape [see Fig. {\ref{PTCMP-1}}(e)], instead of the conventional level anticrossing. Furthermore, the authors observed three peaks in the strong-coupling region ($\Delta\sim0$), in which the ultranarrow central mode corresponds to the dark-state CMP, besides two sideband abnormal Rabi-splitting modes.
An increasing $P$ leads to a coalescence of the peaks. For $P=P_\mathrm{EP3}$, three eigenvalues merge together at $\omega_\mathrm{c}$, and form a flat and wide transparent window shown in Fig. {\ref{PTCMP-1}}(f). When $P$ further increases, i.e., $P>P_\mathrm{EP3}$, the dark-state CMP mode still survives, with its linewidth however being significantly broadened as plotted in Fig. {\ref{PTCMP-1}}(g). This is a Purcell-like effect \cite{PurcellPR1946} induced by the $\mathcal{P} \mathcal{T}$ symmetry breaking. The authors paid special attention to the $P$-dependence of the linewidth of the spectrum at zero detuning, and derived the following asymptotic formulas [solid curves in Fig. {\ref{PTCMP-1}}(i)]
\begin{subequations}\label{Eq-width}
\begin{eqnarray}
P\to0&:&~\kappa_\mathrm{s}/\kappa_\mathrm{c}\simeq (P^2+2)/4,~~~~ \kappa_\mathrm{o}/\kappa_\mathrm{c}\simeq P^2/2,\\
P\gg P_\mathrm{EP3}&:&~\kappa_\mathrm{p}/\kappa_\mathrm{c}\simeq P^2/(P^2-2),
\end{eqnarray}
\end{subequations}
which agree well with the numerical results [symbols in Fig. {\ref{PTCMP-1}}(i)].

The sensitivity is conventionally defined as the splitting of the eigenfrequencies perturbed around the EP. However, this becomes unfeasible due to the significant spectrum broadening near the EP as shown in Fig. {\ref{PTCMP-1}}(f). In the present model, there always exists a real central mode no matter whether the $\mathcal {P}\mathcal {T}$ symmetry is broken or not. A more appropriate definition of the sensitivity is the separation between the always-real central mode and the constant cavity mode. At $P=P_\mathrm{EP3}$,
\begin{equation}\label{Sensitivity}
    \delta\omega_\mathrm{EP3}/g=-\mathrm{sgn}(\Delta)\delta \theta, ~~~~
    \text{with}\ \ \delta \theta=2^{1/3}|\Delta|^{1/3},
\end{equation}
which is consistent with the numerical results in the small detuning regime, as plotted in Fig. {\ref{PTCMP-1}}(j). For detuning $|\Delta|<0.06$ (corresponding to the magnetic field $|\delta B|<2$ mT for $g\sim1$ GHz), exceptional sensitivity on the order $|\Delta|^{1/3}$ is still present.
Considering the cavity frequency resolution $|\delta \omega_\mathrm{EP3}| \sim \kappa_\mathrm{c}$, the magnetic sensitivity is expressed as
\begin{equation}
    |\delta B|\approx \frac{\kappa_\mathrm{c}}{2\gamma C},
\end{equation}
where $C\sim g^2/\kappa_\mathrm{c}^2$ is the strong coupling cooperativity ranging from $10^3\sim 10^7$ \cite{HueblPRL2013,ZhangPRL2014,GorPRAp2014}. For a microwave cavity working at GHz with a MHz resolution and a (sub-)MHz noise, the sensitivity is estimated as $\sim10^{-15}\ \mathrm{T}\ \mathrm{Hz}^{-1/2}$, which is two orders of magnitude higher than that of the state-of-the-art magnetoelectric sensors \cite{AnnSensor2017}.
Hopefully, the EPs discussed here are promising candidate for designing high-precision sensing device.

\begin{figure} %
  \centering
  \includegraphics[width=1.0\textwidth]{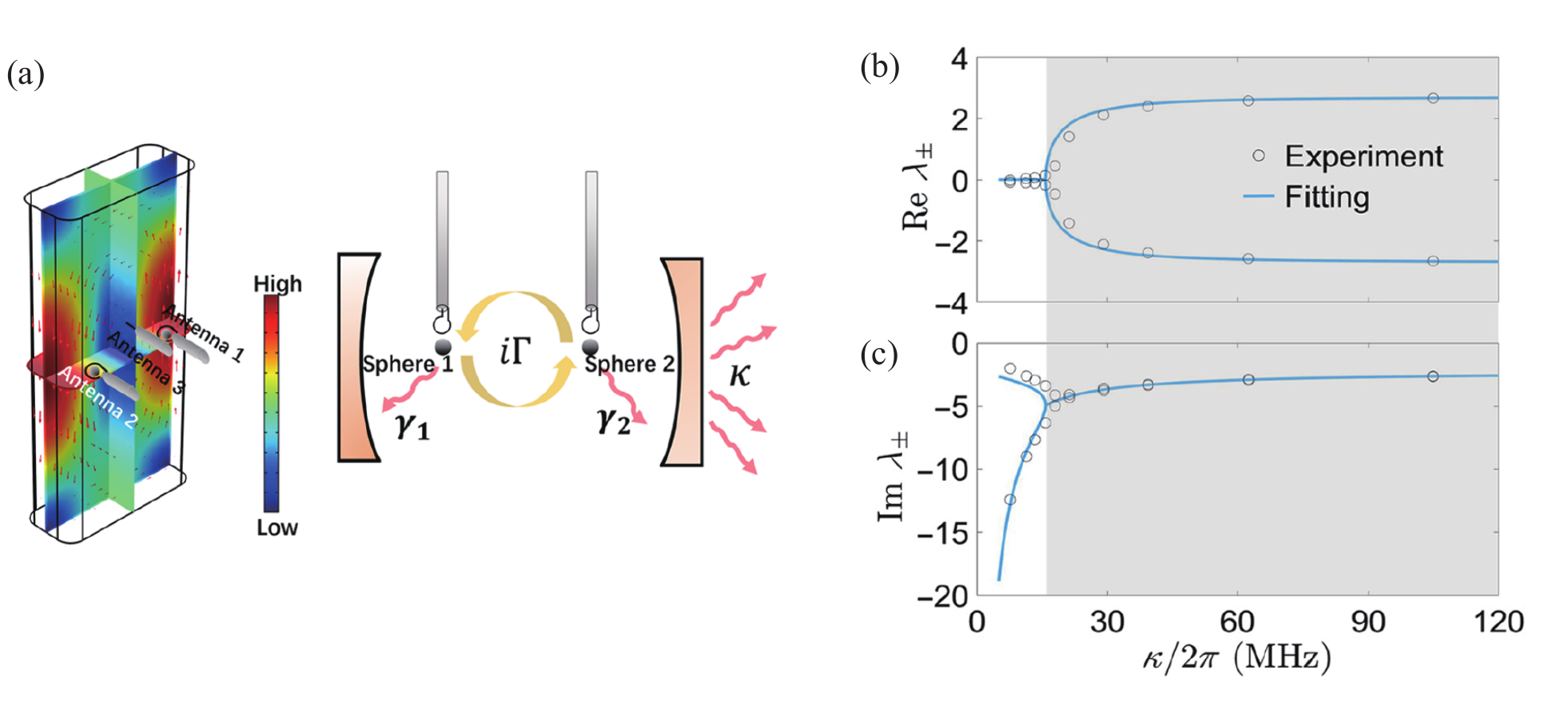}
  \caption{(a)Two YIG spheres with losses coupled to a large dissipative microwave cavity. (b) Real and (c) imaginary parts of the eigenfrequencies as a function of the cavity dissipation rate $\kappa$. The shadowed area is the $\mathcal{APT}$ broken phase. The dots and solid lines represent the experimental data and theoretical prediction, respectively. Source: The figures are adapted from \cite{ZhaoPRAp2020}.}\label{Zhao2020}
\end{figure}

{\it $\mathcal{APT}$ symmetry in CMPs.} The concept of $\mathcal{PT}$ symmetry has been extended to other contexts. $\mathcal{APT}$ symmetry represents such a generalization. The difference with $\mathcal{PT}$ symmetry lies in that $\mathcal{PT}$ symmetry requires the presence of gain to compensate the loss of the system, while $\mathcal{APT}$ symmetry can be realized in all-lossy systems. Zhao et al. \cite{ZhaoPRAp2020} considered the example sketched in Fig. \ref{Zhao2020}(a), in which two YIG spheres with nearly equal losses $\gamma_m$ are placed in a large dissipative microwave cavity. Two YIG spheres are working in the low excitation regime and the magnon excitations of the YIG sphere are simply regarded as harmonic oscillators (Kittle modes only). The two magnons are directly coupled to the microwave photon. By eliminating the photonic degree of freedom, one obtains the effective $\mathcal{APT}$ magnon Hamiltonian
\begin{equation}
  \hat{\mathcal{H}}_\mathrm{eff}=\left(
                             \begin{array}{cc}
                               \Omega-i(\gamma_m+\Gamma) & -i\Gamma \\
                               -i\Gamma & -\Omega-i(\gamma_m+\Gamma) \\
                             \end{array}
                           \right),
\end{equation}
where $i\Gamma$ is the dissipative coupling-rate of the two magnets mediated by the cavity mode and $\Omega$ is the effective frequency detuning between the two magnons. In the regime of $\kappa_c \gg \gamma_m$, we have $\Gamma=g^2/\kappa_c$ with $g$ the direct coupling strength between magnons and photons and $\kappa_c$ the cavity dissipation rate. The eigenvalues are plotted as solid lines in Figs. \ref{Zhao2020}(b) and (c). The eigenvalues are purely imaginary in the $\mathcal{APT}$ exact regime, and bifurcate into complex values with the same imaginary parts in the $\mathcal{APT}$ broken regime.

In the experiments, the magnon states (frequencies) are detected from the output photon spectrum. Experimental measurements of the hybridized magnon modes can be achieved from magnon-read out methods, by exacting the dips and linewidth from the transmission and reflection spectrum. The cavity dissipation rate can be tuned by solely changing the pin length of antenna 3. The measured data shown in Figs. \ref{Zhao2020}(b) and (c) correspond, respectively, to the output frequency dips and the linewidth, and are in good agreement with the theoretical analysis.

\begin{figure}
  \centering
  \includegraphics[width=0.8\textwidth]{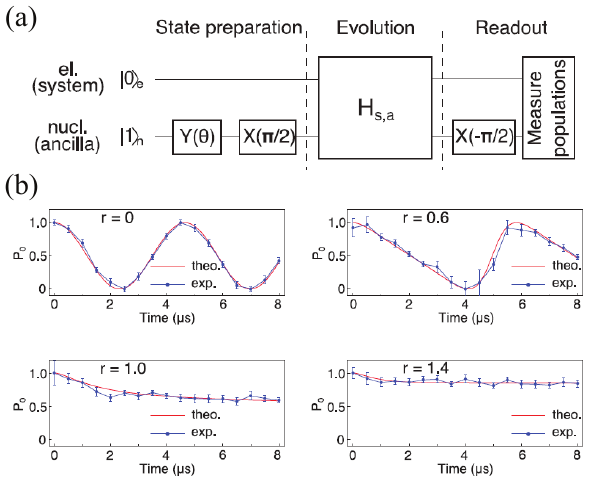}\\
  \caption{(a) Quantum circuit of the experiment to construct $\mathcal{PT}$ symmetric Hamiltonian and to measure the state evolution of a single spin system. $X$ ($Y$) represents the nuclear spin rotation around the $x$ ($y$) axis. (b) Experimental observed evolution of the single spin population in ground state ($P_0$) when the the control parameter $r$ is tuned from $\mathcal{PT}$ exact ($r<1$) to the $\mathcal{PT}$ broken phase ($r>1$). The blue dots and red lines denote experimental and theoretical prediction, respectively. Source: The figures are adapted from Ref. \cite{WuScience2019}.}
  \label{single_spin_PT}
\end{figure}

{\it Purely quantum effects.} In principle, the corresponding physics such as exceptional point is valid in both classical and quantum systems. To study purely quantum effects in the hybrid system, one may go to the low temperature regime to minimize the detrimental influence of temperature on the quantum superposition and entanglement. The magnon-photon entanglement in the $\mathcal{PT}$ exact regime of hybrid magnon-photon system discussed in Section \ref{sec_magnon_photon} is an example along this route. Along this line, Lu et al. \cite{LuPRA2021} showed that the mechanical mode of a cavity magnomechanical system can be perfectly cooled to the ground state when the system approaches the EP. This may open up a possibility to engineer the purely quantum effects. By delicately designing the gain of cavity mode and loss of the magnon mode into the $\mathcal{PT}$ symmetric regions, one can also generate single-magnon states even in the absence of Kerr nonlinearties \cite{WangAP2020}. The second approach to manifest the quantum effects is to reduce the size of magnets, such that the quantum fluctuation of spins can not be neglected any longer. Recently, Wu et al. \cite{WuScience2019} experimentally observed $\mathcal{PT}$ symmetry breaking in a single-spin system. Here the authors studied the $\mathcal{PT}$ symmetric Hamiltonian of the form $\hat{\mathcal{H}}_s=ir\hat{\sigma}_z + \hat{\sigma}_x$ corresponding to $\omega_1=\omega_2=0, \kappa_1=\kappa_2=-r,g=1$ in Eq. \eqref{PT_Ham}, which governs the dynamics of a single electron spin. To generate such an effective Hamiltonian, the electron spin was coupled to a nuclear spin system as shown in Fig. \ref{single_spin_PT}(a), and the dilated Hermitian Hamiltonian $\hat{\mathcal{H}}_{s,a}$ was delicately designed by shining two selective microwave pulses on the system. Population of the electron spin state $P_0$ was measured at different control parameter $r$, as shown in Fig. \ref{single_spin_PT}(b). When $r<1$, the system falls into $\mathcal{PT}$ exact phase and real eigenvalues of $\hat{\mathcal{H}}_s$ determine an oscillating behavior of the population. When $r>1$, the $\mathcal{PT}$ symmetry as well as the oscillation behavior break down.

\section{Applications}\label{sec_application}
In the last few sections, we have reviewed some of the recent developments in quantum magnonics, including the quantum states of magnons and the hybridization of magnons with other quantum platforms, such as microwave photons, superconducting qubits, phonons and electrons. These studies are driven by fundamental interest as well as their promising applications in quantum information science and technologies.
Strictly speaking, part of the proposals work in the classical regime as well, such as non-reciprocical wave propagation \cite{WangPRL2019}, microwave-light conversion mediated by magnons \cite{HisaPRB2016}, microwave circulators \cite{ZhuPRA2020}, gradient memories \cite{ZhangNC2015}, and coherent gate operations \cite{XuPRL2021}. They are, however, expected to work in the quantum regime at low temperature because these applications are mainly based on linear interactions. Some of the proposals for applications, such as quantum memories based on magnons excited in atomic ensembles \cite{TanjiPRL2009}, single-magnon detection by superconducting qubits \cite{LachScience2020}, and weak field sensors for dark matter searches \cite{CresPRL2020} are already in the quantum regime, and require cryogenic conditions to suppress the thermal noise.
In this section, we review a few already existing application directions of quantum magnonics, including quantum memories, high-precision sensors for temperature calibration, weak-field detection for dark matter searching, and coherent gate operations for signal processing.

\subsection{Memory devices}

\begin{figure}
  \centering
  \includegraphics[width=1.0\textwidth]{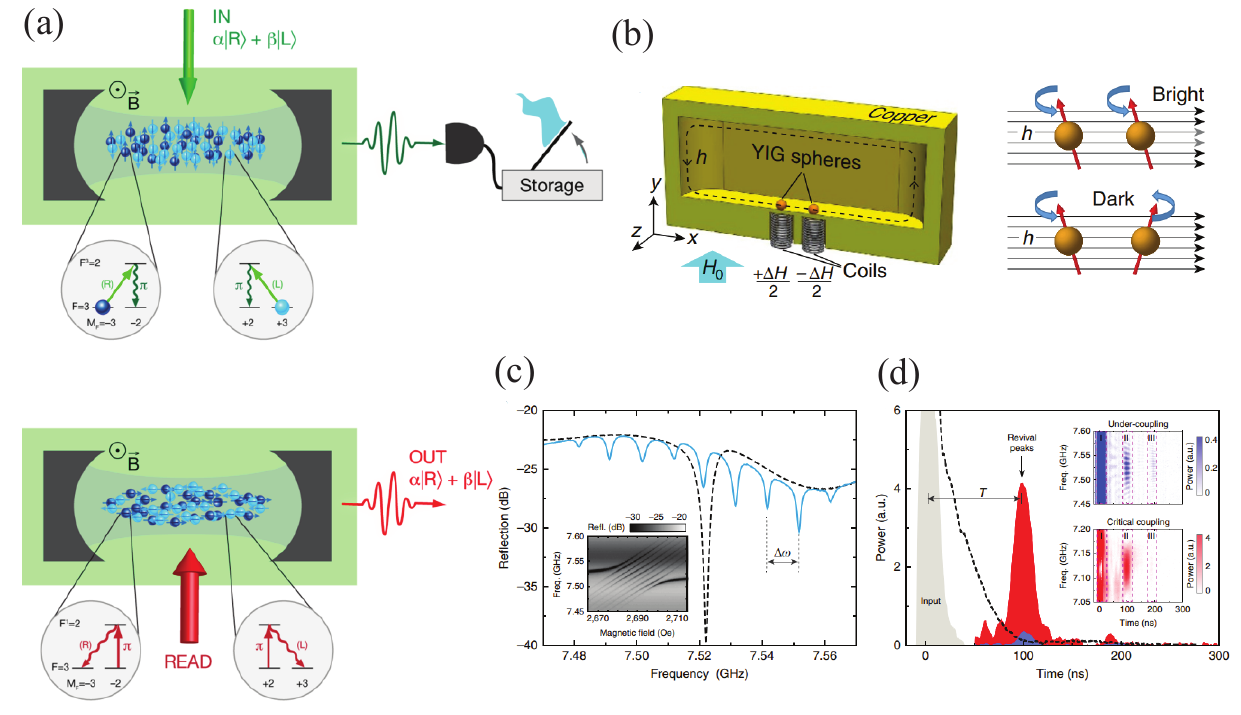}\\
  \caption{(a) Schematics of an atomic ensemble containing two types of oppositely polarized atoms. (top) A write pulse  with arbitrary polarization is fed into the ensemble and generates an entangled magnon state and a heralding photon state. (bottom) A read pulse converts the atomic state into a single photon and is then detected. (b) Magnon bright and dark modes of two YIG spheres in a single cavity. (c) Reflection spectrum of a cavity including 8 YIG spheres and (d) working procedure of the gradient memory in such a hybrid cavity. Source: The figures are adapted from Refs. \cite{Laurat2009,ZhangNC2015}.}
  \label{magnon_memory}
\end{figure}

Tanji et al. \cite{TanjiPRL2009} demonstrated a heralded single-magnon quantum memory for photon polarization states in a resonator containing two types of atomic ensembles as shown in Fig. \ref{magnon_memory}(a). To store a photon state with arbitrary  polarization, i.e. $|\varphi\rangle = \cos \theta |R\rangle + \sin \theta e^{i\phi}|L\rangle$, the authors used two spatially overlapping atomic ensembles A and B inside an optical resonator, where each ensemble consists of approximately 8000 atoms cooled to a temperature of $30~ \mu$K. Here $|L\rangle$ and $|R\rangle$ represent left and right circularly polarized photons, respectively. The energy levels of the atomic ensembles are delicately designed such that the A (B) atoms only absorb $R$-polarized ($L$-polarized) photons, and both atomic ensembles emit photons with the same polarization ($\pi$). The working procedures of the memory are sketched in Fig. \ref{magnon_memory}(a). The authors first apply a write pulse that maps the polarization state of photon $|\varphi\rangle$ into a superposed magnon state $|\Phi \rangle$ as
\begin{equation}
|\Phi \rangle = \cos \theta |1\rangle_A |0 \rangle_B + e^{i\phi} \sin \theta |0\rangle_A |1 \rangle_B,
\end{equation}
where $|n\rangle_\nu$ represents the $n$-magnon state of the ensemble $\nu$ ($\nu=A,B$). With a proper delay, a read pulse (red arrows) is applied to the ensemble to stimulate emission of the $\pi$ photons by both atomic ensembles. The successful detection of one photon indicates the mapping of the light polarization state onto a magnon state. In the experiments, the stored photon is recreated with sub-Poissonian statistics [$g_2(0)=0.24$] and efficiency as high as $50\%$. The original photon polarization is restored with fidelity larger than 0.9.

Zhang et al. \cite{ZhangNC2015} considered $N$ identical YIG spheres loaded in a microwave cavity. This system is described by the following Hamiltonian
\begin{equation}
\hat{\mathcal{H}} = \omega_c \hat{c}^\dagger \hat{c} + \sum_{j=1}^N \left ( \omega_j \hat{a}_j ^\dagger \hat{a}_j + g \hat{c}^\dagger \hat{a}_j + g^* \hat{c} \hat{a}_j ^\dagger\right ),
\end{equation}
where $\hat{c}$ and $\hat{a}_j$ are the annihilation operators of photon mode and $j-$th magnon mode, respectively. The magnon frequency of each YIG sphere $\omega_i$ is controlled by a small coil underneath and the frequencies are tuned to be evenly distributed as
$\omega_j = \omega_c + [j-(N+1)/2] \Delta \omega_m$, with $\Delta \omega_m$ the frequency interval. Typically, the frequency spectrum contains $N+1$ resonant modes with equal spacing $\Delta \omega_m$, as shown in Fig. \ref{magnon_memory}(c) with $N=8$. Among these modes, there are $N-1$ dark modes of magnons $\hat{D}_n$ ($n=1,2,...N-1$) which are not coupled to the cavity mode, while only one bright mode $\hat{B}$ is coupled to the cavity, i.e.,
\begin{equation}
\hat{B}=\frac{1}{\sqrt{N}} \sum_{j=1}^N \hat{a}_j, \hat{D}_n = \frac{1}{\sqrt{N}}\sum_{j=1}^N \hat{a}_j e^{2\pi i (j-\frac{N+1}{2})n/N}.
\end{equation}

Since these modes are uniformly distributed in frequency space, the dark modes and bright modes can dynamically convert from one to another in a period of $T=\pi/(N\Delta \omega)$ with $\Delta \omega \approx \Delta \omega_m$ for large $N$. The proposed gradient memory including a storing process and a readout process works as follows: In the storing process, a pulsed microwave at frequency $\omega_a$ is injected into the cavity and converted into the bright mode of magnons via their coherent coupling. The bright mode temporally transfers into the dark mode, which avoids radiative losses. After one period, the dark mode evolves back to the bright mode and the information coded in magnons is retrieved back by converting to photons. The dynamic output of the memory is found in Fig. \ref{magnon_memory}(d). To improve the signal of the retrieved pulse, the coaxial probe is tuned to meet the critical coupling condition $\kappa_{a,1}=\kappa_{a,0} + \pi g^2/\Delta \omega$ (red peak in the figure), where $\kappa_{a,1}$ is the coupling rate of cavity photons to the probe pulse and $\kappa_{a,0}$ is the intrinsic damping rate of the cavity mode. Compared with the proposals of memory based on spin ensembles, this approach takes advantage of the high spin-density and strong magnon-photon coupling in magnetic insulators, and it avoids the issue of inhomogeneous broadening in the dilute spin ensembles.

Sarma et al. \cite{SarmaNJP2021} proposed to transfer the information from cavity mode to phonon mode in a hybrid cavity-magnet setup with strong magnon-photon-phonon interaction as introduced in Section \ref{sec_magnon_phonon}. Due to the lower damping of the phonon mode compared with magnons and photons, the information coded in the phonon mode may stay available for a longer time. Lai et al. \cite{LaiPRA2018} studied the indirect coupling of a superconducting flux qubit and a NV center assisted by magnons excited in YIG and proposed a scheme to transfer and store the state of a superconducting qubit to the spin of a NV center with fidelity higher than 0.9 and long storage time of 10 ms. Compared with the direct coupling between the flux qubit and NV center, the magnetic mediator enhances the coupling strength and thus suppresses the decoherence of the qubits. Simon et al. \cite{SimonNP2007} demonstrated that cavity photons can serve as a bus to connect two magnon quantum memories.

\subsection{Precision measurements}
\subsubsection{Magnetometers and thermometers}
As introduced in Section \ref{sec_magnon_phonon}, Colombano et al.  \cite{ColomPRL2020} studied the coupling of a breathing microsphere with a magnetic film and demonstrated a magnetometer to detect magnetic fields as low as $\mu$T with a peak sensitivity of $\mathrm{nT\ Hz}^{-1/2}$ up to gigahertz. This finding may be used to build a high-speed sensor of oscillating magnetic fields. The NV center magnetometer is an accomplished application as discussed in Section \ref{sec_magnon_NV}.

Potts et al. \cite{PottsPRAp2020} proposed a thermometer working below 1 K based on  cross-correlation measurements of a hybrid cavity-magnon-phonon system, where the magnon mode is coupled with the phonon mode through magnetostrictive interactions. Specifically, the authors considered the Hamiltonian
\begin{equation}
\hat{\mathcal{H}} = \omega_c \hat{c}^\dagger \hat{c} + \omega_b \hat{b}^\dagger \hat{b} + \omega_r \hat{a}^\dagger \hat{a} + g_{mc}(\hat{a}+\hat{a}^\dagger)(\hat{c}+\hat{c}^\dagger) + g_{mb} \hat{a}^\dagger \hat{a} (\hat{b}+\hat{b}^\dagger) + i\epsilon_d \sqrt{\kappa_P}(\hat{c}e^{i \omega_d t} - \hat{c}^\dagger e^{i \omega_d t})
\end{equation}
where $\kappa_P$ is the coupling rate of the cavity mode to a single external port. By solving the quantum Langevin equation of this Hamiltonian, the cavity field fluctuations are written in terms of the mechanical displacement ($\delta \hat{z} \equiv \delta \hat{b} + \delta \hat{b}^\dagger$)
\begin{equation}
\delta \hat{c} = - \Lambda_{am} \left ( g_{cm} G_{mb} \chi_m \delta \hat{z} + i g_{cm} \chi_m \sqrt{\gamma_m} \delta \hat{\eta}_m - \sqrt{\kappa} \hat{\xi}_P\right ),
\end{equation}
where $\chi_m$ is the magnetic susceptibility, $\gamma_m$ is the magnon dissipation rate, and $\delta \eta_m$ and $\xi_P$ are, respectively, the fluctuation  of the  magnons and cavity photons. This result implies that the thermal mechanical fluctuations ($\delta \hat{z}$) are imprinted onto the cavity microwave mode ($\delta \hat{c}$) through the assistance of the magnon mode ($g_{mc}$). Making use of the input-output relations $\delta \hat{c}_{out}= \hat{\xi}_P -\sqrt{\kappa_P} \delta \hat{c}$, the cavity fluctuations together with the corresponding quadratures $\hat{X}_{\delta \hat{c},\theta}=\cos \theta \hat{X}_{\delta \hat{c}} + \sin \theta \hat{P}_{\delta \hat{c}}$ can be read out. In particular, they calculated the correlation spectrum of the output fields and found that the biased thermal metric relation of the system is
\begin{equation}
\frac{S_{\pi/2,\pi/2}(\omega)}{S_{0,\pi/2}(\omega)} \equiv  \frac{\int_{-\infty}^{\infty} d \omega' \langle \left \{ X_{\delta c,\pi/2}(\omega),X_{\delta c,\pi/2} (\omega')\right \}\rangle }{\int_{-\infty}^{\infty} d \omega' \langle \left \{ X_{\delta c,0}(\omega),X_{\delta c,\pi/2} (\omega')\right \}\rangle }= \frac{4 \coth (\omega/2k_B T)}{2n_{\mathrm{th}}+1}.
\end{equation}
This expression is independent of experimental parameters and it directly determines the temperature of the phonons via the measured correlation spectrum. Note that the above-mentioned thermal metric relation increases with phonon temperature and reaches a plateau for $T>\hbar \omega_{c,r}/k_B$. Therefore, this proposal is most accurate at low temperatures, typically below 1 K, corresponding to the magnon frequency in GHz regime. Compared with existing proposals using high energy optical photons, microwave photons can reduce  heating and thus are compatible with cryogenic temperatures.

\begin{figure}
  \centering
  \includegraphics[width=1.0\textwidth]{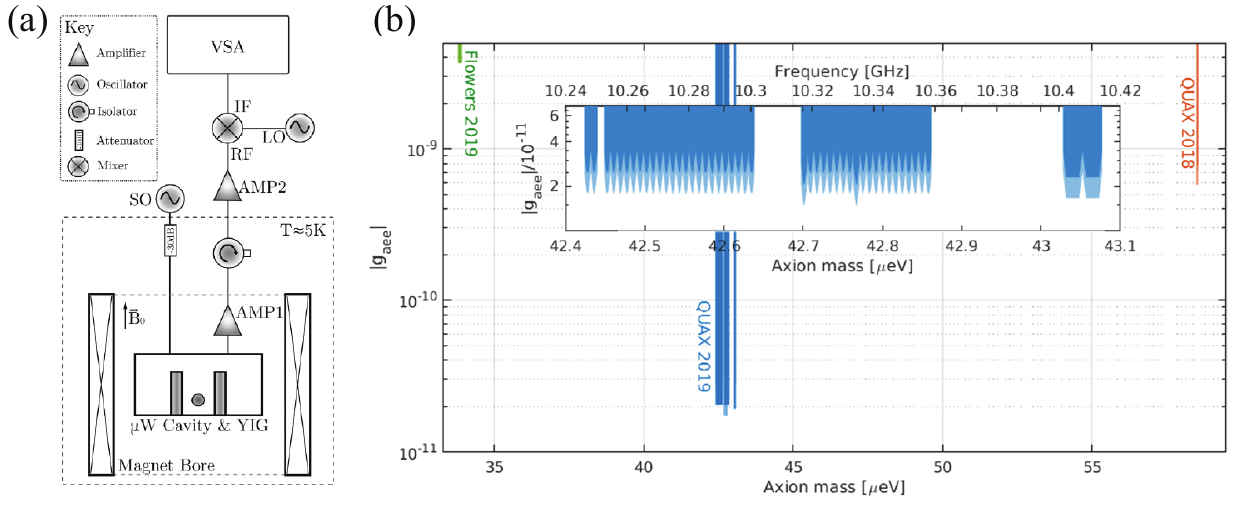}\\
  \caption{(a) Typical experimental setup to detect axionic dark matter using a hybrid cavity-magnet system at low temperature. (b) Overview of the exclusion plot at a 95\% confidence level of the axion-electron coupling and axion mass. The DFSZ axion line is at $g_{aee} \cong10^{-15}$.  Source: The figures are adapted from Refs. \cite{CresPRL2020,FlowerPDL2019,CresciniEPJC2018}.}
  \label{dark_matter}
\end{figure}

\subsubsection{Searching for dark matter}

Dark matter is a hypothetical form of matter that plays an important role in modern cosmology. It is believed to make up 85\% of the matter of our Universe and 27\% of the Universe's total mass-energy density. However, its composition remains a mystery \cite{BertoneRMP2008}. It is generally believed that the known elementary particles of the Standard Model are not dark matter, and thus one has to extend the standard model. Axions, originally proposed to solve the strong-CP problem in quantum chromodynamics \cite{Peccei1977,DuffyNJP2009}, are a leading dark-matter candidate. The detection of the axion mass and its coupling with normal matter is challenging. The detection of axions using magnons was originally proposed by Barbieri et al. \cite{BarbPLB1989}. Using recent developments from cavity spintronics, it becomes achievable to search for axions using hybrid magnet-light systems because of the sensitivity of the output spectrum of the hybrid system to  external small fields \cite{CresPRL2020,FlowerPDL2019,CresciniEPJC2018,RuosoIOP2016,BarbPDU2017}. Such a hybrid system is called a magnetic axion haloscope.

To understand the working principle of magnetic axion haloscopes, we start from the Hamiltonian
\begin{equation}\label{darkham}
\hat{\mathcal{H}} = \omega_c \hat{c}^\dagger \hat{c} + \omega_r \hat{a}^\dagger \hat{a} + g (\hat{c}^\dagger \hat{a} + \hat{c} \hat{a}^\dagger) + \hat{\mathcal{H}}_{aee}.
\end{equation}
where $\hat{c}$ and $\hat{a}$ denote cavity photon and magnon mode, respectively. Here the first three parts of the Hamiltonian are the familiar form describing the hybridization of magnons with photons, while the last term $\hat{\mathcal{H}}_{aee}=-V\mathbf{M}\cdot \mathbf{B}$ is the coupling between the magnetization $\mathbf{M}$ and the pseudo-magnetic field $\mathbf{B}$ generated by the axions, where $\mathbf{B}$ is proportional to the coupling strength between axion and electron spin $g_{aee}$ and the mass of axions $m_a$ \cite{FlowerPDL2019}. Under a HP transformation, the interaction term is quantized as, $\hat{\mathcal{H}}_{aee}=-B_{aee}\sqrt{S/2} \sin\phi (\hat{a} e^{-i\theta } + \hat{a}^\dagger e^{i \theta})$, where $\theta$ and $\phi$ parameterize the orientation of the axion field and $B_{aee}=|\mathbf{B}|$. By solving the quantum Langevin equation corresponding to Hamiltonian (\ref{darkham}) in the rotating frame, one can derive the cavity field in terms of the axion field as, $\hat{c}(\omega) = ig\sqrt{S/8} B_{ace} \sin \phi e^{i \theta}/(\kappa_c \kappa_m \Delta_c \Delta_a-g^2)$, where $\kappa_c$ and $\kappa_m$ are the dissipation rate of the cavity photons and magnons, respectively, while $\Delta_c = (\omega -\omega_c)/\kappa_c + i/2,\Delta_a = (\omega -\omega_r)/\kappa_m + i/2$. Using input-output relations, we solve the output power of the system as
\begin{equation}
P_{out}(\omega_c) \equiv \omega_c \langle \hat{c}_{out}^\dagger \hat{c}_{out}\rangle = \frac{\omega_c \kappa _c^{ext} g^2 S B_{aee}^2 \sin^2\phi}{8|\kappa_c\kappa_m \Delta_c \Delta_a - g^2|^2},
\end{equation}
where $\kappa_c^{ext}$ is the cavity dissipation caused by the coupling of the cavity field to the probe field. Based on this result, one can detect the existence of an axion field by measuring the output power of the hybrid magnet-cavity system, and, furthermore, obtain information on the axion mass $m_a$ and its coupling strength with the electron spin $g_{aee}$. Since the axion field is expected to be very weak, it is crucial to increase the sensitivity of the detectors. One can increase the effective coupling strength of spins and photons $g$ by choosing a proper cavity with large magnetic field, and magnetic materials with high spin density, such as YIG. Cryogenic conditions are required to reduce the influence of thermal fluctuations.

A typical experimental setup is shown in Fig. \ref{dark_matter}(a) and the comparison of three available experimental results is summarized in Fig. \ref{dark_matter}(b). Crescini et al. \cite{CresciniEPJC2018} first set an upper limit on the coupling of axions to electrons at $g_{aee}<4.9 \times 10^{-10}$ for a mass of $58 ~\mathrm{\mu eV}$ with $95\%$ confidence. Flower et al. \cite{FlowerPDL2019} set limits on $g_{aee}> 3.7 \times 10^{-9}$ for a mass range $33.79 ~\mathrm{\mu eV}<m_a<33.94~ \mathrm{\mu eV}$. By further reducing the noise temperature below 1 K and placing 10 YIG spheres to increase the magnetic volume, Crescini et al. \cite{CresPRL2020,CresAPL2020} performed a wide axion-mass scanning and set the coupling limit to $g_{aee}< 1.7 \times 10^{-11}$ in the mass range $42.4-43.1~ \mathrm{\mu eV}$. Such a sensitivity corresponds to a $1-\sigma$ field sensitivity of $5.5 \times 10^{-19}$ T, which is a record value. Nevertheless, all these results are orders of magnitude away from the theoretical prediction of $g_{aee} \cong10^{-15}$ predicted by the DFSZ model \cite{DFSZ1981}. Further upscaling of the hybrid system by using high-quality superconducting cavities and by improving the detector sensitivity to single magnon or photon is on going to test this theoretical value. The sensitivity can also be improved by using long-lived nuclear spins as preamplifier \cite{JiangNP2021}. Recently, following a similar principle, cavity-magnets system were proposed to detect gravitational waves \cite{ItoEPJC2020}.

\subsection{Gate operation for signal processing}

\begin{figure}
  \centering
  \includegraphics[width=1.0\textwidth]{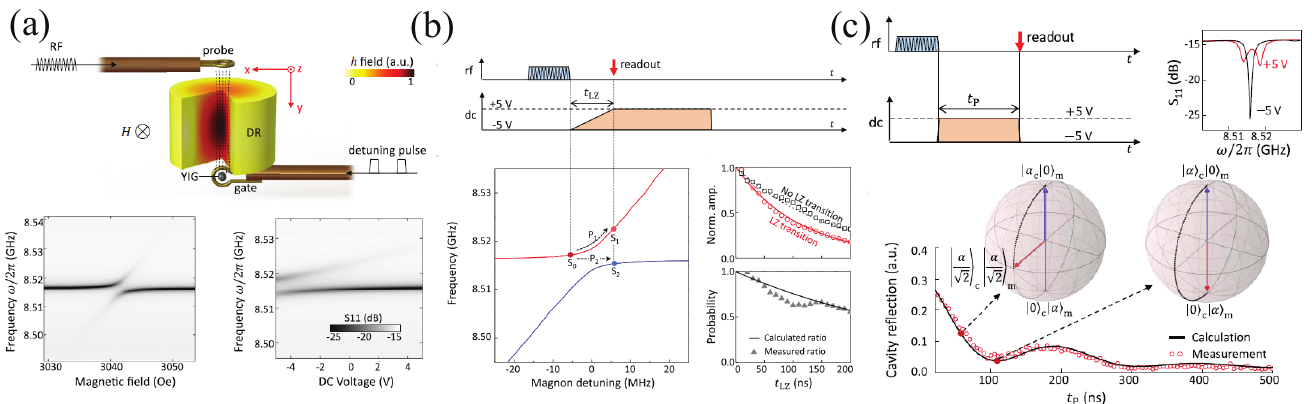}\\
  \caption{(a) Schematic illustration of the device containing a YIG sphere placed under a cylindrical dielectric resonator inside a copper housing. The gate loop wrapping around the YIG allows for a desirable tuning of the magnon frequency with a DC current. (b) Schematic plot of the pulse sequence to measure Landau-Zener transitions and the measured amplitude of the cavity photon as a function of the rise time $t_{LZ}$ of current pulse. (c) Schematic plot of the pulse sequence for measuring the Rabi oscillation and the comparison of measured (red circles) and calculated (black line) time dependence of the cavity reflection signal. The evolution of the system at 56 ns and 111 ns serves as a $\pi/2$ pulse and $\pi$ pulse, respectively. Source: The figures are adapted from Ref. \cite{XuPRL2021}.}
  \label{logic_gate}
\end{figure}
Gate operation is the basic element required for practical signal processing.
Xu et al. \cite{XuPRL2021} demonstrated a series of coherent gate operations using hybrid magnet-light systems, including Landau-Zener transitions, Rabi oscillations, Ramsey interference, and mode swap operations. The schematic of the system is shown in Fig. \ref{logic_gate}(a), where a YIG sphere is placed underneath the dielectric resonator while a probe antenna is placed on top of the resonator to excite and read out the information coded in the cavity photons. A novel feature of this device is that a gate antenna is wrapped around the YIG sphere, which tunes the magnon frequency dynamically via a DC voltage. This system is described by the known Hamiltonian $\hat{\mathcal{H}}=\omega_c \hat{c}^\dagger \hat{c} + \omega_r \hat{a}^\dagger \hat{a} + g(\hat{c}^\dagger \hat{a} + \hat{c} \hat{a}^\dagger)$. Here the magnon and photon modes hybridize to form an anticrossing structure [bottom panel of Fig. \ref{logic_gate}(a)]. The authors first realized an analogue Landau-Zener transition \cite{Zener1932} using the pulse sequence shown in Fig. \ref{logic_gate}(b). Starting from an off-resonant condition ($V=-5$ V), the photon component dominates the upper branch of the eigenstate. When the field is swept slowly to $+5$ V, the hybrid state will stay in the upper branch (path $P_1$), while the cavity photons slowly convert to the magnon mode in this sweeping process. On the other hand, when the field is swept fast, the initial state will jump from the upper branch to the lower branch (path $P_2$), while the hybrid state remains in the photon mode. In the intermediate case, these two paths will interfere and the photon has a certain probability to keep its initial state unchanged, i.e., $P_c=\exp(-\pi g^2 t_{LZ}/\Delta \omega_m)$, where $t_{LZ}$ is the rising time of the current pulse and $\Delta \omega_m$ is the detuning between magnon and photon in the final state. Furthermore, by carefully choosing the pulse length of the dc current, the authors showed that the hybrid system evolves from the initial photon state $|\alpha\rangle_c |0 \rangle_a$ to the superposed state $|\alpha/\sqrt{2}\rangle_c |\alpha/\sqrt{2} \rangle_a$ on a time scale of $\pi/4g$, which corresponds to a $\pi/2$ rotation of the state on the Bloch sphere, while a pulse length  $\pi/2g$ will transform the state to $|0 \rangle_c |\alpha\rangle_a $, which corresponds to a $\pi$ rotation on the Bloch sphere, as shown in Fig. \ref{logic_gate}(c). Therefore, both $\pi/2$ and $\pi$ pulses can be realized in this hybrid system, based on which the authors further demonstrated the design of more complex logic gates, for example, using Ramsey interference. It is expected that these gate operations can be applied to low-temperature quantum measurements.

Skogvoll and coworkers \cite{Skogvoll2021} theoretically proposed a method for generating 3-qubit GHZ states \cite{GHZarxiv2007} required for quantum error correction \cite{ShorPRA1995} using a single-shot gate operation involving the qubits and a squeezed magnon. Such states have previously been generated using multiple two-qubit gate operations \cite{NeeleyNature2010, DiCarloNature2010} which is slower than that in the single-shot proposal and is limited by qubit decoherence. As introduced in Section \ref{sec_bistable}, bistable and multi-stable states in magnet-cavity systems may be used to design logic gates and memory devices, too.

\section{Conclusions and Outlook}\label{sec_outlook}
In conclusion, we have clarified the scope of quantum magnonics and reviewed the recent developments and fruitful applications in the quantum technologies including quantum memories, high-precision sensors and coherent logic gates.  These developments have witnessed quantum magnonics as the intersection of spintronics, quantum optics and quantum information, which not only makes the magnonic system a promising platform to study quantum physics, but also significantly extends the horizon of traditional spintronics. Despite great success, the progress of this field is far from saturation. For example, quite a few nonclassical states of magnons and their usefulness in quantum technologies remain to be verified in the lab. Thanks to the interdisciplinary nature of this field, it has already attracted the interests of very diverse communities, from spintronics to quantum optics, from magnonics to continuous variable quantum information, from condensed matter physics to astrophysics, from theorists to experimentalists and engineers. Therefore, we optimistically believe that quantum magnonics will continue its rapid developing pace in the next few decades that it will synergize all fields involved. Below, we give an outlook over the potential directions of this field in the future.

\begin{itemize}
  \item \textbf{Experimental verification.} The experimental realization of nonclassical states of magnons and the technical applications of hybrid ``magnon+X'' systems have seen considerable advances. However, there remain some nonclassical magnon states that are yet to be realized, for example, single-magnon states caused by nonlinearities and interference effects, Schr\"{o}dinger cat states, and entangled state of magnons and photons. The utilization of these states for testing basic inequalities in quantum information, as well as their technical applications for quantum computing and quantum communication are open problems. The generation of a spin superfluid state for sustainable spin transport is worth studying to clarify the ongoing debates \cite{LebrunNature2018,YuanSA2018}.

  \item \textbf{Squeezed states.} The squeezing of magnons and excitations of magnets in general, present various paths towards interesting phenomena and novel technologies. Some of these stem from nonequilibrium magnon squeezing~\cite{ZhaoPRL2004,ZhaoPRB2006,BossiniPRB2019,Rameshti2021cavity,ElyasiPRB2020} and thus can borrow understanding from quantum optics directly. The path of equilibrium squeezing~\cite{KamraAPL2020,KamraPRL2016}, however, has a clear niche in magnonic systems and might hold the key to unprecedented phenomena. Hence, theoretical efforts are needed to establish further new protocols that can capitalize on this equilibrium squeezing. The double-edged sword of strong interactions in magnets is that they enable large squeezing but may underlie rapid decoherence. Thus, the role of interactions and dissipation needs to be addressed. This will provide realistic estimates for the stability of quantum information encoded in magnonic systems. On the experimental side, in the short term, existing spectroscopic tools could be developed further towards detecting and confirming the nonclassical features of magnons~\cite{FlebusPRL2018,DohertyPR2013,TaylorNatPhys2008,AgarwalPRB2017}. For example, the detection of entanglement between different spins will likely require employing external qubit(s). Spin current correlations~\cite{KamraPRL2016,KamraPRB2017,MatsuoPRL2018,AftergoodPRB2018,AftergoodPRB2019,NakataPRB2019} offer another path where recent experiments show promise~\cite{KamraPRB2014,RumyantsevAPL2019}. In the longer term, the focus of experiments is expected to shift from detection to exploitation. We hope that  ongoing and future theoretical studies will pave the way for many feasible exploitation strategies.

  \item \textbf{Magnonic devices for quantum information processing.} Hybrid magnon-phonon systems are promising for precision metrology of temperature, small magnetic fields, and small mechanical displacements. Even though some of the proposals have been realized in the lab, the state-of-art design and test of these devices needs further investigation. It would also be meaningful to push existing proposals, such as microwave transducers and gradient memories, to the quantum limit, and see whether new physics will emerge. Several groups have already reported on the anomalous increase and saturation of magnon damping with decreasing temperature below 1 K \cite{TabuchiPRL2014,KosenAPL2019,PfirPRR2019}. The authors attributed this phenomenon to the coupling of the magnetic system to an ensemble of two-level atoms, though the physical interpretation of such a bath and its implication for magnetization dynamics remain unclear.
  \item \textbf{Classical analogue of quantum phenomena.} Searching for classical analogues of quantum phenomena or macroscopic quantum phenomena in a magnonic system is another ongoing direction.  This direction will extend the horizon of magnon spintronics and promote it as a diverse and reliable platform. For example, stimulated Raman adiabatic passage is a population transfer process between two quantum states through the assistance of a third quantum state, which is useful when the direct population transfer between the two states is forbidden \cite{BergRMP1998}. Wang et al. demonstrated a magnonic state transfer between two waveguides with the help of a third waveguide \cite{WangAPL2021}. As another example, Klein tunneling (also known as Klein paradox) \cite{KleinZP1929} is an important phenomenon of relativistic particles in quantum physics, but its experimental verification using fundamental particles is nearly impossible because of the extremely high electric fields required. Recently, Harms et al. \cite{JorenarXiv2021} showed that this tunneling can be realized in a driven-dissipative magnonic system, where magnon-antimagnon pairs were produced by delicately designing the balance of driving through spin-orbit torque and intrinsic dissipation of magnons. Yet another example is the recent proposal for quantum computation with magnonic BECs \cite{mohseni2021magnon}.
  \item  \textbf{Noncollinear structures based quantum magnonics.} Until now, most of the existing studies in quantum magnonics focus on magnon excitations in a collinear magnetic structure (either the ferromagnetic state or the antiferromagnetic N\'{e}el state). Magnonics based on noncollinear magnetic structures or spin textures including domain walls, skyrmions, vortices, spirals, hopfions, etc. is itself an interesting field, but most of the studies are still in the classical regime \cite{YuPR2021} while the exploration of their quantum aspects has just began \cite{PsaPRX2017,PsaPRL2020,PsaPRL2021}. It would be interesting to study the nonclassical states of magnons excited in these spin textures and further couple them to other quantum platforms such as qubits and photonic cavities. Recently, magnetic skyrmions have been proposed to serve as a new class of quantum logic qubits \cite{PsaPRL2021}, where qubit information is encoded in the helicity of skyrmions and a single-magnon sensor is required to read out the qubit states. Liensberger et al. \cite{LiensPRB2021} experimentally demonstrated the dispersive coupling between magnons and photons in a chiral magnet with skyrmions, where the coupling rate can be effectively tuned by a magnetic field at the phase boundaries of the skyrmion lattice state. Proskurin et al. \cite{ProskPRB2018,IyaroPRB2021} used the oscillating mode of a pinned domain wall as an effective mechanical subsystem, which mimics the mechanical oscillator from cavity optomechanics. Graf et al. \cite{GrafPRB2018} studied the coupling of a magnetic vortex to an optical whispering gallery mode and Mart\'{i}nez-P\'{e}rez et al. \cite{MariaACS2019} show theoretically that a single photon can be coupled to a gyrating vortex.
\end{itemize}

\section*{Declaration of competing interest}
The authors declare no competing financial interests that could have appeared to influence the work reported in this paper.
\addcontentsline{toc}{section}{Acknowledgments}
\section*{Acknowledgments}
H.Y.Y acknowledges the European Union's Horizon 2020 research and innovation programme under Marie Sk{\l}odowska-Curie Grant Agreement SPINCAT No. 101018193. Y.C. was supported by the National Natural Science Foundation of China (NSFC) (Grant No. 11704060). A.K. acknowledges financial support from the Spanish Ministry for Science and Innovation -- AEI Grant CEX2018-000805-M (through the ``Maria de Maeztu'' Programme for Units of Excellence in R\&D). R.A.D. is member of the D-ITP consortium, a program of the Netherlands Organisation for Scientific
Research (NWO) that is funded by the Dutch Ministry of Education, Culture and Science (OCW). R.A.D. has received funding from the European Research Council (ERC) under the European Union's Horizon 2020 research and innovation programme (Grant No. 725509).  P.Y. was funded by the NSFC under Grants No. 12074057 and No. 11604041.
\bibliographystyle{elsarticle-num}
\bibliography{cas-refs}

\end{document}